\documentclass[11pt]{article}
\usepackage[utf8]{inputenc}
\usepackage{amssymb,amsmath}
\usepackage{bm} %math bold symbols
\usepackage{booktabs} %some alignment stuff
\usepackage{array}
\usepackage{latexsym}
\usepackage{graphicx}
\usepackage{color}
\usepackage{datetime}
\usepackage[nosort]{cite}
\usepackage{verbatim}
\usepackage{enumerate}
\usepackage{chngpage} % allows for temporary adjustment of side margins
\usepackage{mathrsfs}
\usepackage{euscript}
\usepackage{psfrag}
\usepackage{subcaption}
\usepackage{accents}
\usepackage{cancel}

\usepackage{soul}

\usepackage{datetime}

\usepackage[nosort]{cite}
\usepackage{chngpage} % allows for temporary adjustment of side margins
\usepackage{setspace}
\usepackage{tensor}
\usepackage{physics}

\usepackage{tikz}
\usetikzlibrary{decorations.text}
\usetikzlibrary{calc}

\usepackage{mciteplus}

\usepackage[colorlinks=true,      linkcolor=blue,      urlcolor=blue,      
            filecolor=blue,      citecolor=blue,       pdfstartview=FitH,     
						pdfpagemode=UseNone,      bookmarksopen=true]{hyperref}  % LINKS
\usepackage[all]{hypcap}     %should be loaded AFTER hyperref, which should otherwise be loaded last.

\definecolor{cardinal}{rgb}{0.6,0,0}
\definecolor{darkgreen}{rgb}{0,0.4,0}
\definecolor{golden}{rgb}{0.92, 0.7, 0}
\definecolor{midnight}{rgb}{0, 0, 0.5}
\definecolor{darkblue}{rgb}{0, 0, 0.7}
\definecolor{purple}{rgb}{0.5, 0, 0.5}

\def\coeff#1#2{\relax{\textstyle \frac{#1}{#2}}\displaystyle}

\def\IR{\mathbb{R}}

\def\IT{\mathbb{T}}

\def\ZZ{\mathbb{Z}}

\def\cB{{\cal B}}

\def\cD{{\cal D}}
\def\cF{{\cal F}}
\def\cG{{\cal G}}
\def\cH{{\cal H}}
\def\cJ{{\cal J}}

\def\cP{{\cal P}}

\def\cS{{\cal S}}

\numberwithin{equation}{section} 
\newcommand{\rcite}{\cite}

\def\sl{\text{sl}}
\def\su{\text{su}}

\def\vareps{\varepsilon}

\def\sltwo{\ensuremath{SL(2,\bR)}}

\def\sutwo{{SU(2)}}

\def\mhat{m}
\def\nhat{n}

\def\k{\kappa}

\newcommand{\msl}{\mathsf m}

\newcommand{\msu}{\sfm'}
\newcommand{\Msl}{\sfM}
\newcommand{\Msu}{\sfM}
\newcommand{\wsl}{w}
\newcommand{\wsu}{w'}

\newcommand{\bmsl}{\bar{{\mathsf m}}}
\newcommand{\bmsu}{\bar{\sfm}'}
\newcommand{\bMsl}{\bar{\sfM}}
\newcommand{\bMsu}{\bar{\sfM}}

\newcommand{\bwsu}{\bar{w}'}
\newcommand{\jsl}{j}
\newcommand{\jsu}{j'}
\def\tight#1{\! #1 \!}  % tightens annoying spacing in equations

\def\etc{{etc}}

\def\NS{{\sst\rm NS}}

\def\tot{{\rm tot}}

\def\nfive{{n_5}}

\def\sfM{{\mathsf M}}

\def\sfk{{\kappa}}

\def\sfm{{\mathsf m}}		\def\sfn{{\mathsf n}}				
\def\sfq{{\mathsf q}}

\DeclareMathSymbol{\medhatsym}{\mathord}{largesymbols}{"62} % basic symbol

\DeclareMathSymbol{\medtildesym}{\mathord}{largesymbols}{"65}% basic symbol

\newcommand{\rhoo}{\ensuremath{\! \rho \:\! }}

\newcommand{\psig}{\ensuremath{\hat\psi}}
\newcommand{\phig}{\ensuremath{\hat\phi}}

\newcommand{\psie}{\ensuremath{\psi}}
\newcommand{\phie}{\ensuremath{\phi}}

\def\half{\frac12}
\def\coeff#1#2{{\textstyle \frac{#1}{#2}}}
\def\hf{\coeff12}
\def\tr{{\rm Tr}}
\def\One{{\hbox{1\kern-1mm l}}}

\def\barray{\begin{array}}
\def\earray{\end{array}}
\def\be{\begin{equation}}
\def\ee{\end{equation}}
\def\bea{\begin{align}}
\def\eea{\end{align}}
\def\nn{\nonumber}

\newcommand{\bR}{{\mathbb R}}
\newcommand{\bS}{{\mathbb S}}
\newcommand{\bT}{{\mathbb T}}

\newcommand{\bZ}{{\mathbb Z}}

\def\IR{\mathbb{R}}

\def\IT{\mathbb{T}}

\numberwithin{equation}{section}

\mathchardef\mhyphen="2D

 \def\cB{\mathcal {B}} 
\def\cD{\mathcal {D}} \def\cE{\mathcal {E}} \def\cF{\mathcal {F}}
\def\cG{\mathcal {G}} \def\cH{\mathcal {H}} 
\def\cJ{\mathcal {J}}  
  \def\cO{\mathcal {O}}
\def\cP{\mathcal {P}}  
\def\cS{\mathcal {S}}

\def\one{{\hbox{\kern+.5mm 1\kern-.8mm l}}}
\def\zero{{\hbox{0\kern-1.5mm 0}}}

\def\id{{1 \kern-.28em {\rm l}}}

\def\etc{{\it etc}}

\def\sst{\scriptscriptstyle}

\def\coeff#1#2{{\textstyle{\frac{#1}{ #2}}}}
\def\half{\frac12}
\def\hf{{\textstyle\half}}

\def\One{{1\hskip -3pt {\rm l}}}

\begin{document}

\newcommand{\Avg}[2]{\ensuremath{ \big \langle  #1 \big \rangle _{#2}}}

\phantom{AAA}
\vspace{-15mm}

%\begin{flushright}
%
%IPHT-T19/021\\
%
%\end{flushright}
% Report number
\begin{flushright}
YITP-25-125\\
\end{flushright}

\vspace{12mm}

\begin{center}

{\huge {\bf  Effective Microstructure}}\\

{\huge {\bf \vspace*{.25cm}  }}

\vspace{8mm}

{\large{\bf { Iosif Bena$^{1}$,~ Rapha\"el Dulac$^{1}$,~Emil J. Martinec$^{2}$,~Masaki  Shigemori$^{3,4}$, \\ \vskip  7pt  David Turton$^{5}$ and  Nicholas P. Warner$^{1,6,7}$}}}

\vspace{0.7cm}

$^1$Institut de Physique Th\'eorique,  
Universit\'e Paris Saclay, CEA, CNRS,\\
Orme des Merisiers, Gif sur Yvette, 91191 CEDEX, France \\[6 pt]

$^2$Enrico Fermi Institute\ and Department\ of Physics, \\
University of Chicago,  5640 S. Ellis Ave.,
Chicago, IL 60637-1433, USA\\[6 pt]
 
 $^3$Department of Physics, Nagoya University, \\
Furo-cho, Chikusa-ku, Nagoya 464-8602, Japan\\[6 pt]

$^4$\,Center for Gravitational Physics,\\
Yukawa Institute for Theoretical Physics, Kyoto University\\
Kitashirakawa Oiwakecho, Sakyo-ku, Kyoto 606-8502, Japan\\[6pt]

$^5$Mathematical Sciences and STAG Research Centre,\\
University of Southampton, 
Highfield, Southampton SO17 1BJ, UK\\[6 pt]
 
$^6$Department of Physics and Astronomy and $^7$Department of Mathematics, \\
University of Southern California, Los Angeles, CA 90089, USA

\vspace{5mm} 
{\footnotesize\upshape\ttfamily iosif.bena @ ipht.fr, raphael.dulac @ ipht.fr, e-martinec @ uchicago.edu,   \\ masaki.shigemori @ nagoya-u.jp, 
d.j.turton @ soton.ac.uk,  warner @ usc.edu} \\

\vspace{1.0cm}

\thispagestyle{empty}
 
\textsc{Abstract}

\end{center}

\vspace{1mm}
\baselineskip=14.5pt

\noindent

\noindent In AdS$_3$/CFT$_2$ duality, there are large families of smooth, horizonless microstate geometries that correspond to heavy pure states of the dual CFT\@. The metric and fluxes are complicated functions of up to five coordinates. There are also many duals of heavy pure states that cannot be described in supergravity, but only admit a worldsheet description.  Extracting the physical properties of these solutions is technically challenging.
In this paper, we show that there are much simpler \emph{effective descriptions} of these solutions that capture many of their stringy and geometrical features, at the price of sacrificing supergravity smoothness.
In particular, the effective description of some families of superstrata, and of certain worldsheet solutions, is given by easy-to-construct three-center solutions. For example, the effective description of a superstratum with a long AdS$_2$ throat is  a scaling, three-center solution in which the  momentum wave is  collapsed to a singular source at one of the three centers. This also highlights how momentum migrates away from the supertube locus in the back-reacted  geometry. 
Our results suggest that effective descriptions can be extended to more general microstates, and that many singular multi-center solutions can in fact correspond to effective descriptions of smooth horizonless microstructure.

%%%%%%%%%%%%%%%%%%%%%%%%%%

\vspace{2cm}

%%%%%%%%%%%%%%%%%%%%%%%%%%

%%%%%%%%%%%%%%%%%%%%%%%%%%%%%%%%%%%%%

\setcounter{tocdepth}{2}

\tableofcontents

%%%%%%%%%%%%%%%%%%%%%%%%%%%%%%%%%%%%%

\baselineskip=15pt
\parskip=3pt

%%%%%%%%%%%%%%%%%%%%%%%%%%%%%%%%%%%%%
\section{Introduction}
\label{sec:Intro}
%%%%%%%%%%%%%%%%%%%%%%%%%%%%%%%%%%%%%

The fuzzball paradigm calls for replacing the classical black hole of General Relativity with a non-singular, strongly-coupled quantum system, in which no region is causally disconnected from any other; this system is postulated to emerge as matter collapses to the horizon scale of a would-be black hole (for a review, see \cite{Bena:2022rna}).  There are many arguments in favor of this paradigm, not the least of which is that it would resolve the information paradox \cite{Mathur:2009hf}.  String theory has also led to notable advances towards resolving this paradox and, more broadly, to the holographic description of quantum gravity in terms of strongly coupled quantum field theories. Thus string theory is the ideal framework for implementing and describing fuzzballs. However, it is extremely challenging to put detailed computational flesh onto the generic stringy fuzzballs that are proposed to replace black holes having macroscopic horizons. Thus was born the microstate geometry programme \cite{Bena:2007kg}.

At the most basic level, microstate geometries can be viewed as highly coherent states of a fuzzball that can be given a geometric description in a low energy limit of string theory, namely supergravity.  These geometries are thus sourced by the same constituents as the corresponding black hole, and are required to be horizonless and smooth (except for possible physical singularities, such as orbifold singularities), and to closely approximate the black-hole solution, both in terms of charges and exterior geometry.  They are thus smooth, horizonless, solitonic solutions of supergravity whose sources are confined to a region, at high red-shift, with surface area very close to that of the black-hole solution. 

From the fuzzball perspective, General Relativity (GR), despite its incredible successes, is a rather blunt effective field theory that breaks down at the horizon scale.  All the attendant no-hair theorems simply inform us that microstructure is completely invisible to GR\@.  Higher-dimensional supergravity is also a low-energy effective field theory but, with its greater number of degrees of freedom, it is far more successful at resolving at least some of the  ingredients of 
black-hole microstructure.  One of the imperatives of the microstate geometry program is to determine the range of states that can be resolved by supergravity.  This task, however, will not be the focus of this paper (see \cite{Bena:2025pcy} for a recent  discussion).  

Our goal here is somewhat in the opposite direction:  The current state of the art has resulted in extremely complex microstate geometries
\cite{
Bena:2015bea,
Bena:2016agb,
Bena:2016ypk,
Bena:2017xbt,
Bakhshaei:2018vux,
Ceplak:2018pws,
Heidmann:2019zws,Ganchev:2022exf, 
Ceplak:2022pep,Ganchev:2023sth, 
Ceplak:2024dbj}
with a great deal of structure that can be probed and verified by precision holography \cite{Kanitscheider:2006zf,Giusto:2015dfa,Giusto:2019qig,Rawash:2021pik,Ganchev:2021ewa}.  However, for some purposes, this level of complexity is unnecessary and a simpler effective description of the microstructure might suffice to reveal the essential physics of microstructure. 

Black-hole microstructure has also been very successfully explored at the quantum level using string worldsheet methods in relatively simple (pure NS sector) backgrounds \cite{Martinec:2017ztd,Martinec:2018nco,Martinec:2019wzw,Martinec:2020gkv,Bufalini:2021ndn,Bufalini:2022wyp,Bufalini:2022wzu,Martinec:2022okx,Martinec:2025xoy}.  While much of this work has been intrinsically perturbative around two-charge and certain special three-charge backgrounds, it has given us remarkable insights into the phase transitions that are expected to lead to the complex quantum state of a typical fuzzball. In the context of microstates involving NS5-branes, the exact worldsheet constructions show that the throat sourced by the fivebrane caps off at the scale of the fivebrane separation, in a geometry seen as smooth by perturbative string theory (non-perturbatively in $\alpha'$). See~\rcite{Martinec:2024emf} for a recent discussion.  While these results arise from a study of highly coherent backgrounds that make an exact solution possible, an effective description suggests that these features persist for generic two-charge states~\rcite{Martinec:2023xvf,Martinec:2023gte,Martinec:2024emf} as well as a large class of three-charge microstates~\rcite{MZ4} (in a construction that in part builds on and complements our analysis here).  In all of these examples, precisely when the throat becomes deep enough that a horizon would naively begin to form in classical supergravity, non-abelian brane excitations become light enough to compete with and overwhelm supergravity excitations (see also~\rcite{Martinec:2015pfa}).

Thus our goal in the present work is to take a step back from more detailed explorations of microstructure and find effective ways to reveal some of the physics of microstructure in simpler, more universal terms.   In this way, we will also make direct connections between results from world-sheet and supergravity techniques.

We will work with two-charge and three-charge supersymmetric geometries in five and six dimensions, and we will primarily consider solutions with D1, D5 charges, $Q_1,Q_5$ (or the S-dual solutions based on F1, NS5), and a momentum charge, $Q_P$.  The corresponding classical supersymmetric black hole geometry has a horizon area proportional to $\sqrt{Q_1Q_5Q_P-J_L^2\vphantom{|}}$, where $J_L$ is an angular momentum. The black hole solution has an AdS$_2$ throat of infinite proper length, and an infinite redshift as one approaches the horizon.   The corresponding three-charge microstate geometries come in two broad, overlapping varieties:
\begin{itemize}
\item{Deep, or scaling, microstate geometries that have very long AdS$_2$ throats that ultimately cap  off smoothly at a finite but usually very large red-shift. In five dimensions the throat geometry is ${\rm AdS}_2\times S^3$, and in six dimensions the throat geometry is  capped ${\rm extremal~BTZ}\times S^3$.  The microstructure is concentrated in the cap, or at its boundary, at the bottom of the AdS$_2$ throat.  This means that such geometries represent tightly bound states of the system~\rcite{Bena:2006kb,Bena:2007qc,Bena:2018bbd}.} 
 \item{Shallow microstate geometries that have either short, or non-existent, AdS$_2$ throats and cap off at small redshift. Some of these correspond to black holes with small horizon areas as a result of either at least one small charge \cite{Sen:2009bm,Mathur:2018tib}, or very large angular momentum \cite{Bena:2006is, Bena:2006kb}.   
 }
\end{itemize}

In six dimensions one can choose the microstate geometries to be asymptotic (at infinity) to either flat space (with at least one spatial dimension compactified to a circle), or to AdS$_3$ $\times S^3$.  The latter boundary conditions mean that one can apply the holography of the D1-D5 CFT, and try to identify the CFT states dual to these microstate geometries.

The five-dimensional microstate geometries were first constructed 20 years ago \cite{Bena:2005va,Berglund:2005vb,Bena:2007kg}. These solutions replaced the singular charge sources by smooth magnetic cohomological fluxes on 2-cycles, or ``bubbles.'' These bubbles are (orbifolds of) two-spheres created by a $U(1)$ fibration over curves between ``centers'' in an $\IR^3$ base at which the $U(1)$ fiber pinches off. For $n$ bubbles there are $(n+1)$ centers and the locations of centers in $\IR^3$ are defined by $3(n+1)$ vector components, which reduce to $3n$ components once one has removed overall translations.

An essential part of these five-dimensional solutions are the ``Bubble Equations''  \cite{Bena:2005va,Berglund:2005vb, Bena:2007kg} (which are equivalent to  the ``integrability conditions'' of four-dimensional multi-center solutions \cite{Denef:2000nb, Denef:2002ru,Bates:2003vx, Denef:2007vg}).   These equations impose constraints on the relative distances between the centers, and are  required in order to preserve supersymmetry and avoid closed time-like curves. For a selection of state-of-the art constructions, see~\cite{Bena:2017fvm,Avila:2017pwi,Rawash:2022sum}.

Furthermore, one can show that when reducing a five-dimensional smooth solution to four dimensions, each center becomes a D-brane with Abelian world-volume fluxes, which locally preserves 16~supercharges \cite{Balasubramanian:2006gi}.

The five-dimensional solutions are relatively simple, and even though they can have large moduli spaces, they encode rather little microstructure.  
If we set all the charges to be of order $Q$, such that the black-hole entropy scales as $\sim Q^{3/2}$, the five-dimensional solutions account for an entropy at most of order  $\sim Q^{1}$~\cite{deBoer:2009un}.
However, as we will discuss, these geometries enable us to exhibit some interesting, universal aspects of microstructure physics. The six-dimensional solutions, and stringy probes, are more intricate and encode a much higher level of detail.  In particular, six-dimensional microstate geometries can have an entropy of order, at most, $\sim Q^{5/4}$ \cite{Bena:2010gg,Shigemori:2019orj,Mayerson:2020acj}.

Superstrata 
are the most analyzed and best understood families of six-dimensional microstate geometries~\cite{
Bena:2015bea,
Bena:2016agb,
Bena:2016ypk,
Bena:2017xbt,
Bakhshaei:2018vux,
Ceplak:2018pws,
Heidmann:2019zws,Ganchev:2022exf, 
Ceplak:2022pep,Ganchev:2023sth, 
Ceplak:2024dbj,
deLange:2015gca,
Mayerson:2020tcl,
Ganchev:2021iwy}. They start from a circular two-charge supertube solution~\cite{Balasubramanian:2000rt,Maldacena:2000dr,Mateos:2001qs,Emparan:2001ux,Lunin:2001jy,Lunin:2002iz}, which, from a four-dimensional perspective, may be viewed as a singular two-center solution: One center is just the center of space around which the stationary supertube circulates, and the other center is the supertube locus. When seen in five-dimensional supergravity, the center of space becomes smooth, while the supertube center corresponds to a circular singular object -- the supertube itself. In six-dimensional supergravity\footnote{Corresponding to a ten-dimensional duality frame in which the supertube has D1 and D5 (or F1-NS5) charges and KKM dipole charge.}, the supertube locus becomes smooth as well, and the full solution has two $U(1)$ isometries, one pinching off at the center of the space and the other at the supertube locus, giving rise to a topological $S^3$ (or orbifold thereof). Superstrata are fully back-reacted geometries created by particular families of supersymmetric 
momentum waves traveling around this supertube.

If one puts a momentum wave along a circle at the location where this circle pinches off, then the resulting geometry is singular. Hence, one expects this wave to move away from the supertube locus, and become a shape mode on the non-trivial $S^3$ cycle \cite{Niehoff:2013kia}, which is exactly what happens in the superstratum solutions \cite{Bena:2015bea}.

Precision holography shows that this harmonic wave on $S^3$ is indeed the dual  of the original supertube momentum wave.  As we will discuss in this paper, in the back-reacted geometry the momentum wave ``migrates'' off the supertube locus and, at high frequencies, it once again localizes at a particular position in the geometry. We will show that this position is exactly described by using an effective five-dimensional  description in which the localized momentum wave is treated as a (singular) primitive center, whose position can be determined by a five-dimensional bubble equation.

Indeed, we will show how this momentum migration appears in multiple approaches to probing  black-hole microstructure.  
For shallow microstate geometries, high-frequency wavefunctions of string probes localize at a position determined by the quantum numbers of the state.  As one might expect from WKB, or geometric-optics approximations, this localization is also evident from geodesic motion.  Indeed, some simple geodesic analysis of superstrata can be found in \cite{Guo:2024pvv}.  One of the nice aspects of using string probes is that one can use probes that also carry perturbative F1 charge as well as momentum charge. 
Once again, the location of the excitation can be determined from the appropriately modified bubble equations.

\begin{figure}
\centering
\includegraphics[scale=0.3]{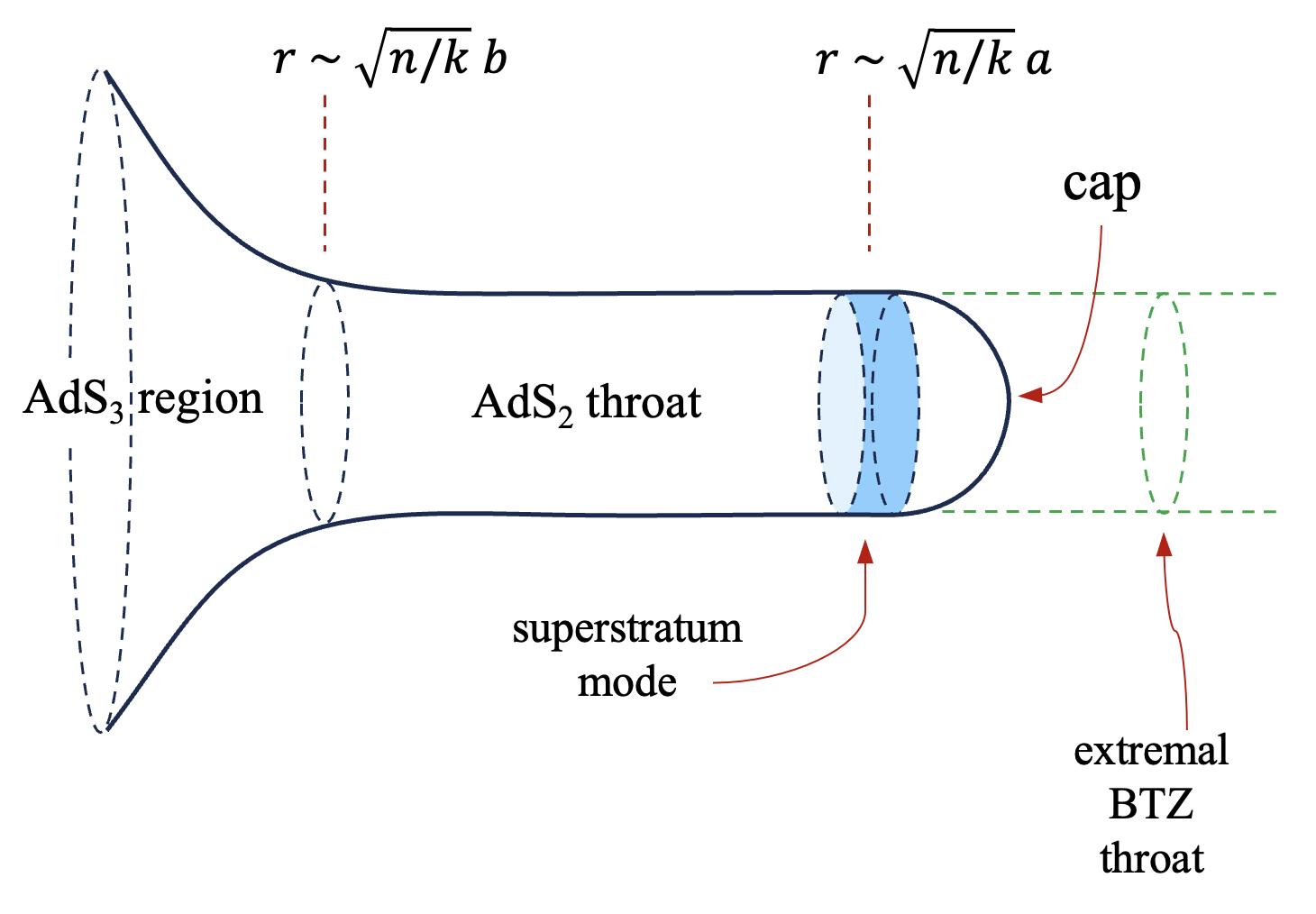}
\caption{\it Single-mode superstratum features.
Here $r$ is a radial coordinate in the $\bR^4$ transverse to the brane sources, while $a$ is the radius of the underlying two-charge (D1-D5) supertube, upon which the superstratum is built by adding a momentum wave.  The latter is supported at the scale $\boldsymbol{\sqrt{n/k}\,a}$ and has charge radius $\boldsymbol{\sqrt{n/k}\,b}$. The quantum numbers $n$ and $k$ are explained in sections \ref{sec:SingleModeSS}, \ref{sec:deep_AdS2} and \ref{sec:BHregime}.
}
\label{fig:supertstratum}
\end{figure}

One of the interesting aspects of this localization story is how it works for the complete range of throat depths in superstrata. One can start with the two centers that correspond to the circular supertube solution which, in the AdS$_3$ decoupling limit, is a rotating global AdS$_3\times S^3$ solution. Adding small amounts of a momentum wave at high enough frequency localizes the momentum wave in the AdS$_3\times S^3$.  Increasing  and back-reacting the momentum charge causes a capped AdS$_2$ $\times S_1$ (or capped BTZ) throat to open up.  The diameter of the $S^1$ is set and centrifugally stabilized by the momentum charge carried by the wave. 
Crudely, the geometry resembles a global AdS$_3$ with a long AdS$_2\times S_1$ region inserted into it, like a plumbing fixture, as follows. From the original global AdS$_3$, imagine that a ``bowl,'' or cap, is cut out.  A long, vertical pipe (the AdS$_2\times S_1$) is glued into the hole, and the cap that has been cut out is now glued to the bottom of this long pipe.  This is the capped BTZ geometry; see Fig.~\ref{fig:supertstratum}. 
The boundary between the cap and the lower end of the AdS$_2\times S^1$ throat is defined by the location of the momentum wave (or by the location of the supertube center, whichever is larger).

%%%%%%%%%%%%%%%%%%%%%%%%%%%%%%%%
%%%%%%%%%%%%%%%%%%%%%%%%%%%%%%%%

In this work we shall demonstrate that in both shallow and  deep scaling superstrata, the locus of the momentum wave is determined \emph{relative to the cap} by bubble equations that are adapted to describe singular momentum-carrying centers.
Our ``effective geometries'' technique thus provides a powerful tool that reveals the features of smooth horizonless solutions without having to handle the details of the exact geometry.   The price one pays is to replace some of the detailed microstructure with an averaged, effective but singular source.  The result is a geometry that reliably captures the location of the source, and yields a background that is reliable on scales larger than the details that have been averaged over.

In Section \ref{sec:Sugra solns} we provide an extended review of the salient details of five-dimensional and six-dimensional microstate geometries.   We apply this technology first in Section~\ref{sec:EffSS} to obtain ``averaged superstrata,'' where one of the centers gives the location of the massless particle, which is the WKB limit of the wave excitation.  In Section \ref{sec:wavefns}, we review geodesic motion and wavefunctions of string probes in backgrounds that can be described by exact world-sheet methods, and show how to use worldsheet spectral flow symmetry to relate particle geodesics to winding string worldsheets.  Then, in Section~\ref{sec:D1-P}, we apply the knowledge of wavefunctions and geodesics to the description of pure momentum, D1-P, or F1-P probes of two-charge circular supertubes and three-charge supersymmetric spectral flowed circular supertube backgrounds~\cite{Lunin:2004uu,Giusto:2004id,Giusto:2004ip,Giusto:2012yz}, where the bubble equations determine the location of the probe. Completing the circle of ideas, in Section~\ref{sec:wound-strings} we match the latter result to a worldsheet analysis of F1-P probes in the S-dual F1-NS5-P backgrounds, where the solution to the bubble equations is reproduced by an analysis of the worldsheet physical state constraints, using the exact worldsheet formulation of these backgrounds in terms of gauged Wess-Zumino-Witten (WZW) models~\rcite{Martinec:2017ztd,Martinec:2018nco,Martinec:2019wzw,Martinec:2020gkv,Bufalini:2021ndn}. This shows that the fundamental string vertex operators  have center-of-mass wavefunctions related to the linearized momentum excitations of supergravitons, thus tying back to the superstratum discussion that begins our analysis. 
Section~\ref{sec:BHregime} contains an application of our effective geometry framework: we calculate the depth of the AdS$_2$ throat and compare the result to other microstates, and a recent analysis of extremal black holes in Jackiw-Teitelboim (JT)  gravity~\rcite{Lin:2022rzw}.
Section \ref{sec:Conclusion} contains concluding remarks. 
%

%%%%%%%%%%%%%%%%%%%%%%%%%%%%%%%%%%%%%
\section{Supergravity BPS equations and solutions}
\label{sec:Sugra solns}
%%%%%%%%%%%%%%%%%%%%%%%%%%%%%%%%%%%%%

As outlined in the Introduction, the story of microstate geometries started in five dimensions by creating smooth structures that had rather limited phase spaces.  This work progressed to ever more sophisticated, six-dimensional solutions, with much larger phase spaces, that could be tested and probed with precision holography.   Despite the limitations of the five-dimensional solutions, they contain  important information about the essential physics of their more sophisticated, higher-dimensional counterparts.  We therefore start this review section with five-dimensional supersymmetric solutions.

%%%%%%%%%%%%%%%%%%%%%%%%%%
 %%%%%%%%%%%%%%%%%%%
\subsection{Five-dimensional BPS equations and solutions}
\label{ss:5dsols}
%%%%%%%%%%%%%%%%%%% 
%%%%%%%%%%%%%%%%%%%%%%%%%%

The five-dimensional BPS solutions have a metric of the local form \cite{Gauntlett:2002nw, Gutowski:2004yv,Bena:2004de,Giusto:2012gt}:
\begin{equation}
ds_5^2 ~=~  - Z^{-2} \, (dt +  {\bf k})^2 ~+~ Z  \, ds_4^2(\cB)\,,
\label{fivemet-1}
\end{equation}
where $\cB$ is some four-dimensional (possibly ambi-polar\footnote{As discussed in, for example,  \rcite{Bena:2005va,Berglund:2005vb, Bena:2007kg, Bena:2015bea,Bena:2017xbt}, the signature is allowed to change from $+4$ to $-4$ on hypersurfaces.}) hyper-K\"ahler base space.

The five-dimensional electromagnetic fields have potentials:
\begin{equation}
\tilde A^I ~=~ - \coeff{1}{2} \,  Z^{-3} C^{IJK} \,  Z_J \,  Z_K \, (dt +  {\bf k})  ~+~\tilde B^I \,, 
\label{fivepots-1}
\end{equation}
where the $Z_I$ are scalars and  $C^{IJK}$ is the totally symmetric tensor that encodes the couplings of the gauge fields.  Note that we have introduced tildes on five-dimensional electromagnetic fields because, as discussed in \cite{Bena:2016agb} the normalizations of these fields  are slightly different in five and six dimensions \eqref{5D6Dtheta}. 

To get the correctly normalized five-dimensional BPS equations, we use the approach of \cite{Bena:2015bea} and introduce five vector fields whose only non-zero structure constants are:  
\begin{equation}
C_{3JK}  ~\equiv~  \widehat C_{JK}  ~=~  \widehat C_{JK} \,, \qquad J, K \in \{1,2,4,5\}
\label{RecC}
\end{equation}
where
\begin{equation}
\widehat C_{JK}  ~=~  
\begin{pmatrix} 
0&1&0&0\\  1&0&0&0\\ 0&0&-1&0\\ 0&0&0&-1
\end{pmatrix}   \,.
\label{hatCform}
\end{equation}
We will ultimately reduce to four vector multiplets by setting  $A^{(5)} = A^{(4)}$  as in \cite{Giusto:2012gt}, however when applying formulae involving structure constants we will apply them using the full index range $I,J,K=1,\dots,5$, and then set all quantities with index $5$ equal to their counterparts with index $4$.

The cubic invariant, which becomes the warp factor in (\ref{fivemet-1}), is  given by: 
\begin{equation}
Z  ~\equiv~  \Big(\coeff{1}{6} \,C^{IJK} Z_I\, Z_J\, Z_K  \Big)^{\frac{1}{3}} ~=~ 
\Big(\big( Z_1 Z_2 -  \coeff{1}{2}\, Z_4^2-  \coeff{1}{2}\, Z_5^2\big)  Z_3\Big)^{\frac{1}{3}}~=~ 
\Big(\big( Z_1 Z_2 -  Z_4^2\big)  Z_3\Big)^{\frac{1}{3}}\,.
\label{tZdefn}
\end{equation}

To write the BPS equations in their canonical linear form (see, for example, \cite{Bena:2004de,Giusto:2012gt}), one introduces magnetic field strengths:
\begin{equation}
 \widetilde \Theta^I ~=~  d \tilde B^I \,. 
\label{fiveThetas-1}
\end{equation}
and then one has
\begin{equation}
\widetilde \Theta^I   ~=~  \star_4 \, \widetilde\Theta^I  \,, \qquad
 \nabla^2  Z_I ~=~  {\half}  \,  C_{IJK} \star_4 ( \, \widetilde\Theta^J  \wedge
\widetilde \Theta^K)  \,, \qquad
 d{\bf k} ~+~  \star_4 d{\bf k} ~=~  Z_I \,  \widetilde \Theta^I \,.
\label{5dBPSeqn}
\end{equation}
More explicitly, the five-dimensional BPS equations are: 
\begin{equation}
\begin{aligned}
\widetilde \Theta^I   & ~=~  \star_4 \, \widetilde\Theta^I  \,, \qquad
 \nabla^2  Z_1 ~=~  \star_4\, (  \widetilde\Theta^2  \wedge
\widetilde \Theta^3)  \,, \qquad \nabla^2  Z_2 ~=~  \star_4 (  \widetilde\Theta^1  \wedge
\widetilde \Theta^3)\,, \\
 \nabla^2   Z_3 & ~=~  \star_4\, (  \widetilde\Theta^1  \wedge
\widetilde \Theta^2 - \coeff{1}{2} \,  \widetilde\Theta^4  \wedge
\widetilde \Theta^4 - \coeff{1}{2} \,  \widetilde\Theta^5  \wedge
\widetilde \Theta^5)~=~  \star_4\, (  \widetilde\Theta^1  \wedge
\widetilde \Theta^2 - \widetilde\Theta^4  \wedge
\widetilde \Theta^4)  \,, \\
 \nabla^2    Z_4 & ~=~  -  \, \star_4\, (  \widetilde\Theta^3  \wedge
\widetilde \Theta^4) \,, \qquad \nabla^2    Z_5  ~=~  -  \, \star_4\, (  \widetilde\Theta^3  \wedge
\widetilde \Theta^5)      \,,  \\
 d{\bf k} ~+~  \star_4 d{\bf k} & ~=~  Z_I \,  \widetilde \Theta^I ~=~  Z_1 \,  \widetilde \Theta^1 ~+~ Z_2 \,  \widetilde \Theta^2~+~ Z_3 \,  \widetilde \Theta^3~+~ 2\, Z_4 \,  \widetilde \Theta^4\,.
 \end{aligned}
\label{5dBPSeqn2}
\end{equation}
Note that the equation for $Z_5$ is compatible with setting $Z_5 = Z_4$ and $\Theta^5 = \Theta^4$.  Also observe that this identification leads to the factor of two for the $Z_4 \,  \widetilde \Theta^4$ in the source of the last BPS equation.  This factor will also reappear in the six-dimensional formulation.  

Having obtained these BPS equations, the fifth vector multiplet will henceforth be dropped from our discussion. 

%%%%%%%%%%%%%%%%%%%%%%%%%%
%%%%%%%%%%%%%%%%%%%%%%%%%%
\subsection{Multi-center solutions}

\label{sec:multi-center}
%%%%%%%%%%%%%%%%%%%%%%%%%%
%%%%%%%%%%%%%%%%%%%%%%%%%%

A  very convenient, non-trivial choice for the base-space metric, $ds_4^2(\cB)$, is a multi-center Gibbons-Hawking  (GH) metric:
\begin{equation}
ds_4^2 ~=~ V^{-1} \, (d \psig + A)^2 ~+~   V  \, d \vec x  \cdot  d\vec x \,,  \label{GHmet}
\end{equation}
where $\vec x\in\mathbb{R}^3$ are the Cartesian coordinates of the flat three-dimensional base and $\psig\cong \psig+4\pi$.

The  BPS equations lead to the following solutions \cite{Gauntlett:2004wh,Bena:2005ni}:
\begin{equation}
Z_I ~=~ \coeff{1}{2}  \, C_{IJK} \, V^{-1}\,K^J K^K  ~+~ L_I \,, \qquad {\bf k} ~=~ \mu\, ( d\psig + A   ) ~+~ \varpi
\label{Zkform}
\end{equation}
where the $K^J$ and $L_I$ are harmonic functions on the $\IR^3$ defined by $\vec x$ (now $I,J=1,2,3,4$).  The pieces of the angular momentum vector, ${\bf k}$, are then determined by:
\begin{equation}
\mu ~=~ \frac {1}{6} \, C_{IJK}\,  \frac{K^I K^J K^K}{ V^2} ~+~
\frac{1}{2 \,V} \, K^I L_I ~+~  M\,,
\label{mures}
\end{equation}
and 
\begin{equation}
\begin{aligned}
\vec \nabla \times \vec \varpi &  ~=~ ( V \vec \nabla \mu ~-~
\mu \vec \nabla V ) ~-~ \, V\, \sum_{I}  \,
 Z_I \, \vec \nabla \bigg(\frac{K^I}{V}\bigg) \\ 
 &~=~  V \vec \nabla M ~-~
M \vec \nabla V ~+~   \coeff{1}{2}\, (K^I  \vec\nabla L_I - L_I \vec
\nabla K^I )\,,
\end{aligned}
\label{varpieqn}
\end{equation}
where $M$ is also harmonic on $\IR^3$.

The solution is thus determined by ten harmonic functions, which one typically chooses to have the form:
\begin{equation}
\label{eq:harm-fns}
    \begin{split}
    V &= \sum_i \frac{\hat q^{(i)}}{|\vec x - \vec x^{(i)}|}\,,\qquad  K^I = \sum_i \frac{\hat{\kappa}_I^{(i)}}{|\vec x - \vec x^{(i)}|},\\
    L_I &= l_I^{(0)} + \sum_i \frac{\hat{Q}_I^{(i)}}{|\vec x - \vec x^{(i)}|}\,,\qquad M ={\hat m}^{(0)}+ \sum_i \frac{{\hat m}^{(i)}}{|\vec x - \vec x^{(i)}|}\,.
    \end{split}
\end{equation}
In the D1-D5 frame, $\hat Q_1$, $\hat Q_2$, and $\hat Q_3$ correspond to D1, D5, and momentum charges, respectively, while $\hat Q_4$ correspond to a certain combination of NS5 and F1 charges.
In the F1-NS5 frame, $\hat Q_1$, $\hat Q_2$, and $\hat Q_3$ correspond to F1, NS5, and momentum charges, respectively, while $\hat Q_4$ correspond to a certain combination of D5 and D1 charges.  In both frames, $\hat m$ corresponds to the left-moving angular momentum $\hat J_L$ in five dimensions.
In the standard, bubbled microstate geometries, $\hat q^{(i)}$ and the $\hat{\kappa}_I^{(i)}$ can be chosen at will (but with appropriate quantization), while the $\hat{Q}_I^{(i)}$ and ${\hat m}^{(i)}$ are fixed by smoothness and the absence of closed time-like curves (CTCs).   For effective superstrata we are going to relax the smoothness conditions but we will still require the absence of CTCs.  

These solutions have a set of gauge invariances  in which the functions can be shifted according to:
\begin{eqnarray}
K^I &~\rightarrow~&   K^I ~+~  c^I\, V \,, \nonumber \\
L_I &~\rightarrow~& L_I ~-~ C_{IJK}\,c^J \,K^K ~-~ \coeff{1}{2} \,
C_{IJK}\, c^J\, c^K\, V  \nonumber \,,\\
M &~\rightarrow~&M - \coeff{1}{2} \,c^I \, L_I +\tfrac{1}{12}\, C_{IJK} \left( V \, c^I \, c^J
\, c^K +3\,  c^I\, c^J\, K^K\right) \,,
\label{eq:fullgauge}
\end{eqnarray}
where the $c^I$ are  arbitrary constants.  The fluxes and metric  are invariant under these transformations.

We also recall the fact that the $\hat q^{(i)} \in \ZZ$ are not necessarily positive: negative $\hat q^{(i)}$ create a singular, ``ambi-polar'' base geometry, but, as has been noted elsewhere (see, for example, \cite{Bena:2005va,Berglund:2005vb, Bena:2007kg, Bena:2015bea,Bena:2017xbt}), this can lead to smooth Lorentzian geometries and well-behaved cohomological Maxwell fluxes in five dimensions. 

Based on (\ref{eq:harm-fns}) we introduce the charge vector:
\begin{equation}
    \Gamma^{(i)}=\left(\hat q^{(i)},{\hat\kappa}_I^{(i)},{\hat Q}_I^{(i)},{\hat m}^{(i)}\right)
\label{Gammavec}
\end{equation}
and define the symplectic product:
\begin{equation}
    \label{eq:symplectic product}   \Gamma^{ij}\,=\,\langle\Gamma^{(i)},\Gamma^{(j)}\rangle\, =\,\hat q^{(i)}{\hat m}^{(j)}-\hat q^{(j)}{\hat m}^{(i)}+ \frac{1}{2} \sum\limits_I
    \left({\hat\kappa}_I^{(i)}{\hat Q}_I^{(j)}-{\hat\kappa}_I^{(j)}{\hat Q}_I^{(i)}\right).
\end{equation}
While  superstrata necessarily involve all four Maxwell fields, our examples of effective geometries will only be sensitive to the fundamental brane charges with $I=1,2,3$ and so we will drop the $I=4$ fields from our discussions of charges and their interactions. In particular, this means that from now on, $ C_{IJK} = C^{IJK} = |\varepsilon_{IJK}|$.

We use (\ref{eq:fullgauge}) to impose a gauge choice in which $\sum_j{\hat\kappa}_I^{(j)}=0$ for all $I$. Then the sum of all the bubble equations implies that the constant in the $M$ harmonic function vanishes, ${\hat m}^{(0)}=0$. 
Having done so, we introduce the following vector containing the moduli that are given by the asymptotic values of the harmonic functions in Eq.\;\eqref{eq:harm-fns}:
\begin{equation}
\label{eq:consts-vec}
h=(0,(0,0,0),(l_1^{(0)},l_2^{(0)},l_3^{(0)}),0)\,,
\end{equation}
where we have dropped the $I=4$ fields as described above.

Regularity at each center was analyzed in~\rcite{Bena:2005va,Berglund:2005vb} and requires that
\begin{equation}
    \hat{Q}_I^{(i)}\,=\,-\frac{|\varepsilon_{IJK}|}{2}\frac{\hat{\kappa}_J^{(i)} \hat{\kappa}_K^{(i)}}{q^{(i)}}\,,\qquad 
    \hat{m}^{(i)}=\, 
    \frac{|\varepsilon_{IJK}|}{12}\frac{\hat{\kappa}_I^{(i)}\hat{\kappa}_J^{(i)} \hat{\kappa}_K^{(i)}}{(q^{(i)})^2}\;.
\label{eq:primitivityGH}
\end{equation}
Or, more generally, (because some of the $\hat q^{(i)}$ could be zero), we impose:
\begin{subequations}
\begin{align} 
    {\rm Q6}^{(i)} &\equiv\, \hat q^{(i)} \,=\; {\bf Q}^{(i)} \cos \alpha^{(i)}_1  \cos \alpha^{(i)}_2  \cos \alpha^{(i)}_3 \,,\\
{\rm Q4}^{(i)}_I &\equiv\, \hat{\kappa}_I^{(i)} \,= \,\sum\limits_{J,K}\frac{|\varepsilon_{IJK}|}{2}  {\bf Q}^{(i)} \sin \alpha^{(i)}_I  \cos \alpha^{(i)}_J  \cos \alpha^{(i)}_K \,,\\
{\rm Q2}^{(i)}_I &\equiv\, \hat{Q}_I^{(i)}\,=\,-\sum\limits_{J,K}\frac{|\varepsilon_{IJK}|}{2} {\bf Q}^{(i)} \cos \alpha^{(i)}_I  \sin \alpha^{(i)}_J  \sin \alpha^{(i)}_K \,, \\
-{\rm Q0}^{(i)} &\equiv\, 2 \hat m^{(i)} \,=\;  {\bf Q}^{(i)}\sin\alpha^{(i)}_1  \sin \alpha^{(i)}_2  \sin \alpha^{(i)}_3 \,,
\end{align} \label{eq:primitivity}%
\end{subequations}%
where ${\bf Q}^{(i)}$ and the $\alpha^{(i)}_I$ are to be determined.  If one reduces the solution to Type IIA String Theory along the Gibbons-Hawking fiber, $\psig$, then the charges
Q0$^{(i)}$, Q2$^{(i)}$,
Q4$^{(i)}$ and Q6$^{(i)}$ are the D0, D2, D4 and D6 (Page) charges\footnote{The sign conventions are chosen such that the mass of a four-dimensional supersymmetric black hole is $M^2 = ({\rm Q6}+\sum_I {\rm Q2}_I)^2 +({\rm Q0}+\sum_I {\rm Q4}_I)^2$.} of the center ${(i)}$ .

Supersymmetric centers with these charges preserve sixteen supercharges and are said to be ``primitive''~\cite{Balasubramanian:2006gi}, since they are simple T-duals of ${\bf Q}^{(i)}$ D3 branes that have been first tilted by $\alpha^{(i)}_I,\alpha^{(i)}_J,\alpha^{(i)}_K$, then smeared and then T-dualized along three orthogonal directions.
The primitivity condition is preserved by gauge transformations \eqref{eq:fullgauge} and generalized spectral flow transformations \cite{Bena:2008wt}.  One can also see that supertube centers (discussed in detail in Section \ref{sec:GH-ST}), which give rise to smooth solutions in six-dimensional supergravity but lead to singular solutions in five-dimensional supergravity, are also primitive centers. Furthermore, bound states of one type of D2 branes and D0 branes, as well as D0 branes alone, are also primitive, although the corresponding geometries are singular. One can also generalize the primitivity condition to centers carrying other charges and dipole charges \cite{Bena:2022fzf}, but we shall not use such centers in this paper.

The absence of closed time-like curves requires the bubble or the integrability equations \cite{Bates:2003vx,
Bena:2005va,Berglund:2005vb, Bena:2008dw}:
\begin{equation}
    \sum_{j(\ne i)} 
    \frac{\Gamma^{ij}}{|\vec x ^{(j)}- \vec x^{(i)}|}=\langle h,\Gamma^{(i)}\rangle \,.
\label{eq:bubble-eqns}
\end{equation}
which are independent of the conditions imposing smoothness or primitivity on the centers  \cite{Balasubramanian:2006gi}.
The angular momenta are given by:
\begin{align}
\label{eq:j-non-q}
    {\hat J}_L&=\sum_i {\hat m}^{(i)},\qquad
    \vec{\hat{J}}_R=\sum_i \langle h,\Gamma^{(i)}\rangle \vec{x}_i\,,
    \qquad
    \hat{J}_R \;=\; |\vec{\hat{J}}_R| \,.
\end{align}

Solutions with five-dimensional-flat-space asymptotics have an asymptotic moduli vector: 
\begin{equation}
    h=\left(0,(0,0,0),(1,1,1),0\right)\,,
\end{equation}
and this will be used in  Section \ref{sec:5DeffSS}. In the majority of this paper, we use the AdS$_3$ decoupling limit for which the asymptotic moduli vector is (for a discussion of this, see, for example,~\rcite{Bena:2018bbd}):
\begin{equation}
    h=\left(0,(0,0,0),(0,0,1),0\right)\,.
\end{equation}

Scaling microstate geometries arise when the bubble equations (\ref{eq:bubble-eqns}) admit a solution in which some subset, $\cS$, of the centers can get arbitrarily close:
\begin{equation}
|\vec x ^{(j)}- \vec x^{(i)}| ~\to~ 0 \,, \qquad i,j \in \cS\,,
\label{scalingx}
\end{equation}
while the charges of the solution remain large. This limit appears to be singular in the $\IR^3$ base space of the solution, but in the full geometry it simply corresponds to the opening of a long, macroscopic, capped AdS$_2$ throat, whose cap remains smooth \rcite{Bena:2006kb,Bena:2007qc,Bena:2018bbd,Denef:2007vg}.  

The condition for scaling  is most simply described for the coincidence of three centers.  Suppose that $\cS = \{i,j,k\}$, then the magnitudes of the three fluxes involved must satisfy the triangle inequalities:
\begin{equation}
\big| \Gamma^{ij} \big| ~\le~ \big| \Gamma^{ik} \big|~+~ \big| \Gamma^{jk} \big| \,, \qquad 
\text{and cyclic permutations.} 
\end{equation}
One can satisfy the bubble equations by arranging the $\Gamma^{ij}$ to have the correct signs and taking: 
\begin{equation}
|\vec x ^{(j)}- \vec x^{(i)}|  ~=~  \pm \lambda \,\Gamma^{ij} ~+~ \cO(\lambda^2) \,,\qquad \lambda \to 0  \,.
\label{eq:triangle}
\end{equation}
For the correct signs of $\Gamma^{ij}$, the divergent terms in (\ref{eq:bubble-eqns}) cancel, and finite terms on the right-hand side come from the $\cO(\lambda^2)$ terms in (\ref{eq:triangle}).  

One can also verify that a scaling cluster, $\cS$, gives a vanishingly small contribution to  $\vec{\hat{J}}_R$ in the limit (\ref{eq:triangle}) \rcite{Bena:2007kg,Bena:2007qc}.

%%%%%%%%%%%%%%%%%%%
\subsection{Six-dimensional BPS equations and solutions}
\label{ss:SSgeom}
%%%%%%%%%%%%%%%%%%% 

The structure and equations that govern superstrata have been given in many places (see, for example \rcite{Giusto:2013rxa,Bena:2015bea, Bena:2016agb, Bena:2017xbt}), and are based on extensions of the most general classes of supergravity solutions in six dimensions \cite{Gutowski:2003rg,Cariglia:2004kk,Bena:2011dd}.    Since we are going to be concerned about reducing the effective superstratum to five dimensions, we will follow the discussion in \rcite{Bena:2016agb}.

The superstratum is defined in IIB supergravity compactified to six dimensions on a $\IT^4$, or K3.  The six-dimensional geometry  has a time coordinate, $t$, and a compact $y$-circle with 
\begin{equation}
y ~\sim~ y \,+\, 2 \pi R_y  \,. 
\label{yperiod}
\end{equation}
The remaining four spatial dimensions define the ``base,'' $\cB$, with metric, $ds_4^2$, and coordinates, $x^\mu$.   In one of the  simpler classes of superstrata, the metric, $ds_4^2$,  is required to be independent of $(t,y)$, however it must be hyper-K\"ahler and it is allowed to be ambi-polar.
The six-dimensional metric then takes the form:
\begin{equation}
ds_6^2 ~=~    -\frac{2}{\sqrt{\cP}} \, (dv+\beta) \big(du +  \omega + \tfrac{1}{2}\, \mathcal{F} \, (dv+\beta)\big) 
~+~  \sqrt{\cP} \, ds_4^2(\cB)\,,  \label{sixmet}
\end{equation}
where $(u,v)$ are related to $t$ and $y$ by:
\begin{equation}
u ~\equiv~   \frac{1}{\sqrt{2}}\, (t-y) \,,  \qquad v ~\equiv~    \frac{1}{\sqrt{2}}\, (t+y)  \,. \label{uvdefn}
\end{equation}
The functions $\cF$ and $\cP$, and the one-forms $\beta$ and $\omega$ can depend on $(v,x^\mu)$, but supersymmetry requires that all fields be independent of $u$.
They will be further constrained by the BPS equations.  

For future reference, we write this metric in a form that is adapted to compactification to five dimensions\footnote{In \rcite{Bena:2016agb} this was referred to as ``Reduction 2.''}: 
\begin{equation}
ds_6^2 ~=  -\frac{1}{Z_3\sqrt{\cP} } \, (dt +   {\bf k})^2 \,+\, 
\frac{Z_3}{\sqrt{\cP}}\, \left[dy  +\left(1- Z_3^{-1}\right)  (dt +   {\bf k}) +\frac{\beta-\omega}{\sqrt{2}} \right]^2   +  \sqrt{\cP} \, ds_4^2(\cB)\,,
\label{sixmet-2}
\end{equation}
where
\begin{equation}
\cP ~\equiv~ Z_1 Z_2 - Z_4^2\,, \qquad  Z_3 ~\equiv~ 1- \frac{\cF}{2} \,, \qquad   {\bf k} ~\equiv~ \frac{\omega+\beta}{\sqrt{2}} \,, 
 \label{Z3kdefn}
 \end{equation}

In the simpler class of superstrata, the $2$-form, $d \beta$, is  independent of both $(u,v)$, and must be self-dual: 
 \begin{equation}
d \beta = *_4 d\beta\,,
\label{eqbeta}
 \end{equation}
 where $*_4$ is the Hodge dual on $\cB$.
 
The full six-dimensional solution involves  three independent $3$-form field strengths,  $G^{(I)}$, whose potentials, $B^{(I)}$, are determined\footnote{See \rcite{Bena:2015bea,Giusto:2013rxa, Bena:2017xbt} for more details.} in terms of  electrostatic potentials, $Z_I$ and $2$-forms, $ \Theta^I$ on $\cB$. For historical reasons that will soon become clear, the index $I$ takes the values $1,2,4$. The BPS equations impose the following linear differential equations\footnote{We define the $d$-dimensional Hodge star $*_d$ acting on a $p$-form to be
$$
 *_d\, (dx^{m_1}\wedge\cdots\wedge dx^{m_p})
 ~=~  \frac{1}{(d-p)!} \, dx^{n_1}\wedge\cdots\wedge dx^{n_{d-p}}\,  \epsilon_{n_1\dots n_{d-p}}{}^{m_1\dots m_p} \,,
$$
where we use the orientation $\epsilon^{+-1234} \equiv \epsilon^{vu1234} = \epsilon^{1234} = 1$. These are the conventions used in~\rcite{Gutowski:2003rg} and note that they differ from the typical conventions for the Hodge dual.}:
\begin{subequations}\label{BPSlayer1}
 \begin{equation}\label{BPSlayer1a}
     *_4 \mathcal{D} \dot{Z}_1 =  \mathcal{D} \Theta^{2}\,,\quad \mathcal{D}*_4\mathcal{D}Z_1 = -\Theta^{2}\wedge d\beta\,,\quad \Theta^{2}=*_4 \Theta^{2}\,,
 \end{equation}
  \begin{equation}\label{BPSlayer1b}
 *_4 \mathcal{D} \dot{Z}_2 =  \mathcal{D} \Theta^{1}\,,\quad \mathcal{D}*_4\mathcal{D}Z_2 = -\Theta^{1}\wedge d\beta\,,\quad \Theta^{1}=*_4 \Theta^{1}\,,
  \end{equation}
  \begin{equation}\label{BPSlayer1c}
 *_4 \mathcal{D} \dot{Z}_4 =  \mathcal{D}  \Theta^4\,,\quad \mathcal{D}*_4\mathcal{D}Z_4 = -\Theta^4\wedge d\beta\,,\quad \Theta^4=*_4 \Theta^4\,.
\end{equation}
\end{subequations}
where the dot denotes $ \frac{\partial}{\partial v}$, $\mathcal{D}$ is defined by
\begin{equation}
\mathcal{D} \equiv \tilde d - \beta\wedge \frac{\partial}{\partial v}\,,
\end{equation}
and $\tilde d$ denotes the exterior differential on the spatial base $\cB$.

The  equations for the function, $\cF$, and the $1$-form, $\omega$, are also linear:\footnote{Note that there is a sign error on the left-hand side of the second equation in \rcite{Bena:2016agb}.}
\begin{subequations}\label{BPSlayer2}
\begin{equation}
\label{BPSlayer2a}
\mathcal{D} \omega ~+~  *_4  \mathcal{D}\omega  ~=~ Z_1 \Theta^{1}+ Z_2 \Theta^{2}  - \mathcal{F} d \beta -2\,Z_4 \Theta^4 \,,
\end{equation}
\begin{equation}
 *_4\mathcal{D} *_4\!\big(\dot{\omega} - \coeff{1}{2}\,\mathcal{D} \mathcal{F} \big) 
~=~\partial_v^2 \big(Z_1 Z_2 - {Z}_4^2\big)  -\big(\dot{Z}_1\dot{Z}_2  -(\dot{Z}_4)^2 \big)-\coeff{1}{2} *_4\!\big(\Theta^{1}\wedge \Theta^{2} - \Theta^4 \wedge \Theta^4\big)\,.
\label{BPSlayer2b}
\end{equation}
\end{subequations} 

When the six-dimensional solution is $v$-independent, one can reduce it to five dimensions using reduction \eqref{uvdefn}, \eqref{sixmet-2}. In this reduction, the five-dimensional field-strengths are related to the six-dimensional ones via (see~\cite[App.~B.2]{Bena:2016agb})
\begin{equation}
\label{5D6Dtheta}
\widetilde \Theta^{I} = {1\over \sqrt{2}} \Theta^{I}\,,~~~~~~I=1,2,4, ~~~~~{\rm and} ~~~~ \widetilde\Theta^3 = \sqrt{2} \;\! d \beta \,.
\end{equation}
Upon flipping the sign of $Z_4$ \cite{Bena:2015bea}, these equations reduce to those of five-dimensional supergravity \eqref{5dBPSeqn2}.

%%%%%%%%%%%%%%%%%%%%%%%%%%%%%%%%%%%%
\subsection{The two-charge circular supertube solution}
\label{sec:circ-ST-soln}
%%%%%%%%%%%%%%%%%%%%%%%%%%%%%%%%%%%%

The two-charge circular supertube solution~\cite{Balasubramanian:2000rt,Maldacena:2000dr,Mateos:2001qs,Emparan:2001ux,Lunin:2001jy,Lunin:2002iz} is the fundamental seed solution upon which all explicitly constructed superstrata have been built. There are also two standard sets of coordinates for defining superstrata: One based on the GH formulation and the other based on the spheroidal coordinates that are frequently used to describe black rings.  The description of the supertube gives us a natural setting for the introduction of these two coordinate systems.

%%%%%%%%%%%%%%%%%%
\subsubsection{The GH formulation of the supertube}
\label{sec:GH-ST}
%%%%%%%%%%%%%%%%%%

The two-charge circular supertube is a two-center solution with a flat four-dimensional base metric. We start by using the GH form in \eqref{GHmet}. We also introduce  spherical polar and cylindrical polar coordinates:
\begin{align}
    x_1= \hat{r}\sin\hat{\theta}\cos{\phig} =  \rho \,  \cos \phig ~~,\qquad  x_2 = \hat{r}\sin\hat{\theta}\sin{\phig}  =   \rho \,  \sin \phig ~~,\qquad
    x_3= \hat{r}\cos\hat{\theta} =  z ~~,
    \label{3Dpolars}
\end{align}
where we put hats on $\hat{r},\hat{\theta}$ to distinguish them from $r,\theta$ that will be introduced later. 
 
In terms of these coordinates, $V$ and $A$ in \eqref{GHmet} are given by
\begin{equation}
V ~=~  \frac{1}{\hat r} \,, \qquad 
A ~=~  \cos\hat{\theta}\,d\phig=\frac{z}{\hat r} \, d\phig \,.
\label{Vrforms}
\end{equation}

The two-charge circular supertube solution~\cite{Balasubramanian:2000rt,Maldacena:2000dr,Mateos:2001qs,Emparan:2001ux,Lunin:2001jy,Lunin:2002iz}, in the AdS$_3$ decoupling limit, is given by the following charge and moduli vectors (see for example \cite{Bena:2017geu}):
 \begin{align}
\begin{split}
    \Gamma_{1}&=(1,(0,0,-\hat{\kappa}_3),(0,0,0),0) \;,\\
    \qquad \Gamma_{2}&=\Bigl(0,(0,0,\hat{\kappa}_3),(\hat Q_1,\hat Q_5,0),\hat m \Bigr) \;, \qquad  \hat m \;=\; \frac{\hat Q_1\hat Q_5}{2\hat\kappa_3} \;,\\
    h&=(0,(0,0,0),(0,0,1),0) \;.
\end{split}
\label{eq:st-gamma}
\end{align}
The charge vectors $\Gamma_{1}$, $\Gamma_{2}$ each satisfy the primitivity condition (\ref{eq:primitivity}).

Center 1 is located at $\hat{r}=0$. Center 2, the supertube center, is located at $\vec x_{\sst S}  \equiv (0,0,-\hat{a})$. Writing the distance from the supertube center as
\be
\hat r_{\sst S} \,\equiv|\, \vec x - \vec x_{\sst S} | 
  \,=\,\sqrt{\rho^2 + (z+\hat{a})^2} \;,
\ee
the harmonic functions are
\begin{align}
\begin{split}
 V&\,=\,\frac{1}{\hat{r}},\qquad K^{(1)}\,=\,K^{(2)}\,=\,0,\qquad K^{(3)}\,=\,\hat{\kappa}_{3}\left(\frac{1}{\hat{r}_{\sst S}}-\frac{1}{\hat r}\right)\,,%\qquad
 \\
L_1&\,=\,\frac{\hat{Q}_1}{\hat{r}_{\sst S}}\,,\qquad 
  L_2\,=\,\frac{\hat{Q}_2}{\hat{r}_{\sst S}},\,\qquad
  L_3\,=\,1,\qquad M\,=\,\frac{\hat{m}}{\hat{r}_{\sst S}} ~.
\end{split}
  \label{eq:LM limit of GLMT}
\end{align}
The one-form, $\beta$, is given by: 
\begin{equation}
\beta ~= ~ \frac{K^{(3)}}{V} \, (d \psig + A) ~+~ \vec \xi \cdot d\vec x \,,  \label{B3field}
\end{equation}
where $\vec  \nabla \times \vec \xi  =  - \vec \nabla K^{(3)}$, which is solved by
\begin{equation}
\qquad \xi  \,=\, {\hat \kappa_3}\bigg(  \frac{z}{\hat r} -\frac{(z+\hat{a})}{\hat r_{\sst S}}  \bigg)\, d\phig  ~. \label{K3form}
\end{equation}
%

%%%%%%%%%%%%%%%%%%
\subsubsection{The spheroidal  form of the supertube}
\label{sec:SBP-ST}
%%%%%%%%%%%%%%%%%%

Let us take $(w_1, w_2, w_3, w_4)$ to be Cartesian coordinates on  $\mathbb{R}^4$. Then spheroidal coordinates $(r, \theta, \phie, \psie)$ are defined, and related to GH coordinates, by:
 \begin{equation}
  \begin{aligned}
 w_1 + i w_2 ~=~ & \sqrt{r^2+a^2}\,\sin\theta\,e^{i \phie}~=~ 2 \,\sqrt{\hat r} \,  \sin \bigg(\frac{\hat \theta}{2}\bigg)  \  e^{i (\psig - \phig)/ 2}\,, \\
 w_3 + i w_4 ~=~ & r \,\cos\theta\,e^{i \psie}~=~ 2 \,\sqrt{\hat r}\,  \cos \bigg(\frac{\hat \theta}{2}\bigg) \  e^{i (\psig + \phig)/ 2}\,, 
\end{aligned}
\label{eq:coords1}
 \end{equation}
for a parameter $a$ which will be related to $\hat{a}$ momentarily. The  coordinates have the ranges and identifications
 \begin{align}
  \begin{aligned}
r, \hat r &\in [0, \infty), \quad~ \theta\in[0,\pi/2], \quad~ \hat\theta\in[0,\pi], \quad~
\phie,\psie,  \phig \in [0, 2\pi), \quad~ \psig \in [0, 4\pi)\,, \\[1mm]
& \phie\sim \phie+2\pi, \quad~ \psie\sim \psie+2\pi,\quad~ (\psig, \phig) \sim  (\psig+4\pi, \phig) \sim  (\psig+2\pi, \phig+2\pi). 
  \end{aligned}
\label{eq:coords2identifications}
 \end{align}
Recall that the $\mathbb{R}^3$
base coordinates for the multi-center solutions are defined in (\ref{3Dpolars}).

The locus $r=0$ thus describes a disk of radius $a$ lying in the $w_1$-plane at $w_2=0$, parameterized by $\theta$ and $\phi$ with the origin of $\mathbb{R}^4$  at $(r=0,\theta=0)$. The supertube lies at the perimeter of this disk, at  $(r=0,\theta=\pi/2)$.   
 
 In these coordinates the flat $\mathbb{R}^4$ metric is
 \begin{equation}\label{ds4flat}
 d s^2_4 = \Sigma\, \Bigl(\frac{d r^2}{r^2+a^2}+ d\theta^2\Bigr)+(r^2+a^2)\sin^2\theta\,d\phie^2+r^2 \cos^2\theta\,d\psie^2\,,
\end{equation}
where
\begin{equation}
\Sigma ~\equiv~ r^2 + a^2 \cos^2\theta\,.
\end{equation}

The AdS$_3$ decoupling limit of the supertube solution has $\Theta^I \equiv 0$ for $I =1,2,4$, and
\begin{equation}
 Z_1~=~ \frac{Q_1}{\Sigma} \,, \quad  Z_2~=~ \frac{Q_5}{\Sigma} \,, \quad Z_4 ~=~ 0   \,,  
   \label{BSTcharges}
\end{equation}
where we use $Q_5$ and $Q_2$ interchangeably, and where the conversion between GH charges $\hat{Q}_I$ and six-dimensional supergravity charges $Q_I$ is
\begin{equation}
 \hat Q_I~=~ \frac{Q_I}{4}  \,.  
   \label{Qcharges1}
\end{equation}
The one-form $\beta$ is given by
 \begin{equation}
\beta ~=~  \frac{R_y\, \kappa \, a^2 }{\sqrt{2}\,\Sigma}\,(\, \sin^2\theta\, d\phie - \cos^2\theta\,d\psie\,)   \,.
 \label{betadefn}
\end{equation}
The remaining six-dimensional ansatz quantities are then:
\begin{equation}
 \qquad \cF ~=~ 0   \,, \qquad \omega ~=~  \frac{R_y\, \kappa  \, a^2}{ \sqrt{2}\,\Sigma}\,  (\sin^2 \theta  d \phie + \cos^2 \theta \,  d \psie ) \,.
\label{Fomega-ST}
\end{equation}
For future reference, we note that (\ref{eq:coords1}) imply the following identities:
\begin{equation}
\hat r = |\vec x | = \coeff{1}{4}\, (r^2\tight+a^2 \sin^2\theta)  ~~, \qquad \hat r_{\sst S} = |\vec x - \vec x_{\sst S}| \equiv \coeff{1}{4}\,  \Sigma = \coeff{1}{4}\, (r^2\tight+a^2 \cos^2\theta)  ~~, \qquad \hat{a} = \coeff{1}{4}\, a^2~,
 \label{eq:coords2}
\end{equation}
 \label{}
\begin{equation}
\cos^2 \theta = \frac{\hat a - \hat r + \sqrt{\hat a^2 + \hat r^2 + 2 \hat a \hat r \cos \hat \theta}}{2\hat a}~~, \quad 
r^2 = 2\left(-\hat a + \hat r + \sqrt{\hat a^2 + \hat r^2 + 2 \hat a \hat r \cos \hat \theta}\right)~.
\label{thrhatunhat}
\end{equation}

One can also match the proper radius of the $y$-circle to the normalization of $K^{(3)}$. One must remember that in making the reduction to five dimensions, there is a relative factor of $\sqrt{2}$  between $\widetilde \Theta_3$ and $d \beta$ \eqref{5D6Dtheta}.  We therefore have:
 \begin{equation}
K^{(3)} = \frac{\kappa R_y}{2} \, \bigg(   \frac{1}{\hat r_{\sst S} } - \frac{1}{\hat r}\bigg)  \qquad \Rightarrow \qquad \hat \kappa =  \frac{\kappa R_y}{2}  ~, \qquad~~ \kappa\in\mathbb{Z} \,,
\label{K3norm}
\end{equation}
where $\kappa$ is the winding number of the supertube (taken to be positive, as usual). 
Furthermore, regularity requires
\be
\label{a-Q-Ry}
a^2 \,=\, \frac{Q_1 Q_5}{\kappa^2 R_y^2} ~.
\ee
Defining the rescaled coordinates
\be
\label{eq:coords3}
\tilde{t} \,=\, \frac{t}{R_y} \,, \qquad
\tilde{y} \,=\, \frac{y}{R_y} \,, \qquad
\sinh \rho \,=\, \frac{r}{a} \,, \qquad
\ee
the six-dimensional metric then takes the form of rotating orbifolded global AdS$_3 \times $S$^3$,
\begin{align}
\label{adslim}
    \frac{1}{\sqrt{Q_1 Q_5}}ds^2 \,=\;\:\! &{}-\frac{1}{\k^2}\cosh^2\rhoo \;\! d\tilde{t}^2 + d\rho^2 +\frac{1}{\k^2}\sinh^2\rhoo \;\! d\tilde{y}^2  \cr
    &{}\qquad+d\theta^2 + \cos^2\theta\left( d\psi + \frac{1}{\k}d\tilde{y} \right)^2
    + \sin^2\theta\left( d\phi - \frac{1}{\k} d\tilde{t} \right)^2 \,.~~~~
\end{align}
In the D1-D5 frame, the supergravity charges $Q_I$ and angular momenta $\hat{J}_L$, $\hat{J}_R$ \eqref{eq:j-non-q} are related to integer quanta $n_I\in\bZ$, $\left(J_{L},J_{R}\right)\in\half\mathbb{Z}$ via (see e.g.~\rcite{Giusto:2012yz})
\begin{align}
\label{eq:q-quant}
            Q_1 \,=\, \frac{g_s(\alpha')^3}{V_4}n_1~,&\qquad 
   Q_2 \,=\, Q_5\,=\, g_s\alpha' n_5~,\qquad 
  Q_3 \,=\, Q_p\,=\,\frac{g_s^2(\alpha')^4}{R_y^2V_4}n_p ~,\\[1mm]
\label{eq:j-quant}
    &\hat{J}_{L} =\frac{g_s^2(\alpha')^4}{
    8 V_4 R_y} J_{L} ~~,\qquad 
 \hat{J}_{R}\;=\;\frac{g_s^2(\alpha')^4}
    {8 V_4 R_y} J_{R}~.
\end{align}
Thus the quantized angular momenta take the values
\begin{align}
\label{eq:j-quant-ST}
    {J}_{L} &\,=\, J_R \,=\, \frac{n_1 n_5}{2 \kappa} ~,
\end{align}
implying the known fact that $\kappa$ must divide $n_1 n_5$ for this configuration to have correctly quantized angular momenta. 

%%%%%%%%%%%%%%%%%%%%%%%%%%%%%%%%%%%%
\subsection{Three-charge supersymmetric spectral flowed supertube solutions}
\label{sec:GLMT soln}
%%%%%%%%%%%%%%%%%%%%%%%%%%%%%%%%%%%%

If one performs a fractional spectral flow with parameter $s/\kappa$ with $s\in \mathbb{Z}$ on the circular supertube solution with winding number $\kappa$, one obtains the GLMT solution~\rcite{Giusto:2012yz}, the most general two-center solution with smooth GH centers and $\mathbb{R}^4$ asymptotics. We take $s>0$ without loss of generality. We work in the AdS$_3$ limit, in which we have two centers with the charge vectors and the asymptotic moduli\footnote{If the spectral flow is an integer, such that $s$ (or $s-1$) is a multiple of $\kappa$, this solution reduces to that of \cite{Giusto:2004ip}; see also~\cite{Lunin:2004uu,Giusto:2004id}.}:
\begin{align}
\begin{split}
    \Gamma_{1}&=\biggl(s+1,(-\hat{\kappa}_1,-\hat{\kappa}_2,-\hat{\kappa}_3),\Big(-\frac{\hat{\sfq}_1}{s+1},-\frac{\hat{\sfq}_2}{s+1},-\frac{\hat{\sfq}_3}{s+1}\Big),-\frac{\tilde{m}}{2(s+1)^2}\biggr),\\
    \Gamma_{2}&=\biggl(-s,(\hat{\kappa}_1,\hat{\kappa}_2,\hat{\kappa}_3),\Big(\frac{\hat{\sfq}_1}{s},\frac{\hat{\sfq}_2}{s},\frac{\hat{\sfq}_3}{s}\Big),\frac{\tilde{m}}{2s^2}\biggr)\,,\\
    h&=\left(0,(0,0,0),(0,0,1),0\right)\,.
\end{split}
\label{bgGamma}
\end{align}

The primitivity condition \eqref{eq:primitivity} requires the relations
\begin{align}
\label{eq:primitivity GLMT}
    \hat{\sfq}_1=\hat{\kappa}_2 \hat{\kappa}_3~,\quad \hat{\sfq}_2=\hat{\kappa}_1 \hat{\kappa}_3 ~,\quad\hat{\sfq}_3=\hat{\kappa}_1 \hat{\kappa}_2~,\quad \tilde{m}=\hat{\kappa}_1\hat{\kappa}_2\hat{\kappa}_3~.
\end{align}
The charges $\hat\kappa_I$ are related to the quantized charges, $\kappa_I$, as follows (see e.g.~\rcite{Giusto:2012yz}). In the D1-D5-P duality frame, when $Q_1$ denotes the D1 charge, $Q_2$ denotes the D5 charge, and $Q_3$ denotes the momentum charge, we have:
\begin{equation}
\label{k-quant-D1-P}
    \qquad\qquad
    \hat{\kappa}_1=\frac{g_s\alpha'}{2R_y}\kappa_1~,\qquad\hat{\kappa}_2=\frac{g_s(\alpha')^3}{2R_y V_4}\kappa_2~,\qquad\hat{\kappa}_3=\frac{R_y}{2}\kappa_3 \equiv \frac{R_y}{2}\kappa~, \qquad
    \kappa_I ~\in~\mathbb{Z} ~.
\end{equation}

On the other hand, in the NS5-F1-P duality frame, when $Q_1$ denotes the F1 charge, $Q_2$ denotes the NS5 charge, and $Q_3$ denotes the momentum charge, we have  
\begin{equation}
\label{k-quant-F1-P}
\qquad\qquad
    \hat{\kappa}_1=\frac{\alpha'}{2R_y}\kappa_1~,\qquad
    \hat{\kappa}_2=\frac{g_s^2(\alpha')^3}{2R_y V_4}\kappa_2~,\qquad
    \hat{\kappa}_3=\frac{R_y}{2}\kappa_3\equiv \frac{R_y}{2}\kappa~, \qquad
    \kappa_I ~\in~\mathbb{Z} ~.
\end{equation}

The charge vectors of the two-charge circular supertube, \eqref{eq:st-gamma}, can be obtained by taking the combined scaling limit $s\rightarrow 0$, $\hat{\kappa}_1,\hat{\kappa}_2\rightarrow 0$ with $\hat{\kappa}_1/s$ and $\hat{\kappa}_2/s$ held fixed; c.f.~\eqref{eq:primitivity}.
To match onto the circular supertube in this limit, we place $\Gamma_1$ at $\vec x=0$ and $\Gamma_2$ at $\vec x=\vec x_{\sst S}$. 

If we were working in asymptotically flat space, the relation controlling the distance $a$ between the centers, \eqref{a-Q-Ry}, would be modified compared to that of the two-charge supertube. However, since we are working in the AdS$_3$ limit, the relations 
\eqref{eq:coords2} and \eqref{a-Q-Ry} from the two-charge circular supertubes carry over unchanged to the GLMT configurations.\footnote{The parameter $\eta$ in~\cite{Giusto:2004kj,Giusto:2012yz} becomes equal to 1 in the AdS$_3$ limit in which we work.} Note however that since this configuration involves two GH centers, the 4D base is not flat~\cite{Giusto:2004kj}. 

The six-dimensional metric then takes the form of a more general rotating orbifolded global AdS$_3 \times $S$^3$,
\begin{align}
\label{ads-glmt}
    \frac{1}{n_5}ds^2 \,=\;\:\! &{}-\frac{1}{\k^2}\cosh^2\rhoo \;\! d\tilde{t}^{\,2} + d\rho^2 +\frac{1}{\k^2}\sinh^2\rhoo \;\! d\tilde{y}^2  \cr
    &{}+d\theta^2 + \cos^2\theta\left( d\psi + \frac{s}{\k} d\tilde{t} + \frac{s+1}{\k}d\tilde{y} \right)^2
    + \sin^2\theta\left( d\phi - \frac{s+1}{\k} d\tilde{t} - \frac{s}{\k}d\tilde{y} \right)^2.~~~~
\end{align}
The following large coordinate transformation maps this decoupled geometry to an orbifold of global AdS$_3 \times$S$^3$, and is known as (fractional) spacetime spectral flow~\rcite{Giusto:2012yz}, see also~\rcite{Lunin:2004uu,Giusto:2004id,Giusto:2004ip}:
\begin{align}
\label{eq:sf-GLMT-sug}
    \psi_{\NS}  \,=\;\:\! & \psi + \frac{s}{\k} \tilde{t} + \frac{s+1}{\k}\tilde{y}  ~,\qquad
    \phi_{\NS}
    \,=\, \phi - \frac{s+1}{\k} \tilde{t} - \frac{s}{\k} \tilde{y} ~.
\end{align}
Flux quantization in the bulk, and quantization of momentum per strand (symmetric group cycle) in the holographic CFT, impose the requirement that~\cite{Giusto:2012yz}
\be
\frac{s(s+1)}{\sfk} \, \in \, \mathbb{Z} \,.
\ee
Then, for $\sfk>1$, the coordinate identification $\tilde{y} \sim \tilde{y} + 2 \pi $ induces orbifold singularities via the identification~\rcite{Giusto:2012yz,Jejjala:2005yu}
\begin{align}
\label{eq:sug-orbact-GLMT}
\big(\tilde{y},\psi_{\NS},\phi_{\NS}\big) 
\; \sim \;
\big(\tilde{y},\psi_{\NS},\phi_{\NS}\big)  +
2\pi \Big(1,\frac{s+1}{\k},-\frac{s}{\k}\Big) \,. 
\end{align}
Although the $s\to 0$ limit of the multi-center charge vectors requires a careful combined scaling limit, at the level of the six-dimensional geometry the limit can be taken straightforwardly.

%%%%%%%%%%%%%%%%%%%%%%%%%%%%%%%%%%%%%
\subsection{Superstrata}
\label{sec:superstrata-review}
%%%%%%%%%%%%%%%%%%%%%%%%%%%%%%%%%%%%%

%%%%%%%%%%%%%%%%%%%%%%%%%%%%%%%%%%%%%
\subsubsection{General superstrata}
\label{sec:GenSS}
%%%%%%%%%%%%%%%%%%%%%%%%%%%%%%%%%%%%%

The superstratum~\cite{
Bena:2015bea,
Bena:2016agb,
Bena:2016ypk,
Bena:2017xbt,
Bakhshaei:2018vux,
Ceplak:2018pws,
Heidmann:2019zws,Ganchev:2022exf, 
Ceplak:2022pep,Ganchev:2023sth, 
Ceplak:2024dbj,
Heidmann:2019xrd,
Shigemori:2020yuo} 
involves adding excitations on the $S^3$
and left-moving on the AdS$_3$.  The most general mode dependence consistent with supersymmetry involves \begin{align}
    v_{k,m,n} &\equiv (m+n) \,\frac{\sqrt{2}\,v}{R_y} + (k-m)\,\phie - m\,\psie \,,
 \label{v_kmn_def}    
\end{align}
for some quantum numbers $k,m,n$ that are non-negative integers with $0 \le m \le k$.

To perform the harmonic analysis one also needs the functions that govern the $r$ and $\theta$ dependence:
 \begin{equation}
 \Delta_{k,m,n} ~\equiv~  \bigg(\frac{r}{\sqrt{r^2+a^2}}\bigg)^n  \, \bigg(\frac{a}{\sqrt{r^2+a^2}}\bigg)^k  \, \sin^{k-m}\theta \,  \cos^{m}\theta\,.
  \label{Deltadefn}
 \end{equation}
To describe the two-form fluxes we  define an unnormalized basis of self-dual 2-forms on $\IR^4$:
\begin{equation}
\label{selfdualbasis}
\begin{gathered}
\Omega^{(1)} \equiv \frac{dr\wedge d\theta}{(r^2+a^2)\cos\theta} + \frac{r\sin\theta}{\Sigma} d\phie\wedge d\psie\,,\\
\Omega^{(2)} \equiv  \frac{r}{r^2+a^2} dr\wedge d\psie + \tan\theta\, d\theta\wedge d\phie\,,\qquad
 \Omega^{(3)}\equiv \frac{dr\wedge d\phie}{r} - \cot\theta\, d\theta\wedge d\psie\,.
\end{gathered}
\end{equation}

The complete set of modes 
are then:
\begin{align}
\widetilde{z}_{k,m,n} &=\,R_y \,\frac{\Delta_{k,m,n}}{\Sigma}\, \cos{v_{k,m,n}},\label{ztilde}\\
\widetilde{\vartheta}_{k,m,n}& ~\equiv~ -\sqrt{2}\,
\Delta_{k,m,n}
\biggl[\left((m+n)\,r\sin\theta +n\left(\frac{m}{k}-1\right)\frac{\Sigma}{ r \sin\theta}  \right)\Omega^{(1)}\sin{v_{k,m,n}}  \nonumber \\
&\hspace{20ex}
 +\left(m\left({n\over k}+1\right)\Omega^{(2)} +\left({m\over k}-1\right)n\, \Omega^{(3)}\right) \cos{v_{k,m,n}} \biggr]
\,,
\label{thetatilde}
\end{align}
and
\begin{equation}
\widehat{\vartheta}_{k,m,n} ~\equiv~
\sqrt{2}\, \Delta_{k, m, n}\left[\,\frac{\Sigma}{r\sin\theta}\, \Omega^{(1)}\, \sin{v_{k,m,n}}+ \left(\Omega^{(2)} + \Omega^{(3)}\right)\cos{v_{k,m,n}}\,\right]\,.
\label{thetahat}
\end{equation}
These modes satisfy:
\begin{equation}
*_4\cD\,\dot{\widetilde{z}}_{k,m,n} = \cD\,\widetilde{\vartheta}_{k,m,n}, \qquad \cD *_4 \cD\, \widetilde{z}_{k,m,n} = - \widetilde{\vartheta}_{k,m,n} \wedge d\beta,\qquad \widetilde{\vartheta}_{k,m,n} = *_4 \widetilde{\vartheta}_{k,m,n},\\
\end{equation}
and
\begin{equation}
\cD\,\widehat{\vartheta}_{k,m,n} = 0, \qquad \widehat{\vartheta}_{k,m,n} \wedge d\beta=0,\qquad \widehat{\vartheta}_{k,m,n} = *_4 \widehat{\vartheta}_{k,m,n}\,,
 \label{schg1stlayer}
\end{equation}
and so the general solution to the first layer of the BPS system, \eqref{BPSlayer1}, can be built out of superpositions of the modes.

In particular, the superstratum involves using the Ansatz \rcite{Heidmann:2019zws}:
\begin{equation}
\begin{aligned}
Z_1 & ~=~  \frac{Q_1}{\Sigma} ~+~ \frac{R_y}{2\, Q_5} \,  \sum_{k_1,m_1,n_1}\, b^{(1)}_{k_1,m_1,n_1} \, \widetilde{z}_{k_1,m_1,n_1}     \,,   
\qquad
Z_4  ~=~  \sum_{k_1,m_1,n_1}\,   b^{(4)}_{k_1,m_1,n_1} \, \, \widetilde{z}_{k_1, m_1, n_1} \,,  \\
Z_2  &~=~ \frac{Q_5}{\Sigma} \,, \qquad \Theta^{1}  ~=~  0 \,,\\  \Theta^4   & ~=~ \sum_{k_1,m_1,n_1}\,  \Big[ \, b^{(4)}_{k_1,m_1,n_1}\, \widetilde{\vartheta}_{k_1, m_1, n_1} ~+~ c^{(4)}_{k_1,m_1,n_1}\, \widehat{\vartheta}_{ k_1, m_1, n_1} \Big] \,,   \\
\qquad \Theta^{2} & ~=~\frac{R_y}{2\, Q_5}\,\sum_{k_1,m_1,n_1}\,  \Big[ \,b^{(2)}_{k_1,m_1,n_1} \, \widetilde{\vartheta}_{  k_1, m_1, n_1}   ~+~ c^{(2)}_{k_1,m_1,n_1} \,\widehat{\vartheta}_{ k_1,m_1, n_1}   \Big]\,, 
\end{aligned}
\label{eq:Zansatz}
\end{equation}

The final layer of BPS equations, (\ref{BPSlayer2a}) and (\ref{BPSlayer2b}), can then be solved provided that some of the Fourier coefficients are locked to one another through ``coiffuring conditions'' \rcite{Bena:2016ypk,Bena:2017xbt,Ceplak:2018pws,Heidmann:2019zws,Heidmann:2019xrd}.  These conditions determine the modes of $(Z_1, \Theta^{2})$ in terms of the modes of $(Z_4, \Theta^4)$, and the latter are unconstrained \rcite{Heidmann:2019zws}.   Thus the most general superstratum is parametrized by two arbitary functions of three variables that are encoded  in the Fourier coefficients $b^{(4)}_{k_1,m_1,n_1}$ and $c^{(4)}_{k_1,m_1,n_1}$.  The end result of solving the last layer of BPS equations is that the function, $\cal F$, and angular momentum vector, $\omega$, are determined in terms of quadratics in these two sets of Fourier coefficients.

We have been rather cursory in our discussion here because the details of all the actual modes are going to wash out of the effective superstrata, and the only details that will survive are the components that are independent of all three Fourier angles: $(v, \phie, \psie)$. Such terms  appear as the ``zero modes'' in the squares of each of the Fourier series that contribute to the sources in (\ref{BPSlayer2a}) and (\ref{BPSlayer2b}). These feed into the zero-modes of the momentum function, $\cF$, and angular momentum vector, $\omega$.  We will not need the detailed solutions for these quantities: what is important is how they localize within the geometry and this can be inferred from their sources in (\ref{BPSlayer2a}) and (\ref{BPSlayer2b}).  From our discussion it should be evident that this localization is controlled by appropriate squares of (\ref{Deltadefn}).  

Indeed, because the effective superstratum reduces to the contribution from the sum of the squares of the Fourier modes, the essential features are well-illustrated by focusing on a ``single-mode'' superstratum.  As we will see, the effective superstratum reduces all the details of the two arbitrary functions of three variables to a few quantum numbers.   This is essentially why the semi-classical entropy drops from $Q^{5/4}$ to $Q^{1}$ in going from six to five dimensions.

%%%%%%%%%%%%%%%%%%%%%%%%%%%%%%%%%%%%%
\subsubsection{Single-mode superstrata}
\label{sec:SingleModeSS}
%%%%%%%%%%%%%%%%%%%%%%%%%%%%%%%%%%%%%

Our primary focus is to see how high-frequency superstratum modes localize within the geometry, and this is most easily demonstrated by restricting to a single Fourier mode.  Here we summarize the results of one of the analyses in \rcite{Heidmann:2019zws}. The basic single-mode superstratum starts from:
\begin{equation}
\begin{aligned}
Z_1 & =  \frac{Q_1}{\Sigma} + \frac{R_y\,  b_1}{2\, Q_5}   \, \widetilde{z}_{2k,2m,2n}    ~~,  
\qquad Z_2  = \frac{Q_5}{\Sigma} ~~, \qquad
Z_4  =   b_4  \, \widetilde{z}_{k, m, n} ~,  \\
 \Theta^{1}  & =  0 ~~, \qquad  \Theta^{2} = \frac{R_y \,b_2}{2\, Q_5}\,  \widetilde{\vartheta}_ {2k,2m,2n} ~~, \qquad \Theta^4    =     b_4\, \widetilde{\vartheta}_{k, m, n}  ~~.  
\end{aligned}
\label{eq:Zansatz2}
\end{equation}
Note that the mode numbers of $(Z_2, \Theta^{1})$ are twice those of $(Z_4, \Theta^4)$.  Indeed the coiffuring constraint requires:
\begin{equation}
b_1 ~=~ b_4^2 \,.
\end{equation}
This constraint removes all the source terms in (\ref{BPSlayer2a}) and (\ref{BPSlayer2b}) that depend on  $v_{2k,2m,2n}$, leaving only terms that are independent of $(v, \phie, \psie)$. In general, coiffuring removes all terms that involve sums of frequencies, leaving the ``beat,'' or difference, frequencies.\footnote{This has a nice characterization in terms of the holomorphic forms of superstratum waves \rcite{Heidmann:2019xrd}.}

The canonical normalization of the Fourier modes is given by defining:
\begin{equation}
b^2 ~=~  \genfrac(){0pt}{0}{k}{m}^{-1}    \,  \genfrac(){0pt}{0}{k +m -1}{n}^{-1}    \, b_4^2   ~. 
\label{sschg2}
\end{equation}
One then finds that regularity at the original supertube locus requires:
\begin{equation}
\frac{Q_1 \,Q_5 }{R_y^2 }  ~=~ a^2  + \coeff{1}{2} \, b^2
\,\equiv \, a_0^2 \,,  \label{ssReg}
\end{equation}
where we have introduced $a_0$ for later convenience. The conserved charges are given by: 
\begin{equation}
Q_P =   \frac{m+n}{2 k  } \, b^2    ~~,  \qquad \hat{J}_R  = \coeff{1}{2}\,   R_y \,a^2 ~~, \qquad  \hat{J}_L - \hat{J}_R =   \coeff{1}{2}\, R_y  \, \frac{m}{ k  } \, b^2   ~.  \label{sschg1}
\end{equation}
The general expressions for $\cF$ and $\omega$ are  complicated but can be found in several places (see, for example \rcite{Heidmann:2019zws}), however, all we will need is to observe is that, apart from overall constants,  the sources in (\ref{BPSlayer2a}) and (\ref{BPSlayer2b}) are proportional to 
 \begin{equation}
 \Delta_{k,m,n}^2 ~=~  \Delta_{2k,2m,2n} ~=~  \bigg(\frac{r^2}{r^2+a^2}\bigg)^{n}  \, \bigg(\frac{a^2}{r^2+a^2}\bigg)^k  \, \sin^{2(k-m)}\theta \,  \cos^{2m}\theta\,.
  \label{Deltafactors}
 \end{equation}

It is useful to note here that for $b=0$, one has $Q_P =0$, and the solution reduces to the AdS$_3 \times S^3$ of the supertube. As one increases $b$, and hence $Q_P$, one must decrease $a$ in accordance with (\ref{ssReg}).  In the dual CFT this reflects the ``strand budget'' constraint that the total strand length must be equal to $n_1n_5$.  As $b$ increases, the geometry develops a capped AdS$_2 \times S^1$ throat as studied in \cite{Bena:2016ypk,Bena:2017xbt} and depicted in Fig.~\ref{fig:supertstratum}.  For $ b \gg a$, this throat becomes very long and there can be a large redshift between the top and bottom of the throat. The story is the same in general superstrata with $b^2$ replaced by a weighted (as in (\ref{sschg2})) sum of squares of all the Fourier coefficients of the momentum modes.  
The AdS$_2$ scaling of the geometry as a function of $ b/a$ is the superstratum analog of the scaling solutions discussed at the end of Section \ref{sec:multi-center}.  This structure will be important throughout this paper and we will return to the issue of the depth of the throat in Section \ref{sec:BHregime}.  

%%%%%%%%%%%%%%%%%%%%%%%%%
%%%%%%%%%%%%%%%%%%%%%%%%%

%%%%%%%%%%%%%%%%%%%%%%%%%%%%%%%%%%%%%
\section{Effective Superstrata}
\label{sec:EffSS}
%%%%%%%%%%%%%%%%%%%%%%%%%%%%%%%%%%%%%

 %%%%%%%%%%%%%%%%%%%
\subsection{The averaged geometry}
\label{ss:AvgGeom}
%%%%%%%%%%%%%%%%%%% 

One can, in principle, average over all the fluctuations of a superstratum, but if one wants to preserve the generic five-dimensional structure of the solution one can start  by averaging over the $y$-dependence, or, equivalently, the $v$-dependence.  To that end, define: 
\begin{equation}
\Avg {{\cal X}}{v}~\equiv~  \frac{1}{2\pi R_y}\, \int_{0}^{2\pi R_y}\,  {\cal X} \, dy ~=~  \frac{1}{\sqrt{2} \,\pi R_y}\, \int_{0}^{\sqrt{2} \,\pi R_y}\,  {\cal X} \, dv \,.
 \label{avgDefn1}
\end{equation}
Since everything is periodic in $v$, such averaging of equations of motion kills the terms that are pure $v$-derivatives, and the BPS equations \eqref{BPSlayer1} and \eqref{BPSlayer2}  reduce to:
 \begin{equation}
 \begin{aligned}
\Avg{ \Theta^I } {v} ~=~ &  *_4 \Avg{ \Theta^I } {v} \,,  \qquad \tilde d \, \Avg{ \Theta^I } {v} ~=~ 0\,,   \qquad I=1,2,4\\
 \tilde d *_4\tilde d \,\Avg{Z_1} {v}  = &  -\Avg{\Theta^{2}}{v}\wedge d \beta  \,, \quad \tilde d *_4\tilde d \,\Avg{Z_2}{v} = -\Avg{\Theta^{1}}{v}\wedge d \beta \,, \quad \tilde d *_4\tilde d \,\Avg{Z_4}{v} = -\Avg{\Theta^4}{v}\wedge d \beta  \,,
  \end{aligned}
 \label{avgBPS1}
\end{equation}
and
 \begin{equation}
  \begin{aligned}
\tilde d   \omega  ~+~ *_4\tilde d  \omega  & ~=~ \Avg{Z_1 \Theta^{1}+ Z_2 \Theta^{2}  -2\,Z_4 \Theta^4} {v} ~-~  \Avg{ \mathcal{F}} {v}\,d \beta\,,\\
 *_4 \tilde d  *_4 \tilde d \, \Big( - \coeff{1}{2}\,\Avg{ \mathcal{F}} {v}   \Big)  
&~=~  -\Avg{ \dot{Z}_1\dot{Z}_2  -(\dot{Z}_4)^2 } {v}   ~-~  \coeff{1}{2} *_4\!\Avg{  \Theta^{1}\wedge \Theta^{2} - \Theta^4 \wedge \Theta^4} {v}    \,.
  \end{aligned}
  \label{avgBPS2a}
\end{equation}

Note that we average over harmonic functions, and their sources, rather than over metric  coefficients or gauge potentials, which involve non-linear combinations of harmonic functions.  This prescription for averaging preserves the BPS property, and guarantees that the averaged geometry is a solution of the field equations. As we described in Section \ref{sec:superstrata-review}, the equations for the $Z_I$ and $\Theta^I$ are all linear and homogeneous, the solutions are all expressed as Fourier modes in $v$, and hence the averages of  $Z_I$ and $\Theta^I$ reduce to the zero modes with respect to $v$. 

Similarly, the expansions on the right-hand sides of the equations for  $\cF$ and $\omega$ can be expressed in terms of Fourier modes, and the procedure replaces these by their averaged values. It is these averaged sources that create the non-trivial angular momentum in $\omega$, and the non-trivial momentum charge in $\cF$.  In general, these sources will involve factors of $\Delta_{k,m,n}^2$, which will act as bump functions, reflecting the distributed sources of momentum and angular momentum.  As we will explain in Section \ref{sec:stratumred},  for large values of $k$, these functions become highly localized and in such effective superstrata one can  replace the bump-function sources by $\delta$-functions that carry the corresponding charges. 

As described in Section \ref{sec:superstrata-review}, the standard superstratum \rcite{Bena:2015bea,Bena:2017xbt,Heidmann:2019zws} has $\IR^4$ as the base geometry and $\Theta^{1} \equiv 0$, $\partial_v Z_2 \equiv 0$, while  $\Theta^{2}$ is purely oscillatory.  The BPS equations (\ref{avgBPS2a}) can then be recast as 
 \begin{equation}
  \begin{aligned}
\tilde d   \omega  ~+~ *_4\tilde d  \omega  & ~=~  -2\,\Avg{ Z_4 \Theta^4} {v} ~-~  \Avg{ \mathcal{F}} {v}\,d \beta\,,\\
 *_4 \tilde d  *_4 \tilde d \, \Big( - \coeff{1}{2}\,\Avg{ \mathcal{F}} {v}   \Big)  
&~=~   \Avg{  (\dot{Z}_4)^2}  {v}   ~+~  \coeff{1}{2} *_4\!\Avg{ \Theta^4 \wedge \Theta^4} {v}    \,.
  \end{aligned}
  \label{avgBPS2b}
\end{equation}
From this one can see that the only non-trivial averaged sources in  (\ref{avgBPS2b}) come from $Z_4$ and $\Theta^4$, which encode the NS-NS fluxes sourced by the open-string excitations.  

If one uses the five-dimensional expressions  (\ref{Z3kdefn}) and (\ref{eqbeta}) {and the relation between five-dimensional and six-dimensional self-dual fluxes in (\ref{5D6Dtheta}),
one can recast (\ref{avgBPS2a}) as}
 \begin{equation}
 \begin{aligned}
\tilde d  {\bf k}  ~+~ *_4\tilde d {\bf k}  & ~=~\frac{1}{\sqrt{2}}\,  \Avg{Z_1 \Theta^{1}+ Z_2 \Theta^{2} + 2\, Z_3\,d \beta    -2\,Z_4 \Theta^4} {v}\,, \\
 \nabla^2_{(4)} \,  \Avg{ Z_3} {v}  & ~\equiv~ -   *_4 \tilde d  *_4 \tilde d \,  \Avg{ Z_3} {v}   ~=~  \Avg{ \dot{Z}^{(1)}\dot{Z}^{(2)}  -(\dot{Z}_4)^2 } {v}   ~+~  \coeff{1}{2} *_4\!\Avg{  \Theta^{1}\wedge \Theta^{2} - \Theta^4 \wedge \Theta^4} {v}    \,,
  \end{aligned}
  \label{avgBPS2c}
\end{equation}
and we can see that the equations governing averaged solutions reproduce (upon flipping the sign of $Z_4$) the five-dimensional BPS equations \eqref{5dBPSeqn2}.

Finally, note that we have focused on reducing the geometry to an effective superstratum in five dimensions by smearing over the $v$-circle.  However, if  we consider a solution in which the base metric, $ds_4^2$,  is actually a Gibbons-Hawking (GH) metric, it can also be convenient to average fields over both $v$ and the GH fiber, $\psig$. We  therefore define:
\begin{equation}
\Avg {{\cal X}}{v,\psig}~\equiv~    \frac{1}{4 \sqrt{2} \,\pi^2  R_y}\, \int_{0}^{4 \pi}  d\psig \int_{0}^{\sqrt{2} \,\pi R_y} dv\,  {\cal X} \, \,.
\label{avgDefn2}
\end{equation}
The result may then be thought of as an effective multi-centered solution defined on the three-dimensional base of the GH metric.

%%%%%%%%%%%%%%%%%%%%%%%%%%%%%%%%%%%%%
\subsection{Five-dimensional effective superstrata}
\label{sec:5DeffSS}
%%%%%%%%%%%%%%%%%%%%%%%%%%%%%%%%%%%%%

Since one has averaged over the $v$ (or $y$) circle, the effective superstratum can be  realized in five-dimensional supergravity but not as a smooth solution:  There will be singular sources for the supertube and the momentum.  This means that
we need to carefully analyze the regularity of the solution from scratch.
We start from the harmonic functions that define the five-dimensional solution and then show how they are obtained as an effective description from the six-dimensional solution.

The five-dimensional solution starts from the two-charge supertube solution described in Section~\ref{sec:circ-ST-soln}, with $V$ and $K^{(3)}$ given in \eqref{Vrforms} and \eqref{K3form}, and we now add a third center with a singular momentum source:
\begin{equation}
 L_1 ~=~ 1 ~+~  \frac{\hat Q_1}{\hat r_{\sst S}}  \,, \quad   L_2 ~=~ 1 ~+~  \frac{\hat Q_2}{\hat r_{\sst S}}  \,, \quad L_3~=~ 1 ~+~  \frac{\hat Q_{\sst P}}{\hat r_{\sst P}}  \,, \quad L_4 ~\equiv~ 0  \,, \label{LIfns}
\end{equation}
where we define
\begin{equation}
\hat r_{\sst P} ~\equiv~ | \vec x - \vec x_{\sst P}| \,, \qquad \vec x_{\sst P}  ~\equiv~ (\rho_0  \,  \cos \phig_0 \,,   \rho_0 \, \sin \phig_0 \,,    z_0)   \,.  \label{momsrc}
\end{equation}
The charges, $\hat Q_1$ and $\hat Q_2$, when suitably normalized, will be the D1 and D5 charges associated with the supertube, while $\hat Q_{\sst P}$ will be the singular source of the momentum charge of an effective superstratum placed at a separate point $\vec x_{\sst P}$.  The constants in $L_I$ have been chosen so that the eleven-dimensional geometry goes to $R^{4,1} \times \IT^6$ at infinity.

In the same vein, we take the angular momentum harmonic function to have possible sources at the supertube and at the momentum source:
\begin{equation}
M ~=~ {\hat m}^{(0)} ~+~  \frac{ {\hat m}_{\sst S}}{\hat r_{\sst S}}  ~+~  \frac{{\hat m}_{\sst P}}{\hat r_{\sst P}}  \,.  \label{Mfn}
\end{equation}
With $K^I$ given by (\ref{eq:LM limit of GLMT}) and $K^4=0$, and using the foregoing $L_I$ and $M$, we have: 
\begin{equation}
\begin{aligned}
Z_1 &=~ 1 ~+~  \frac{\hat Q_1}{\hat r_{\sst S}}  \,, \qquad   Z_2 =~ 1 ~+~  \frac{\hat Q_2}{\hat r_{\sst S}}  \,, \qquad Z_3 =~ 1 ~+~  \frac{\hat Q_{\sst P}}{\hat r_{\sst P}}  \,,
\\   \quad Z_4 &  = 0 \,, \qquad  \mu ~=~ \frac{\hat \kappa_3}{2}\, \bigg(\frac{ \hat r}{\hat r_{\sst S}} ~-~1\bigg)\bigg( 1 ~+~  \frac{\hat Q_{\sst P}}{\hat r_{\sst P}}\bigg)  ~+~  \frac{ {\hat m}_{\sst S}}{\hat r_{\sst S}}  ~+~  \frac{{\hat m}_{\sst P}}{\hat r_{\sst P}} ~+~{\hat m}^{(0)}\,. \label{Zmufns}
\end{aligned}
\end{equation}

For regularity at infinity, we require $\mu \to 0$ as $\hat r \to \infty$, and this means that
\begin{equation}
\hat{m}^{(0)} ~=~ 0 \;.   \label{minfty}
\end{equation}

One can use the bubble equations (\ref{eq:bubble-eqns}) to ensure appropriate regularity conditions, but it is also easy to follow \cite{Bena:2008dw} and check directly. At the origin, the absence of CTCs requires $\mu \to 0$ as $\hat r \to 0$, and hence:
\begin{equation}
-  \frac{\hat \kappa_3}{2}\, \bigg( \frac{\hat Q_{\sst P}}{|\vec x_{\sst P}|} ~+~ 1 \bigg )  ~+~  \frac{ \hat{m}_{\sst S}}{\hat a}  ~+~  \frac{\hat{m}_{\sst P}}{|\vec x_{\sst P}|}  ~=~ 0 \,.  \label{mPeqn1}
\end{equation}
We use this to solve for  $\hat{m}_{\sst P}$:
\begin{equation}
\hat{m}_{\sst P}~=~\frac{\hat \kappa_3 }{2} \,  \hat Q_{\sst P} ~+~\frac{ |\vec x_{\sst P}| }{\hat a} \,   \bigg(\frac{ \hat \kappa_3 \,\hat a}{2} ~-~ \hat{m}_{\sst S}\bigg)  \,.  \label{mPeqn2}
\end{equation}

Regularity of the metric at the original supertube locus means that  the coefficient of $(d \psig + A)^2$ must be regular as  $\hat r_{\sst S}\to 0$.  This requires:
\begin{equation}
\lim_{\hat r_{\sst S}\to 0}\,  \hat r_{\sst S}^2 \, \big[\, Z_3\, (K^{(3)})^2 ~-~ 2 \,\mu \, V K^{(3)} ~+~  \cP \, V\, \big] ~=~ 0 \,. \label{regconda}
\end{equation}
Finally, if the solution to (\ref{varpieqn})  leads to Dirac strings in $\varpi$ then the solution will have CTCs.  Removing the Dirac strings at the supertube requires:  
\begin{equation}
\lim_{\hat r_{\sst S}\to 0}\,  \hat r_{\sst S} \, \big[\, V  \mu ~-~ Z_3\,  K^{(3)}\,  \big] ~=~ 0
.\label{regcondb}
\end{equation}
These two equations are precisely the constraints imposed by the bubble equations, and they give:
\begin{equation}
\hat Q_1 \, \hat Q_5  ~=~2 \hat \kappa_3 \,\hat{m}_{\sst S}~=~ \hat \kappa_3 \, \hat{a} \, \bigg[ \, \hat \kappa_3  \,\bigg(   \frac{  \hat Q_P}{|\vec x_P|} ~+~ 1 \bigg )    ~-~  \frac{2\, \hat{m}_P}{|\vec x_P|}  \, \bigg]  \,,  \label{regcondc}
\end{equation}
and
\begin{equation}
\frac{\hat Q_P}{|\vec x_P -\vec x_{\sst S}|} ~=~ \frac{2\,\hat{m}_{\sst S}}{\hat \kappa_3 \, \hat{a}} ~-~ 1~=~ \frac{\hat Q_1 \, \hat Q_5}{\hat \kappa_3 ^2 \, \hat{a}} ~-~ 1 \,.   \label{regcondd}
\end{equation}

While we are  allowing a singular momentum source, we must make sure that there are no CTCs at this source, which means that there must be no Dirac strings ending at the source. Using the second expression in (\ref{varpieqn}), the absence of $\vec \nabla \frac{1}{r_{\sst P}}$ sources leads to the condition:
\begin{equation}
 \frac{\hat{m}_{\sst P}}{|\vec x_{\sst P}|} ~+~  \frac{\hat \kappa_3 \, \hat Q_{\sst P}}{2}\, \bigg( \frac{1}{|\vec x_{\sst P} -\vec x_{\sst S}| } ~-~  \frac{1}{|\vec x_{\sst P}|}  \bigg )     ~=~ 0   \,.   \label{regconde}
\end{equation}
This last equation, which is the third bubble equation, is satisfied as a consequence of  (\ref{regcondc}) and (\ref{regcondd}), which is to have been expected as a Dirac string must begin and end somewhere, and we have already eliminated all other possible ends for the Dirac string. 

We will satisfy all these regularity conditions by taking $\hat \kappa_3 $, $\hat Q_1$, $\hat Q_5$ and $\hat Q_{\sst P}$ as fundamental parameters with $\hat{m}_{\sst S}$ and  $\hat{m}_P$ determined by    (\ref{mPeqn2}) and (\ref{regcondc}):
\begin{equation}
\hat{m}_{\sst S}~=~ \frac{\hat Q_1 \, \hat Q_5 }{2 \,\hat \kappa_3 }  \,, \qquad \hat{m}_{\sst P}  ~=~  \frac{\hat \kappa_3 }{2}\, \bigg[ \, \hat Q_{\sst P} ~-~   |\vec x_{\sst P}| \, \bigg(\frac{\hat Q_1 \, \hat Q_5 }{\hat \kappa_3^2 \hat{a}}  ~-~1 \bigg) \bigg]   \,,  \label{mparams}
\end{equation}
and where the positions of the charges, $\hat a$ and $\vec x_P$, must satisfy the constraint~\eqref{regcondd}.

%%%%%%%%%%%%%%%%%%%
\subsection{Charges, positions and peaks}
\label{ss:Charges}
%%%%%%%%%%%%%%%%%%% 

The supergravity charges of the solution can be read off using the results in Section 6.4 of \rcite{Bena:2007kg}.%
\footnote{Note that in that paper $J_L$ denotes the angular momentum along the $\IR^3$ base of the solution, which is denoted by $J_R$ here.} First, the electric charges are given by:
\begin{equation}
Q_1 ~=~ 4 \, \hat Q_1 \,, \qquad Q_5 ~=~ 4 \, \hat Q_5 \,, \qquad Q_P ~=~ 4 \, \hat Q_P  \,.  \label{Qcharges}
\end{equation}
The angular momenta can then be read off from the expansion of $\mu$ at infinity:
\begin{equation}
\mu ~\sim~   \frac{1}{8\, \hat r} \, \big( \, \hat{J}_L ~+~ \hat{J}_R \cos \hat \theta \,  \big) ~+~ \dots   \,.  \label{muasymp}
\end{equation}
where $\cos \hat \theta \equiv \frac{z}{\hat r}$ in the cylindrical polar coordinates (\ref{3Dpolars}).

One then finds:
\begin{equation}
  \hat{J}_R ~=~ 4\, \hat \kappa_3 \, \hat{a}  \,, \qquad \hat{J}_L ~=~  8 \, ({\hat m}_P + {\hat m}_{\sst S})  \,.  \label{Jcharges1}
\end{equation}
Using (\ref{eq:coords2}), (\ref{K3norm}), (\ref{mparams}) and (\ref{regcondd})  we therefore obtain: 
\begin{equation}
 \hat{J}_R ~=~ \coeff{1}{2}\,  \kappa \, R_y \,a^2 \,, \qquad  \hat{J}_L ~=~   \coeff{1}{2}\,   \kappa\, R_y  \, \bigg[ \frac{Q_1 \,Q_5 }{\kappa^2 R_y^2}  ~+~ 4\, \big(|\vec x_P| - |\vec x_P - \vec x_{\sst S}|  \big)\bigg( 1~-~ \frac{Q_1 \,Q_5 }{\kappa ^2 R_y^2\, a^2 } \bigg) \bigg]\,,  \label{Jcharges2}
\end{equation}
and the constraint in \eqref{regcondd} becomes: 
\begin{equation}
\frac{Q_P}{4\, |\vec x_P -\vec x_{\sst S} |}  ~=~  \frac{Q_1 \,Q_5 }{\kappa^2 R_y^2\, a^2 } ~-~ 1    \,.  \label{beqn2}
\end{equation}

From  (\ref{eq:coords2}), one sees that 
\begin{equation}
  4\, \big( |\vec x_P - \vec x_{\sst S}|  - |\vec x_P| \big) ~=~ a^2 \, \cos( 2 \theta_P),
  \label{pointseps}
\end{equation}
 where $\theta_P$ is the coordinate of the singular source.  It follows that 
\begin{equation}
 \hat{J}_L - \hat{J}_R ~=~   \kappa \, R_y  \,a^2 \,  \bigg(\frac{Q_1 \,Q_5 }{\kappa^2 R_y^2\, a^2 } ~-~1  \bigg)    \,  \cos^2\theta_P \,.  \label{Jcharges3}
\end{equation}
The constraint,  (\ref{beqn2}), can be written 
\begin{equation}
Q_P ~=~    \bigg(\frac{Q_1 \,Q_5 }{\kappa^2 R_y^2\, a^2 } ~-~1  \bigg)    \, (r_P^2 + a^2  \cos^2\theta_P) \,.  \label{beqn3}
\end{equation}
The important point is that these last two equations determine the location, $(r_P, \theta_P)$ of the singular source in terms of the conserved charges, $Q_1, Q_5, Q_P, \hat{J}_L, \hat{J}_R$ and the positive integer $\hat \kappa_3$.

We now equate these charges to those of the single-mode superstratum with winding number one ($\hat \kappa_3 = \kappa =1$) and use (\ref{ssReg}) and (\ref{sschg1}).  One then obtains:
\begin{equation}
  \coeff{1}{2} \, R_y  \,b^2 \,   \cos^2\theta_P ~=~  \hat{J}_L - \hat{J}_R ~=~   \coeff{1}{2}\, R_y  \, \frac{m}{ k  } \, b^2 \,,  \label{Jchgmatch}
\end{equation}
\begin{equation}
   \frac{b^2}{2\, a^2 }  \, (r_P^2 +a^2   \cos^2\theta_P) ~=~ Q_P ~=~  \frac{(m+n )}{2 k  } \, b^2   ,  \label{QPmatch}
\end{equation}
and hence 
\begin{equation}
 \cos^2\theta_P ~=~   \frac{m}{ k }  \,, \qquad    r_P^2 ~=~   \frac{n}{ k } \, a^2  \,.\label{positions1}
\end{equation}
Note that, from the  five-dimensional viewpoint, the quantum number $k$ is not something directly visible as the coefficient in the harmonic functions \eqref{LIfns}, \eqref{Mfn} but a parameter inherited from the six-dimensional modes and introduced via \eqref{sschg1}. In five-dimensions one can only see the coarse-grained data $\frac{m}{k}$ and $\frac{n}{k}$.
  
On the other hand, the function that localizes the superstratum excitations, and gets squared in the averaged source function, is:
 \begin{equation}
 \Delta_{k,m,n} ~\equiv~  \bigg(\frac{r}{\sqrt{r^2+a^2}}\bigg)^n  \, \bigg(\frac{a}{\sqrt{r^2+a^2}}\bigg)^k  \, \sin^{k-m}\theta \,  \cos^{m}\theta\,.
  \label{Deltadefn-2}
 \end{equation}
Differentiating this with respect to $\theta$ and setting the result to zero gives 
 \begin{equation} 
\cos^2\theta_* ~=~   \frac{m}{ k } \,.
  \label{thpeak}
 \end{equation}
Similarly, locating the peak in $r$ gives
 \begin{equation} 
\frac{r_*^2}{a^2} ~=~   \frac{n}{ k }\,.
  \label{rpeak}
 \end{equation}
Comparing \eqref{positions1} with \eqref{thpeak} and \eqref{rpeak}, one sees that the regularity conditions of the effective superstratum localizes the singular source exactly at the peak of the bump function of the superstratum wave in the exact solution.  In particular, we see that the momentum and angular momentum of the superstratum are effectively moving away from the actual supertube locus. This is what we mean by \emph{momentum migration}.

The foregoing analysis is, in principle, sufficient for showing that, upon averaging over the $v$ direction, the six-dimensional superstratum reduces to the five-dimensional multi-center solution, giving effective localized source terms for momentum and angular momentum.
In the next section~\ref{sec:stratumred}, we explicitly demonstrate that the six-dimensional superstratum solution gives a delta-function source with the correct strength on the right hand of the second-layer equation
\eqref{avgBPS2c} in the limit where $k\sim m \sim n\gg 1$.

It is useful to underline the change in perspective created by going from the full six-dimensional superstratum to the five-dimensional effective superstratum. 
In scaling solutions in five dimensions, we require (\ref{scalingx}), which for the superstratum  means $a \to 0$.  This necessarily implies that $
\hat{J}_R = \frac{1}{2}\,  \kappa  R_y a^2$ is becoming small.  
On the other hand $Q_P$ and $\hat{J}_L$ can remain large. Indeed, recall that $\hat{J}_L$ is given by (\ref{Jcharges1}) and note that that we can re-write (\ref{mparams}) as
\begin{equation}
8\,\hat{m}_{\sst S}~=~  \frac{1}{2}\,\kappa \, R_y \, a^2\, \bigg(\frac{Q_1 \,Q_5 }{\kappa^2 R_y^2\, a^2 }  \bigg)  \,, \qquad 8\,\hat{m}_{\sst P}  ~=~   \frac{1}{2}\,\kappa \, R_y \, a^2\, \bigg(\frac{Q_1 \,Q_5 }{\kappa^2 R_y^2\, a^2 } ~-~1  \bigg) \,  \cos 2\theta_P   \,. \label{mparams2}
\end{equation}

In scaling superstrata, the quantities in parentheses are large, $\sim \frac{b^2}{a^2}$. Thus, from 
the five-dimensional perspective, the supertube center has a very large $\hat{J}_L$ (compared to  $\hat{J}_R$), that grows with $\frac{b^2}{a^2}$.  However, one also has $0 \le \theta_P \le  \frac{\pi}{2}$, which means that  the momentum center has a $\hat{J}_L$ that can be positive or negative. Indeed, for $\theta_P = \frac{\pi}{2}$ this angular momentum almost cancels the angular momentum of the supertube, as is evident from (\ref{Jchgmatch}) and (\ref{positions1}).

From the six-dimensional and CFT perspectives, $\hat{J}_L$ and $\hat{J}_R$ of the supertube are being modified by trading  maximally spinning modes ($|++ \rangle$ states) with  density modes  ($|00 \rangle$ states). One therefore would expect $\hat{J}_L = \hat{J}_R$ for such a supertube.  However, 
the momentum-carrying density modes can be non-spinning ($L_{-1}^n |00 \rangle$ states), or spinning even more strongly (${J^+_{-1}}^m |00 \rangle$ states).  This leads to the dependence on $\theta_P$, or $\frac{m}{k}$.  

In the scaling solution, the five-dimensional supertube center, somewhat counter-intuitively, has $\hat{J}_L \gg \hat{J}_R$, while the momentum center also carries a large  $\hat{J}_L$ that can add to, or almost cancel the total $\hat{J}_L$ of the system.  The six-dimensional picture is faithfully reproducing the CFT, while the five-dimensional effective theory  migrates the momentum excitations, and creates a somewhat counterintuitive split of  $\hat{J}_L$ between the centers.

%%%%%%%%%%%%%%%%%%%
\subsection{Delta function sources for the effective superstratum}
\label{sec:stratumred}
%%%%%%%%%%%%%%%%%%%

It is instructive to see how the $v$ averaging, and the large mode number limit,  of the superstratum solution gives rise to a delta-function source with the correct strength on the right hand of the second-layer equation in the large quantum number limit.

If we substitute the six-dimensional single-mode superstratum data
given in \eqref{eq:Zansatz2} into the averaged second-layer equation \eqref{avgBPS2c}, we find
\begin{align}
    \nabla_{(4)}^2\ev{Z_3}_v \;=\;
    -\ev*{(\dot{Z}_4)^2}_v-{1\over 2}*_4\ev{\Theta^4\wedge\Theta^4}_v
    \;=\; -b_{k,m,n}^2 \Delta_{k,m,n}^2 f(r,\theta) \,,
    \label{LapZ3=eff}
\end{align}
where we used the fact that $\Theta^{1}=0$, $\partial_v Z_2=0$ 
in the standard superstratum \rcite{Bena:2015bea,Bena:2017xbt}.  The function $f(r,\theta)$ is found to be
\begin{align}
    f(r,\theta)\;=\;2\frac{(k-m)^2 n^2 \Sigma +k (m+n) (k (m-n)+2 m n) r^2 \sin ^2\theta}{k^2 r^2 \left(r^2+a^2\right) \Sigma  \sin ^2\theta 
   \cos ^2\theta} \,.
\end{align}
As we saw in section \ref{ss:Charges}, the function $\Delta_{k,m,n}$ appearing in \eqref{LapZ3=eff} has a maximum at $(r_*,\theta_*)$  satisfying
\eqref{thpeak} and \eqref{rpeak}. The expansion of $\Delta_{k,m,n}(r,\theta)$ around that point is
\begin{align}
\label{Delta*}
    \Delta_{k,m,n}(r,\theta)
    \;\approx\; \Delta_{k,m,n}^* \exp[-{1\over 2A_{k,m,n}}(r-r_*)^2-{1\over 2B_{k,m,n}}(\theta-\theta_*)^2],
\end{align}
where
\begin{align}
\label{Delta coeffs}
    \Delta_{k,m,n}^* \,=\, {(k-m)^{k-m\over 2}m^{m\over 2}n^{n\over 2}\over (k+n)^{k+n\over 2}},\qquad
    A_{k,m,n}\,=\,{k+n\over 2k^2}a^2,\qquad    B_{k,m,n}\,=\,{1\over 2k} \,.
\end{align}
So, $\Delta_{k,m,n}$ is very sharply peaked ($A_{k,m,n}\ll a_2 $, $B_{k,m,n}\ll 1$), and can be regarded as a delta function if 
\be
\label{eq:kmnscal}
k \,\sim \, m \, \sim \, n \,\gg \, 1 \,,
\ee
where $k\sim m\sim n$ is required for the position $(r_*,\theta_*)$ to remain finite.

In order to find the strength of the delta function, one can integrate the right-hand side of \eqref{LapZ3=eff} over $\mathbb{R}^4$. We find
\begin{align}
    -(2\pi)^2 b_{k,m,n}^2 \big[f(r,\theta) \Sigma\, r \sin\theta \cos\theta\big]\Big|_{r_*,\theta_*}\int dr d\theta\, \Delta_{2k,2m,2n}(r,\theta)\;=\;-4\pi^2 Q_P,
    \label{intRHS}
\end{align}
where $(2\pi)^2$ is from the $\phie,\psie$ integrals and $\Sigma\, r\sin\theta\cos\theta$ is from the volume form of the metric \eqref{ds4flat}. We also used relations such as
\begin{align}
\begin{gathered}
f(r_*,\theta_*)\,=\,{2k^2\over a^4}\,,\qquad
\int dr d\theta\, \Delta_{2k,2m,2n} \,\approx\, 2\pi \sqrt{A_{2k,2m,2n}B_{2k,2m,2n}}\,\Delta_{2k,2m,2n}^*\,,\\
 b_{k,m,n}^2
 \,\approx\, 
 Q_P
 {k^2\over \pi \sqrt{(k-m)(k+n)mn\,}(m+n)}\,{(k+n)^{k+n}\over (k-m)^{k-m}m^m n^n}\,,\label{fswx15Apr25}
\end{gathered}
\end{align}
where we can derive the last expression from \eqref{sschg1} and \eqref{sschg2} using Stirling's formula.
Comparing \eqref{LapZ3=eff} and \eqref{intRHS}, we find
\begin{align}
    \int_{\mathbb{R}^4} d^4x\, \nabla_{(4)}^2\ev{Z_3}_v\;=\;-4\pi^2 Q_P \,.
\end{align}
Because of the relation \eqref{GHmet} between four-dimensional and three-dimensional bases, this implies that $Z_3$ has a pole with the correct coefficient:
\begin{align}
    \ev{Z_3}_v \sim {Q_P\over 4\hat r_P}\;=\;{\hat{Q}_P\over \hat r_P}\,.
    \label{Z3fromStrata}
\end{align}
Namely, in the large $k,m,n$ limit, the superstratum wave localizes in $r,\theta$, and is effectively described by a pointlike center in a five-dimensional multi-center solution.

To be more precise, because the superstratum wave is a plane wave delocalized along the $\phie,\psie$ directions, the center in the three-dimensional base is also delocalized along the azimuthal direction $\phig$.
To obtain a harmonic function localized in the $\phig$ direction as in \eqref{Z3fromStrata},  in six dimensions, one would have to take a superposition of the superstratum wave \eqref{eq:Zansatz2} with different values of $k$ and $m$ to localize the wave, which should be possible for $k,m\gg 1$.   

%%%%%%%

%%%%%%%%%%%%%%%%%%%%%%%%%%%%%%%%%
\section{Geodesics and Wavefunctions}
\label{sec:wavefns}

In the WKB limit, supergraviton wavefunctions localize along null geodesics.  We can understand the localization of the modes $\Delta_{k,m,n}$ of  \eqref{Deltadefn} at large $k,m,n$ by studying BPS null geodesics in the underlying supertube geometry.  Similarly, the F1-P solutions of Section~\ref{sec:wound-strings} will follow from the underlying group symmetry.

In Section~\ref{sec:GLMT soln}, we reviewed the fact that in the AdS$_3$ limit, the GLMT geometry is an orbifold of the group manifold $\sltwo\times\sutwo$, together with a spacetime spectral flow large coordinate transformation, \eqref{ads-glmt}--\eqref{eq:sug-orbact-GLMT}.

The null geodesics in the AdS$_3$ GLMT solutions can therefore be studied by first performing an analysis of null geodesics in $\sltwo\times\sutwo$, then  implementing the orbifold quotient, and then making the fractional spectral flow coordinate transformation to go back to the rotating spacetimes \eqref{ads-glmt}.

In this section, we consider such null geodesics on the $\sltwo\times\sutwo$ group manifold and reproduce the features~\eqref{thpeak}, \eqref{rpeak} of the $\Delta_{k,m,n}$.  We then show that these functions are simply the BPS harmonics (Wigner $d_{jm\bar m}$-functions) on the group manifold, from which it is straightforward to understand the localization in the large $k$ limit.  

Passing to fundamental strings on the group manifold, a classical \emph{worldsheet} spectral flow transformation in the WZW model on the group turns the BPS geodesic trajectories of a massless particle into primitive winding strings, which are the classical limit of the quantized string worldsheet states analyzed in Section~\ref{sec:wound-strings}.%
\footnote{Thus we consider the NS5-F1 duality frame S-dual to that of the previous and following sections, so that we can use string worldsheet methods.  The round NS5-F1 supertube admits an exact worldsheet description as a gauged Wess-Zumino-Witten model~\rcite{Martinec:2017ztd}, or as a group orbifold~\rcite{Martinec:2023zha}, and so we can compare classical probe properties to corresponding string vertex operators, as we do in Section~\ref{sec:wound-strings}.}
These transformations ``spin up'' the string along a combination of the worldsheet coordinates, typically yielding a string trajectory that winds around a particular cycle in the $(y,\phi,\psi)$ torus of the target $\sltwo\times\sutwo$ spacetime.

These solutions will be reproduced by yet another approach in the S-dual R-R flux background in Section~\ref{sec:D1-P}, by considering the bubble equations that arise when the back-reaction of the string is taken into account.

%%%%%%%%%%%%%%%%%%%%%%%%%%%%%%%%%
\subsection{BPS geodesics and classical strings on \texorpdfstring{$\sutwo$}{}}
\label{sec:SU2 geod}

At high momentum, string wavefunctions are concentrated along semi-classical trajectories, which are geodesics on the group manifolds $\sltwo$ and $\sutwo$. We follow the recent discussion in~\cite{Martinec:2025xoy}. Consider $\sutwo$, parametrized by Euler angles
\be
\label{gsumat1}
g_\su \;=\; e^{-i(\phi-\psi)\sigma_3/2}\, e^{i(\pi/2-\theta)\sigma_2}\, e^{-i(\phi+\psi)\sigma_3/2} \;=\; 
\left(\begin{matrix} e^{-i\phi}\sin\theta ~&~  e^{i\psi}\cos\theta \\ - e^{-i\psi}\cos\theta ~&~ e^{i\phi} \sin\theta \end{matrix}\right) ~.
\ee
Classical solutions to the WZW model take the form
\be
\label{gclassical}
g(z,\bar z) =  g_\ell(z) g_r(\bar z) ~,
\ee
where $z,\bar z=\xi_0\pm\xi_1$ are worldsheet coordinates.  The classical solution
\be
\label{suhwtgeod}
g_\ell(z) = e^{-i\nu' z \sigma_3/2}
~~,~~~~
g_r(\bar z) = e^{-i\nu' \bar z \sigma_3/2} 
\ee
describes geodesic motion along the $\phi$ circle at $\theta=\frac\pi2$.  One can then rotate this to some other geodesic on $\bS^3$ via
\be
\label{sugengeod}
g_\ell(z) =  e^{-i\alpha'_\ell \sigma_1/2}  \,  e^{-i\nu' z \sigma_3/2} 
~~,~~~~
g_r(\bar z) =  e^{-i\nu' \bar z \sigma_3/2}  \,  e^{-i\alpha'_r \sigma_1/2} ~.
\ee
The $\sutwo$ conserved charges of this geodesic motion are
\begin{align}
\label{suqnums}
\cE &= \frac{\nfive}2\tr\bigl[\partial g \partial g^{-1}\bigr] + \frac{\nfive}2\tr\bigl[\bar\partial g \bar\partial g^{-1}\bigr] = \frac{\nfive}2 (\nu')^2,
\nn\\[.2cm]
 J^3 &= -\frac{i\nfive}2 \tr\bigl[(\partial g)g^{-1}\sigma_3\bigr] = -\frac{\nfive}2\,\nu'\cos(\alpha'_\ell),
\\[.2cm]
\bar J^3 &= -\frac{i\nfive}2 \tr\bigl[g^{-1}(\bar\partial g)\sigma_3\bigr] = -\frac{\nfive}2\,\nu' \cos(\alpha'_r)
  ~~;
\nn
\end{align}
in the quantum theory, these quantities are related to the (half) integer quanta $j',\msu,\bmsu$ of $\sutwo$ representation theory via
\be
\label{jmsu}
j'\sim \frac\nfive2\nu'
~~,~~~~
\frac{\msu}{\jsu} \sim -\cos\big(\alpha'_\ell\big)
~~,~~~~
\frac{\bmsu}{\jsu} \sim -\cos\big(\alpha'_r\big)  ~.
\ee

We will be interested in BPS trajectories associated to lowest weight states on the right, and so we set $\alpha'_r=0$.
One then finds the geodesic motion 
\be
\label{geodesic gsu}
g(\xi_0) = 
\left(\begin{matrix} 
e^{-i\nu'\xi_0}\cos\frac{\alpha'_\ell}2  
~~&~  
-ie^{+i\nu'\xi_0}\sin\frac{\alpha'_\ell}2 
\\[.3cm] 
-ie^{-i\nu'\xi_0}\sin\frac{\alpha'_\ell}2 
~~&~~ 
e^{+i\nu'\xi_0}\cos\frac{\alpha'_\ell}2  
\end{matrix}\right) ~.
\ee
Comparing this matrix to \eqref{gsumat1}, one finds a trajectory that sits at a fixed value of $\theta$ 
\be
\label{thetamotion}
-\cos \theta_* = 
\cos(\alpha'_\ell/2)  ~,
\ee
or in other words
\be
\label{thetapm}
\theta_* =\half\big( \pi - |\alpha'_\ell| \big) ~.
\ee
The geodesic~\eqref{suhwtgeod} with $\alpha'_\ell=\alpha'_r=0$ sits at the pole $\theta=\frac\pi2$ and corresponds to waveforms with $\msu=\bmsu=-j'$.

The unitary range of allowed values is $0\le\nu'\le1$,%
\footnote{The unitarity bound for quantized strings on the $\sutwo$ and $\sltwo$ group manifolds is seen in the classical theory as a bound on stationary solutions~-- when the momentum exceeds a particular value, the Lorentz force of the background B-field exceeds the string tension and pulls the string apart.  See for instance~\rcite{Martinec:2020gkv} Section 5 for a discussion.}
while $0\le\alpha'_{\ell,r}\le\pi$ code $\msu,\bmsu$ (or rather coherent states).%
\footnote{The classical solution corresponds to the limit of large $\nfive$, with $\nu'$ held fixed, and so does not distinguish between $(\nu')^2=(\frac{j'}\nfive)^2$ and $(\nu')^2=\frac{j'(j'+1)}{\nfive^2}$.}
In particular, $\alpha'_\ell=0$ corresponds to $\msu=-j'$, and the solution~\eqref{suhwtgeod} corresponds to the lowest weight state.  Holding $\alpha'_r=0$ and dialing $\alpha'_\ell$ coherently excites larger values of $\msu$, resulting in circular trajectories concentrated at fixed latitude lines~\eqref{thetapm}.  

Combining~\eqref{thetapm}, \eqref{jmsu}, and using $\msu=-j'+m$, one identifies
\be
\label{thetamap}
\frac{m}{2j'} = \half\Big(\frac{\msu}{j'}+1\Big) = \cos^2 \theta_*  ~.
\ee

%%%%%%%%%%%%%%%%%%%%%%%%%%%%%%%%%
\subsection{BPS geodesics and classical strings on \texorpdfstring{$\sltwo$}{}}

Similarly, for $AdS_3$ one has the Euler angle parametrization in terms of global coordinates $(\tau,\sigma,\rho)$ 
\be
\label{gslmat}
g_\sl \;=\; e^{i(\tau+\sigma)\sigma_3/2}\, e^{\rho\sigma_1}\, e^{i(\tau-\sigma)\sigma_3/2} \;=\; 
\left(\begin{matrix} e^{i\tau}\cosh\rho ~&~  e^{i\sigma}\sinh\rho \\ e^{-i\sigma}\sinh\rho ~&~ e^{-
i\tau} \cosh\rho \end{matrix}\right) ~.
\ee
Highest weight states in the discrete series representation of $\sltwo$ correspond to geodesics
\be
\label{slhwtgeod}
g_\ell(z) = e^{i\nu z \sigma_3/2}
~~,~~~~
g_r(\bar z) = e^{i\nu \bar z \sigma_3/2} ~.
\ee
The matrix $g_\sl$ is diagonal, and thus the geodesic sits at $\rho=0$, the center of $AdS_3$, running up the time axis at a velocity $\nu$.

The boost transformation
\be
\label{slgengeod}
g_\ell(z) =  e^{\alpha_\ell \sigma_1/2}  \,  e^{i\nu z \sigma_3/2} 
~~,~~~~
g_r(\bar z) =  e^{i\nu \bar z \sigma_3/2}  \,  e^{\alpha_r \sigma_1/2} ~,
\ee
leads to a geodesic trajectory with the $\sltwo$ conserved quantum numbers
\begin{align}
\label{slqnums}
\cE &= -\frac{\nfive}2\tr\bigl[\partial g \partial g^{-1}\bigr] - \frac{\nfive}2\tr\bigl[\bar\partial g \bar\partial g^{-1}\bigr] = -\frac{\nfive}2 \, \nu^2,
\nn\\[.2cm]
J^3 &= \frac{i\nfive}2 \tr\bigl[(\partial g)g^{-1}\sigma_3\bigr] = \frac{\nfive}2\,\nu\cosh(\alpha_r)  ,
\\[.2cm]
\bar J^3 &= \frac{i\nfive}2 \tr\bigl[g^{-1}(\bar\partial g)\sigma_3\bigr] = \frac{\nfive}2\,\nu \cosh(\alpha_\ell)
~~; 
\nn
\end{align}
Again we have 
\be 
\label{jmsl}
j=\frac{\nfive}2\nu
~~,~~~~
\frac\msl\jsl \sim \cosh\alpha_\ell
~~,~~~~
\frac\bmsl\jsl \sim \cosh\alpha_r
~,
\ee 
with $\alpha_r=0$ for BPS geodesics.
Multiplying out the group elements~\eqref{slgengeod}
\be
\label{geodesic gsl}
g(\xi_0) = 
\left(\begin{matrix} 
e^{+i\nu\xi_0}\cosh\frac{\alpha_\ell}2 
~~&~  
e^{-i\nu\xi_0}\sinh\frac{\alpha_\ell}2 
\\[.3cm] 
e^{+i\nu\xi_0}\sinh\frac{\alpha_\ell}2 
~~&~~ 
e^{-i\nu\xi_0}\cosh\frac{\alpha_\ell}2 
\end{matrix}\right) ~,
\ee
one finds that the geodesic sits at the fixed radial position
\be
\label{rhopm}
\cosh \rho_* = \cosh(\alpha_\ell/2)
~~\Longrightarrow~~
\rho_* = \half\,\alpha_\ell ~.
\ee
To compare to \eqref{rpeak}, we recall from \eqref{eq:coords3}
\be
\label{rhomap}
\frac{r}{a} = \sinh\rho ~.
\ee
Combining~\eqref{rhopm}, \eqref{jmsl}, one identifies (for $D^+$ representations with $\msl=j+n$) 
\be
\label{rhomap2}
\frac{n}{2j} = \half\Big(\frac{\msl}{j}-1\Big) = \sinh^2\rho_*  ~.
\ee

%%%%%%%%%%%%%%%%%%%%%%%%%%%%%%%%%
\subsection{Wavefunctions}

The eigenfunctions of the scalar Laplacian on the $\sutwo$ group manifold (Wigner functions) reflect the foregoing semi-classical features.  These (unnormalized) wavefunctions are
\begin{align}
\label{su2wavefns}
D_{j' \msu \bmsu}(\theta,\phi,\psi) &= e^{-i \msu (\phi+\psi)}\,e^{-i\bmsu(\phi-\psi)}\, d_{j'\msu\bmsu}(\theta),
\nn\\[.2cm]
d_{j'\msu\bmsu}(\theta) &= %\cN_{j',\msu,\bmsu} 
(\cos\theta)^a (\sin\theta)^b P_{q}^{(a,b)}(\theta),
\\[.2cm]
P_{q}^{(a,b)}(\theta) &= \sum_{p=0}^q \bigg(\begin{matrix}{q+a}\\{q-p}\end{matrix}\bigg)\bigg(\begin{matrix}{q+b}\\{p}\end{matrix}\bigg) \big(\sin\theta\big)^{2p}\big(\cos\theta\big)^{2q-2p},
\nn
\end{align}
where $a=|\msu-\bmsu|$, $b=|\msu+\bmsu|$, $q=j'-\mu$, with $\mu=\max(|\msu|,|\bmsu|)$.

\begin{figure}[ht]
\centering
\includegraphics[width=0.45\textwidth]{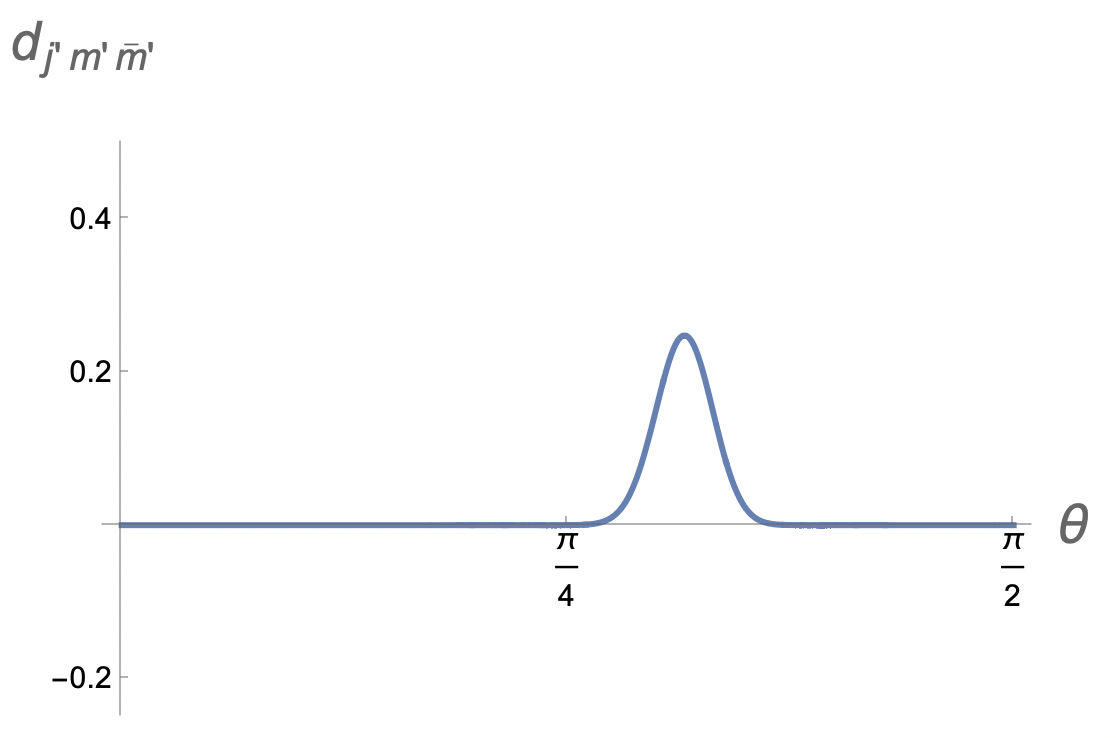}
\caption{ \it Example of the $\theta$-dependent part of a Wigner function, $d_{j'\msu \bmsu}(\theta)$, for $\sutwo$. 
For $|\bmsu|=j'$ or $|\msu|=j'$, the wavefunction is peaked at a particular polar angle $\theta_*$, and has a width of order $1/\sqrt{j'}$.
}
\label{fig:YtildeTubeS3}
\end{figure}

For the BPS states, one has  $\bmsu=-j'$ ($\alpha'_r=0$) and thus $\mu=j'$, and the sum over $p$ in the Jacobi polynomial $P_q^{(a,b)}$
collapses to a constant since $p=q=0$.  One then has $a=\sfm'\tight+j'=m$, $b=j'\tight-\sfm'=k\tight-m$; the trigonometric polynomial $d_{j'\msu\bmsu}(\theta)$ has a single peak at $\theta_*=\theta_+=\theta_-$ determined the value of $\msu$ given by $\alpha'_{\ell}$ via~\eqref{thetapm}, \eqref{suqnums}.%  
\footnote{If one were to consider non-BPS waveforms, the Jacobi polynomial would be non-trivial, having several nodes corresponding to a standing wave that oscillates back and forth in the effective potential for $\theta$ (similarly for $\rho$).}
The resulting wavefunctions are simply the $\theta$-dependent factors in $\Delta_{k,m,n}$, Eq.~\eqref{Deltadefn}.  
At large $j'$, these are just the WKB wavefunctions associated to the classical trajectories~\eqref{sugengeod}, \eqref{thetamotion}.
A representative such wavefunction is plotted in Figure~\ref{fig:YtildeTubeS3}.

The eigenfunctions of the scalar Laplacian on $AdS_3$ again reflect the properties of the geodesics.  The $S^3$ and $AdS_3$ metrics are related (up to an overall sign) by the analytic continuation
\be
\label{anglemap}
(\coeff{\pi}2-\theta)\to -i\rho
~~,~~~~
\phi\to -\tau
~~,~~~~
\psi\to \sigma+\coeff\pi2 ~.
\ee
The discrete series representations of $\sltwo$ result from the continuation of the $\sutwo$ representations if we let
\be
\label{sutosl}
j' \to -j
~~,~~~~
\msu \to \msl
~~,~~~~
\bmsu \to \bmsl
~~,~~~~
\mhat \to \nhat
\ee
(in particular $\msu=-j'+\mhat$ maps to $\msl=j+\nhat$).
The lowest weight state wavefunction of the $\sutwo$ representation of spin $j'$ continues to the lowest weight state wavefunction of the positive discrete series $D^+_j$, 
\be
\label{highest wt}
e^{-2ij'\phi}(\sin\theta)^{2j'} ~\longrightarrow~
e^{-2ij\tau}(\cosh\rho)^{-2j} ~~,
\ee
and the fact that the raising operators map from those of $\sutwo$ to those of $\sltwo$ guarantees that the rest of the wavefunctions~\eqref{su2wavefns} continue appropriately from $\sutwo$ to $\sltwo$.  The continuation of $j$ means that the representation has no highest weight.  In particular, for $j\tight-1=j'=2k$, $\bar m=j$ and $m=j\tight+n$, one finds, using \eqref{rhomap}, the $r$-dependent contribution to $\Delta_{kmn}$, defined in \eqref{Deltadefn}.  

The highest weight wavefunctions~\eqref{highest wt} exhibit the narrowing of the spread of support as $j,j'$ grow large.  The width is of order $1/\sqrt{j}$ in units of the AdS$_3$ curvature radius for the $\sltwo$ harmonics, and $1/\sqrt{j'}$ in units of the $S^3$ radius for $\sutwo$ harmonics, as one sees for instance from~\eqref{Delta*}, \eqref{Delta coeffs}.

With the identification~\eqref{rhomap},
and the mass shell condition $j\approx j'$ for massless wavefunctions (with $O(1)$ shifts depending on the polarization state that are irrelevant in the semi-classical limit), the wavefunctions~\eqref{Deltadefn} are none other than the Wigner functions~\eqref{su2wavefns} (and their $\sltwo$ continuations) that are BPS on the right, $\bmsu=-j'$ and $\bmsl=j$.  The spectral flow~\eqref{eq:sf-GLMT-sug} that maps the factorized $\sltwo\times\sutwo$ group manifold to the $AdS_3$ limit~\eqref{adslim} of the GLMT geometry only mixes the coordinates $\tau,\sigma,\phi,\psi$, hence does not alter the location of the wavefunctions/geodesics in $\rho,\theta$.

These wavefunctions are then related to the massless geodesics with $\nu\approx \nu'$ for the classical solutions~\eqref{sugengeod}, \eqref{slgengeod} in the WKB limit of large $j,j'$.  The BPS condition sets $\alpha_r=\alpha'_r=0$, and thus from~\eqref{rhopm} $\rho_*=\hf\alpha_\ell$, and from~\eqref{thetapm} $\theta_*=\hf(\pi\tight-|\alpha'_\ell|)$; the wavefunctions are thus strongly peaked at these values.
From~\eqref{suqnums}, \eqref{slqnums} we identify
\be
\mhat = \frac\nfive2 \nu'(1-\cos\alpha'_\ell)
~~,~~~~
\nhat = \frac\nfive2 \nu  (\cosh\alpha_\ell-1).
\ee
Of course, these values are none other than those found by differentiating the waveforms to find their peaks, equations~\eqref{thpeak} and~\eqref{rpeak}.

%%%%%%%%%%%%%%%%%%%%%%%%%%%%%%%%%%%%%
\subsection{Spectral flow to generate winding strings}
\label{WS specflow}

The classical version of $\sutwo$ spectral flow spins the string trajectory around the center-of-mass geodesic motion~\eqref{gclassical} according to~\rcite{Martinec:2020gkv}
\be
\label{cl SU flow}
g^{(\wsu,\bwsu)} (z,\bar z) = e^{-i\wsu z\,\sigma_3/2}g_\ell(z)\, g_r(\bar z) e^{-i\bwsu\bar z\,\sigma_3/2}  ~.
\ee
This transformation creates a string winding along the various Euler angles $\phi,\psi$ (and correspondingly $\sigma$ for $\sltwo$ spectral flow).
In general $\wsu\ne\bwsu$, in which case spectral flow extends the string along both Euler angles $\phi$ and $\psi$ in a correlated manner; for $\wsu=\bwsu$, the string winds only along $\phi$; while for $\wsu=-\bwsu$, the string winds only along $\psi$.  For BPS states on the right, we have $\bmsu=-j'$ and so $\alpha'_r=0$; the spectrally flowed matrix $g_\su$ is
\be
\label{flowed gsu}
\left(\begin{matrix} 
e^{-i[(2 \nu'+w' + \bar w' )\xi_0 + (w' - \bar w' ) \xi_1]/2}\cos\frac{\alpha'_\ell}2 ~&~  
e^{i[ (2 \nu'-w' + \bar w' )\xi_0 - (w' + \bar w' ) \xi_1]/2}\sin\frac{\alpha'_\ell}2 \\ 
- e^{-i[ (2 \nu'-w' + \bar w' )\xi_0 - (w' + \bar w' ) \xi_1]/2}\sin\frac{\alpha'_\ell}2 ~&~ 
e^{i[(2 \nu'+w' + \bar w' )\xi_0 + (w' - \bar w' ) \xi_1]/2} \cos\frac{\alpha'_\ell}2 
\end{matrix}\right) ~,
\ee
where again $z=\xi_0+\xi_1, \bar z = \xi_0-\xi_1$.

Similarly, spectral flow in $\sltwo$ is the transformation
\be
\label{cl SL flow}
g^{(\wsl)} (z,\bar z) = e^{i\wsl z\,\sigma_3/2}g_\ell(z)\, g_r(\bar z) e^{i\wsl\bar z\,\sigma_3/2}  ~;
\ee
the left/right transformations are equal because the timelike direction is non-compact.  The result is the matrix~\eqref{flowed gsu} with the primes dropped, $\wsl=\bar\wsl$, and trigonometric functions replaced by hyperbolic functions according to~\eqref{anglemap}.

Processing this solution through the coordinate transformation~\eqref{eq:sf-GLMT-sug} and noting that the BPS condition also imposes $\bwsu=\wsl$,%
\footnote{The latter follows from the worldsheet physical-state constraints, to be discussed in Section~\ref{sec:backgds}.}
recalling that the Euler angles in Section~\ref{sec:SU2 geod} are actually $\phi_\NS,\psi_\NS$, one finds the classical string trajectory
\begin{align}
\phi &= +\big((2s+1)\wsl+\wsu\big)(\xi_0 + \xi_1),
\nn\\
\psi &= -\big((2s+1)\wsl+\wsu\big)(\xi_0 + \xi_1),
\\
\sigma &= -\nu\xi_0 + w\xi_1,
\nn\\
\tau &= (\nu+w)\xi_0 .
\nn
\end{align}
Strings with $\wsl\ne 0$ are $AdS_3$ giant gravitons which puff up along the azimuthal direction parametrized by $\sigma$ as they evolve along the cap time coordinate $\tau$, which because of the mixing of coordinates~\eqref{eq:sf-GLMT-sug} in the background also extends along $S^3$, while those with $\wsu\ne 0$ puff up along $S^3$.
A key feature of these spectral flow transformations is that, while they puff up the string by making it wind the various Euler angles in the geometry, they do not move its location in $\rho,\theta$; it remains at the location $\rho_*,\theta_*$, given respectively by~\eqref{rhomap2} and~\eqref{thetamap}.

Furthermore, the center-of-mass wavefunctions don't change under spectral flow, and thus wavefunctions of the quantized winding string states remain the same $\Delta_{kmn}$ as one has for the unwound strings.

%%%%%%%%%%%%%%%%%%%%%%%%%%%%%%%%%%%%%
\section{Effective description of momentum waves and supertube probes}
\label{sec:D1-P}
%%%%%%%%%%%%%%%%%%%%%%%%%%%%%%%%%%%%%

In Section~\ref{sec:EffSS} we saw that averaging the superstratum led to a five-dimensional effective description with three Gibbons-Hawking centers: the two background centers which describe the seed two-charge circular supertube, and a momentum center coming from the superstratum mode. The position of the momentum center in the Gibbons-Hawking base exactly matches the position of the maximum of the function that describes the superstratum mode in the full six-dimensional solution.  

One can also use this philosophy to obtain effective descriptions of the string excitations on ${\rm AdS}_3 \times S^3 \times T^4$ geometries.  We will treat both F1-P probes in the GLMT geometry sourced by purely NS-NS fluxes, and the S-dual situation of D1-P probes in the geometry sourced by R-R fluxes.%
\footnote{The NS-NS duality frame tends to be more appropriate in the regime $n_1\gg n_5$, while the R-R frame tends to be more appropriate when $n_1\approx n_5$.}   
We will also take  the three-dimensional perspective developed in \cite{Denef:2000nb, Denef:2002ru,Bates:2003vx, Denef:2007vg} in which the bubble equations are integrability conditions of four-dimensional multi-center solutions. In more practical terms, this means we will focus on the geometry of the $\IR^3$ of the GH base space.

%%%%%%%%%%%%%%%%%%%%%%%%%%%%%%%%%%%%
\subsection{Adding a momentum center to the round supertube }
%%%%%%%%%%%%%%%%%%%%%%%%%%%%%%%%%%%%

\begin{figure}[htbp]
\begin{center}
\begin{tikzpicture}[scale=0.5]

\coordinate (A) at (0, 5);
\coordinate (B) at (2,4);
\coordinate (C) at (0, 0);

\draw[thick] (A) -- (B) -- (C) -- cycle;

\fill[black] (A) circle (3pt);
\fill[black] (B) circle (3pt);
\fill[black] (C) circle (3pt);

\node[anchor=east] at (A) {(1)};
\node[anchor=west] at (C) {LM, (2)};
\node[anchor=west] at (B) {P, (3)};

\node[anchor=south] at ($ (A)!0.5!(B) $) {$\hat b$};
\node[anchor=west]  at ($ (B)!0.5!(C) $) {$\hat c$};
\node[anchor=east]  at ($ (C)!0.5!(A) $) {$\hat a$};
\end{tikzpicture}
\caption{\it There are three centers corresponding to the center of space,``(1)'', a supertube center, ``(2)'', and the momentum center, ``(3)''.
\label{fig:3ctrPP}}
\end{center}
\end{figure}
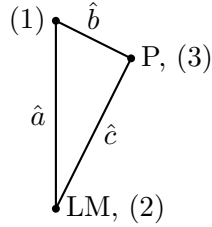

In order to explain the philosophy of the computation in a simple context, we first revisit 
the solution analyzed in Section \ref{sec:5DeffSS}, which amounted to adding a momentum center to the round supertube. The conventions for the distances between the centers are given in Figure \ref{fig:3ctrPP}.

We add to the round supertube solution, described by the two-center solution of \eqref{eq:LM limit of GLMT}, a third center with a charge vector given by:
\begin{equation}
    \Gamma_{P}=\bigl(0,(0,0,0),(0,0,\hat{Q}_3'),\hat{m}'\bigr)\,.
\end{equation}
This center corresponds to the center (with a slightly different notation) at $\hat r_P =0$  in (\ref{LIfns}) and~(\ref{Mfn}).  In all, the charge vector of the centers at $\hat r =0$, $\hat r_S =0$ and $\hat r_P =0$, and the asymptotic moduli vector are
\begin{align}
\begin{split}
\label{eq:3charges P center}
\Gamma^{(1)}&=\bigl(1,(0,0,-\hat{\kappa}_3),
(0,0,0),
0\bigl),\\
\Gamma^{(2)}&=\bigl(0,(0,0,\hat{\kappa}_3),(\hat{Q}_1,\hat{Q}_2,0),
\tilde{m}\bigr),
\\
\Gamma^{(3)}&=\Gamma_{P}
=
\bigl(0,(0,0,0),(0,0,\hat{Q}_3'),\hat{m}'\bigr),\\
h&=\big(0,(0,0,0),(0,0,1),0\big)~,
\end{split}
\end{align}
where we have dropped the $1$'s in $L_1$ and $L_2$ because we are interested in AdS$_3$ asymptotics.

%%%%%%%%%%%%%%%%%%%
\subsubsection{A probe approximation}
%%%%%%%%%%%%%%%%%%%
\label{sec:P-probe}

Let us first employ a probe approximation in which the momentum center with charge $\Gamma^{(3)}$ is much lighter than the other centers, and determine its position. Namely, we treat primed charges $\hat{Q}'_3,\hat{m}'$ to be much smaller than other, unprimed charges. This corresponds to a superstratum in the $b/a \ll 1$ regime discussed in section \ref{sec:SingleModeSS}, where a light momentum center sits in the AdS$_3\times S^3$ background created by $\Gamma^{(1)},\Gamma^{(2)}$.  We will discuss the opposite regime $b/a\gg 1$, where the momentum center is heavy and creates a long AdS$_2$ throat, in section \ref{sec:deep_AdS2}.

In this probe approximation, the distance $\hat{a}$ between the two background centers is close to its unperturbed value $\hat{a}_0$. We denote the change in $\hat{a}$ from the unperturbed value by $\hat{a}'=\hat{a}-\hat{a}_0$, which is of the order of the primed charges.

The bubble equations are given by:
\begin{equation}
    \frac{\Gamma^{12}}{\hat a}+\frac{\Gamma'^{13}}{\hat b}=\frac{\hat{\kappa}_3}{2}
    ,\qquad \frac{\Gamma'^{13}}{\hat b}+\frac{\Gamma'^{23}}{\hat c}=0. \label{eq:first bubble ST}\,
\end{equation}
We put primes on symplectic products that are proportional to primed charges and are thus small. The leading terms in \eqref{eq:first bubble ST} gives the unperturbed separation 
\begin{equation}
 \label{eq:ahat0}   \hat{a}_0=\frac{2\Gamma^{12}}{\hat\kappa_3}=\frac{\hat{Q}_1\hat{Q}_2}{\hat{\kappa}_3^2}\,
\end{equation}
while the subleading terms lead to
\begin{align}
    \frac{\hat{\kappa}_3}{2}\left(\frac{\hat a'}{\hat 
    a_0}\right)=\frac{\Gamma'^{13}}{\hat b},\qquad
    \frac{ \hat{\kappa}_3 \, \hat{Q}_3'}{\hat c} + \frac{ - \hat{\kappa}_3 \, \hat{Q}_3' + 2 \, \hat{m}'}{\hat b} = 0.
    \label{eq:linearized_bbl_eqs}
\end{align}

The asymptotic charges \eqref{Qcharges} are given by 
\begin{align}
    Q_1= 4\hat{Q}_1 \,,
    \qquad
    Q_{\sst P}= 4\hat{Q}_3'
    \,,\qquad
    Q_5=4\hat{Q}_2 \,.
\end{align}
The angular momenta \eqref{eq:j-non-q} are given by
\begin{align}
    \hat{J}_L=\tilde{m}+\hat{m}',\qquad
    \vec{\hat{J}}_R=\frac{\hat{\kappa}_3}{2} \hat{a} \, \vec{z} \quad~ \Rightarrow \quad~ \hat{J}_R\,=\,\frac{\hat{\kappa}_3}{2}\hat a \,=\, \frac{Q_1 Q_5}{32 \hat{\kappa}_3}+\frac{\hat{\kappa}_3}{2} \hat{a}'\,,
\end{align}
where $\vec{z}=(0,0,1)$.
The (small) change in $\hat J_R$ by the addition of the probe center is
\begin{align}
\hat{J}_R'=
\frac{\hat{\kappa}_3}{2} \hat{a}'\,,
\end{align}
which we interpret as the angular momentum of the probe.
Then the bubble equations \eqref{eq:linearized_bbl_eqs} allow us to express
the distances $\hat{b},\hat{c}$ in terms of the probe charges as follows:
\begin{equation}
\label{abc relation}
    \hat{b}\,=\, \hat a_0\frac{\Gamma_{13}'}{ \hat{J}_R'} \,, \qquad
    \hat{c}\,=\,\hat a_0\frac{ \Gamma_{32}'}{\hat{J}_R'}\,.
\end{equation}
Note that these distances are of order one, the numerator and denominator containing primed quantities.

Setting $\hat{r}\to \hat{b}$ in \eqref{thrhatunhat}
and using the relation \(\hat c = \sqrt{\hat{a}^2 + \hat{b}^2 + 2\hat{a}\hat{b} \cos \hat\theta}\), we find the position of the new center in terms of the six-dimensional coordinates $r,\theta$:
\begin{equation}
\label{eq:cossq-abc}
\cos^2 \theta_* = \frac{\hat a - \hat b + \hat c}{2\hat a}, \qquad r_*^2 = 2(-\hat a + \hat b + \hat c)\,.
\end{equation}
Substituting \eqref{abc relation} into these relations, we obtain the expression for the position of the momentum center as a function of the physical charges:
\begin{align}
\label{eq:position superstrata probe}
\cos^2 {\theta}_* =\frac{1}{2}\left(1-\frac{\hat{m}'}{ \hat{J}_R'}\right)
,\qquad
    {r}_*^2=2 \hat{a}_0\left(-1+\frac{\hat{m}' -\hat{\kappa}_3 \hat{Q}_3'}{ \hat{J}_R'}\right)
\,.
\end{align}
It will be useful to express these in terms of quantized numbers. We thus convert the coefficients of the poles from hatted quanties to unhatted quantized parameters, using \eqref{eq:q-quant} and \eqref{eq:j-quant}. 
We also replace $\hat{a}_0$ by $\hat{a}$, which is valid to leading order in the probe approximation, and use $4\hat{a}=a^2$.
We thereby obtain
\begin{equation}
    \cos^2 \theta_* =\frac{1}{2}\left(1-\frac{J_L'}{ J_R'}\right),\qquad
    \frac{r_*^2}{a^2}=\frac12 \left(-1+\frac{ J_L' -\kappa\, n_p'}{J_R'}\right)\,.
    \label{itrt2Aug24}
\end{equation}
Importantly, we observe that all the moduli have canceled out of these equations, and the position of the momentum-carrying center is fixed only in terms of quantized numbers.

To connect explicitly to the superstratum analysis, we now specialize to the background of AdS$_3\times S^3$ with no orbifold singularities, and
therefore set  $\kappa=1$. Since we are in the probe limit, we must specify the charges carried by the third center to be those of the single-particle wavefunction on global AdS$_3 \times$S$^3$ that arises in the small $b/a$ limit of a 
single-mode superstratum with mode dependence 
\eqref{v_kmn_def}. The quantized charges of such a single-particle wavefunction are (see \cite[Eqs.\;(3.8)--(3.9)]{Bena:2017xbt})
\begin{align}
\label{eq:superstrata momentum number}
 n_P'&={m+n} \,, \\[2mm]
% \end{align}
%and 
% \begin{align}
\label{eq:superstrata angular momentum number}
 J_L'={m}-{{k}\over 2} &\,,\qquad 
 J_R'=-{{k}\over 2} \,.
\end{align}
Upon substituting these into \eqref{itrt2Aug24}, we find
\begin{align}
 \cos^2{\theta}_*
 ={m\over k}\,,\label{fkcq3Aug24}
\end{align}
in precise agreement with \eqref{thpeak}. Similarly, substituting for the radial position, we obtain
\begin{align}
 \frac{r_*^2}{a^2} = {n\over k},\label{fkcs3Aug24}
\end{align}
in precise agreement with \eqref{rpeak}.

Thus, we see that for small $b/a$, the bubble equations determine the location of the third center to be exactly at the location determined in Section~\ref{sec:EffSS}, as they should.

%%%%%%%%%%%%%%%%%%%
\subsubsection{The deep AdS$_2$ scaling regime}
%%%%%%%%%%%%%%%%%%%
\label{sec:deep_AdS2}

Next, we consider the regime of large $b/a$, in which the $y$-momentum and angular momenta carried by the third center are no longer taken to be small parameters. We keep AdS$_3$ asymptotics, but arrange a sufficiently large $Q_P$ so that there is a long AdS$_2$ throat in the deep interior of the solution.

When the AdS$_2$ throat is very long, we have the hierarchy of scales (recall that $a_0$ was defined in \eqref{ssReg})
\be
Q_1 \,\sim\, Q_5 
\,\gg \, Q_P 
= \frac{m+n}{k}\:\! b^2 
\,\sim\, a_0^2 
\,=\, 4 \hat{a}_0 
\,\gg \, \hat{a} \,,
\ee
where the hierarchy between $Q_{1,5}$ and $Q_P$ is to have an AdS$_3$ region, we have taken $m\sim n \sim k$ as in \eqref{eq:kmnscal}, and the remaining relations are to have an AdS$_2$ throat which is as long as it can be, as predicted by the holographic CFT\@. We will discuss this further in Section \ref{sec:BHregime}.

We make a mild genericity assumption that, in $\Gamma^{13}$, given by:
\be
\Gamma^{13} 
\,=\,
\hat{m}'
- \frac{\kappa \hat{Q}_P R_y}{2} \,,
\ee
we assume there is no cancellation between the two terms on the right-hand side. This is to ensure that the ratio $\Gamma^{13}/\Gamma^{23}$ is of order one, as we shall use momentarily.

The bubble equations \eqref{eq:first bubble ST} can be written as
\begin{equation}
    \frac{2\Gamma^{12}}{\hat{\kappa}_3 \hat a}+\frac{2\Gamma^{13}}{\hat{\kappa}_3 \hat b}\,=\,1\,
    ,\qquad \frac{\Gamma^{13}}{\hat b}+\frac{\Gamma^{23}}{\hat c}\,=\,0 \,. \label{eq:Bubble equations scaling regime}
\end{equation}
We recall from \eqref{eq:ahat0} that $\hat{a}_0=2 \Gamma^{12}/{\hat{\kappa}_3}$. The first bubble equation then becomes
\begin{equation}
     \qquad\quad
     \frac{\hat{a}_0}{\hat{a}}+\gamma \frac{\hat{a}_0}{\hat{b}}=1
    \,,
    \qquad\quad
    \gamma \,\equiv \, \frac{2 \Gamma^{13}}{\hat{a}_0\hat{\kappa}_3} \sim 1 \,,
\end{equation}
where $\gamma\sim 1$ follows from the mild assumption above. 

To have a long AdS$_2$ throat, we require ${\hat{a}}\ll {\hat{a}_0}$. 
Then the first bubble equation  implies that ${\hat{b}}\ll {\hat{a}_0}$. Then both terms on the left-hand side are much bigger than one, and we are in a scaling regime in which the leading-order solution is obtained by setting the right-hand side to zero.
In particular, this imposes that:
\begin{equation}
    \hat{a}\sim \hat{b}\,.
\end{equation}

Examining now the second bubble equation in  \eqref{eq:Bubble equations scaling regime}, recall that the mild assumption above implies that $\Gamma^{13} \sim \Gamma^{23}$. This equation is already homogeneous. Then we see that 
\begin{equation}
    \hat{a}\,\sim \,\hat{b} \,\sim \,\hat{c} \,\ll \,\hat{a}_0\,,
\end{equation}
so all of the centers lie deep inside an AdS$_2$ throat. It is still possible to have a modest hierarchy between any of the distances $\hat{a}, \hat{b}, \hat{c}$, provided that any such hierarchies do not compete with the hierarchy between   $\hat{a}$ (say) and $\hat{a}_0$.

We now demonstrate that the expressions for the position of the momentum center relative to the cap, \eqref{itrt2Aug24}, \eqref{fkcq3Aug24}, and \eqref{fkcs3Aug24}, are valid also in the scaling regime of large $b/a$. Physically, even though the solution is now scaling with a deep AdS$_2$ throat, the microstructure is deep inside the AdS$_2$ throat in a region in which the cap can be described as a deformation of global 
AdS$_3 \times$S$^3$.

 In the scaling regime, the total right-moving angular momentum is small. It is useful to write this as the sum of two large contributions of opposite sign, from the original two centers and the momentum-carrying center respectively, 
\begin{equation}
   \hat{J}_R\,=\,\frac{\hat{\kappa}_3}{2}\hat a \,=\, \frac{Q_1 Q_5}{2 \hat{\kappa}_3}+\frac{\hat{\kappa}_3}{2}( \hat{a}-\hat{a}_0)\,,
\end{equation}
such that the (large) angular momentum assigned to the momentum-carrying center is
\begin{equation}
    \hat{J}_R'=\frac{\hat{\kappa}_3}{2}( \hat{a}-\hat{a}_0)\approx\frac{\hat{\kappa}_3}{2}( -\hat{a}_0)\,.
\end{equation}
Note that this quantity is \emph{not} any contribution of the superstratum wave to the right angular momentum, but is rather a book-keeping device. 

To leading order, the scaling solution is given by:
\begin{equation}
    \hat{b}=-\hat{a}\,\frac{\Gamma^{13}}{\Gamma^{12}}\,,\qquad \hat{c}=-\hat{b}\,\frac{\Gamma^{23}}{\Gamma^{13}}=+\hat{a}\,\frac{\Gamma^{23}}{\Gamma^{12}}\,.
\end{equation}
Also, remember that, by definition:
\begin{equation}
    \Gamma^{12}=\frac{\hat{\kappa}_3}{2} \hat{a}_0=-\hat{J}_R'\,,
\end{equation}
where we used the approximation $\hat{a}\ll \hat{a}_0$.  Then a direct specialization of \eqref{eq:cossq-abc} leads to:
\begin{align}
\begin{split}
     \frac{r_*^2}{a^2} &\,=\, \frac12\left(-1+\frac{-\Gamma^{13}+\Gamma^{23}}{\Gamma^{12}}\right)\,=\,\frac12\left(-1+\frac{\hat{\kappa}_3\hat{Q}_3'-\hat{m}'}{-\hat{J}_R'}\right)\,,\\[.2cm]
    \cos^2 \theta_* &\,=\,\frac{1}{2}\left(1+\frac{\Gamma^{13}+\Gamma^{23}}{\Gamma^{12}}\right)\,=\,\frac{1}{2}\left(1-\frac{\hat{m}'}{\hat{J}_R'}\right)\,,
\end{split}
\end{align}
which are exactly the same formulae as \eqref{eq:position superstrata probe}.
We reiterate that, since $b/a$ is now large, the primed quantities are now  generically large. Using the results of \cite{Bena:2016ypk}, we have, in the regime where $\frac{b^2}{2} \simeq a_0^2 =\frac{Q_1Q_5}{R_y^2}$: 
\begin{align}
    n_P&\,\simeq\,\frac{m+n}{k} n_1n_5\nonumber ~,\\
    J_R&\,=\,\frac{n_1 n_5 R_y^2}{Q_1Q_5}\frac{a^2}{2}
    \,=\,
    \frac{n_1 n_5 R_y^2}{{ 2 Q_1Q_5}}a_0^2+\frac{n_1 n_5 R_y^2}{{2} Q_1Q_5}\left(-\frac{b^2}{2}\right)\,, \\
    J_L&\,=\,\frac{n_1 n_5 R_y^2}{2 Q_1Q_5}\left(a^2+\frac{m}{k}b^2\right)
    \,=\,
    \frac{n_1 n_5 R_y^2}{2 Q_1Q_5} a_0^2+\frac{n_1 n_5 R_y^2}{2 Q_1Q_5}\left(\frac{m}{k}-\frac{1}{2}\right)b^2\,,  \nonumber
\end{align}
where we extensively used $a_0^2=\frac{b^2}{2}+a^2$ and wrote the expressions to make explicit the round supertube contributions plus the superstratum ones. To first non-trivial order, the relations for the primed angular momenta and  momentum are:

\begin{equation}
    n_P^{}\,=\,\frac{m+n}{k} n_1 n_5\,,\qquad J_L'\,=\, \frac{m-\frac{k}{2}}{k} n_1n_5\,,\qquad J_R'\,=\,-\frac{1}{2} n_1n_5\,.
\end{equation}
Using these relations, the position of the momentum center relative to the cap can be written in terms of $(k,m,n)$ so that we again have:
\begin{equation}
    \cos^2{\theta}_*
 \,=\,\frac{m}{k}\,,\qquad 
 \frac{r_*^2}{a^2} \,=\, {n\over k}\,.
\end{equation}
We have thus obtained exactly the same relations in the deep AdS$_2$ regime as in the probe approximation. This is as it should be, since the analysis leading to Eq.~\eqref{positions1} is valid for general values of the ratio $a/b$. It is nevertheless satisfying to see how the result emerges from a direct multi-center analysis, in both the probe regime and the deep AdS$_2$ scaling regime.

\subsection{Adding a D1-P center to the GLMT background}
\label{sec:D1-P_in_GLMT}

There are two natural ways to generalize the foregoing analysis. The first is to take the probe to be a two-charge D1-P supertube center. The second is to generalize the background from the circular supertube to the GLMT solution \eqref{bgGamma}.  The analysis is similar, so we shall proceed directly to the most general D1-P probe in the GLMT background.  At the end of the section, we shall specialize the analysis to the D1-P probe in the two-charge circular supertube background.

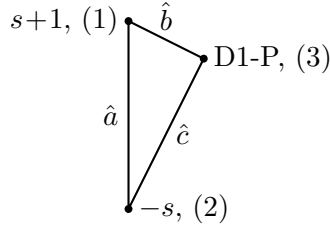
\begin{figure}[htbp]
\begin{center}
\begin{tikzpicture}[scale=0.5]

\coordinate (A) at (0, 5);
\coordinate (B) at (2,4);
\coordinate (C) at (0, 0);

\draw[thick] (A) -- (B) -- (C) -- cycle;

\fill[black] (A) circle (3pt);
\fill[black] (B) circle (3pt);
\fill[black] (C) circle (3pt);

\node[anchor=east] at (A) {$s\tight+1$, (1)};
\node[anchor=west] at (C) {$-s$, (2)};
\node[anchor=west] at (B) {D1-P, (3)};

\node[anchor=south] at ($ (A)!0.5!(B) $) {$\hat b$};
\node[anchor=west]  at ($ (B)!0.5!(C) $) {$\hat c$};
\node[anchor=east]  at ($ (C)!0.5!(A) $) {$\hat a$};
\end{tikzpicture}
\caption{\it The 3-center configuration with GLMT centers and a D1-P center.  Here {\rm``(1)''}, {\rm``(2)''}, and {\rm``(3)''} label the centers;  the quantities $s+1$ and $-s$ associated to the first two centers refer to their $V$ charges.  
\label{fig:3ctr}}
\end{center}
\end{figure}

As described in Section \ref{sec:GLMT soln}, the supersymmetric three-charge spectral flowed supertubes are the general two-centered bubbling solutions, with charge vectors given in \eqref{bgGamma}. As in the previous subsection, we add a third center, but now we take it to be a two-charge  supertube center with D1-brane and momentum charges (or F1 and momentum charges, in the F1-NS5-P duality frame).  We call this the  D1-P center.
This center is a singular center in supergravity, but again this corresponds to a physical source in string theory. Our conventions for the distances between the centers are given in Figure \ref{fig:3ctr}.

We will be interested in matching to the worldsheet analysis in the NS5-F1-P frame in Section~\ref{sec:wound-strings}. For this purpose, it suffices to restrict the analysis in this subsection to the probe approximation, in which the charges of the third center are small, as was done in Section \ref{sec:P-probe}.
We write the charge vector of the D1-P probe center as
\begin{equation}
    \Gamma_{\text{D1-P}}=(0,(0,\hat{\kappa}_2',0),(\hat{Q}_1',0,\hat{Q}_3'),{\hat m}')\,,
\end{equation}
where the primed quantities are small. The primitivity condition for this center is
\begin{equation}
\label{F1-P primitivity-pre}
     {\hat m}'=\frac{\hat{Q}_1'\hat{Q}_3'}{2\hat{\kappa}_2'}~,
\end{equation}
or in terms of quantized charges, using Eqs.~\eqref{k-quant-D1-P}--\eqref{eq:j-quant}, 
\begin{equation}
\label{F1-P primitivity}
     J_L' \;=\; \frac{n_1' n_P'}{2{\kappa}_2'} \quad \in ~ \frac12 \, \mathbb{Z}.
\end{equation}

Adding a third probe center \eqref{F1-P primitivity} can back-react on the background charges \eqref{bgGamma} by changing them.  Part of the change can be determined by requiring that the total charges in $K^I$ remain unchanged by the addition of the third center. This is because these charges, given by the sum of pole residues in these harmonic functions, are quantized Page charges that cannot change under smooth physical processes \rcite{Park:2015gka}. This fixes the charge vectors $\Gamma^{(1)}$, $\Gamma^{(2)}$ to the following form:
\begin{align}
\Gamma^{(1)}&=\biggl(s+1,\Big(-\hat{\kappa}_1,-\hat{\kappa}_2-(s+1)\hat{\kappa}_2',-\hat{\kappa}_3\Big),
\nn\\[-.2cm]
&\hskip 1cm
\Big(-\frac{\hat{\sfq}_1+(s+1)\hat{\kappa}_2'\hat{\kappa}_3}{s+1},-\frac{\hat{\sfq}_2}{s+1},-\frac{\hat{\sfq}_3+(s+1)\hat{\kappa}_2'\hat{\kappa}_1 }{s+1}\Big),
-\frac{\tilde{m}+(s+1)\hat{\kappa}_2'\hat{\kappa}_1 \hat{\kappa}_3}{2(s+1)^2}\biggl),
\nn\\[.2cm]
\Gamma^{(2)}&=\biggl(-s,\Big(\hat{\kappa}_1,\hat{\kappa}_2+s \hat{\kappa}_2',\hat{\kappa}_3\Big),\Big(\frac{\hat{\sfq}_1+s \hat{\kappa}_2'\hat{\kappa}_3}{s},\frac{\hat{\sfq}_2}{s},\frac{\hat{\sfq}_3+s \hat{\kappa}_1\,\hat{\kappa}_2'}{s}\Big),
\frac{\tilde{m}+s\hat{\kappa}_1\,\hat{\kappa}_2'\,\hat{\kappa}_3}{2\,s^2}\biggl),
\nn\\[.2cm]
\Gamma^{(3)}&=\Gamma_{\text{D1-P}}
=
\Big(0,\big(0,\hat{\kappa}_2',0\big),\big(\hat{Q}_1',0,\hat{Q}_3'\big),\hat{m}'\Big),\label{Gamma_D1-P}
\\[.2cm]
h&=\big(0,(0,0,0),(0,0,1),0\big)\,,
\nn
\end{align}
where the $\hat{\sfq}_I$ were defined in~\eqref{eq:primitivity GLMT}.
In fact, there is an apparent arbitrariness to shift the second dipole charge in $\Gamma_{1,2}$ above so that the total dipole charge vanishes, but this has been fixed by the following argument.
Between the two background centers, there is a non-trivial $S^2$ through which there are fluxes
\begin{equation}
    \Pi_I=\frac{\hat{\kappa}^{(1)}_I}{q_{(1)}}-\frac{\hat{\kappa}^{(2)}_I}{q_{(2)}}\,
\end{equation}
that cannot change by the continuous process of bringing in a third center.
Requiring that these fluxes remain unchanged fixes the arbitrariness. Alternatively, one can require that we should compensate the addition of the D1-P center by a gauge transformation of the initial centers, which also fixes the arbitrariness. %; this holds beyond the probe approximation. 

Let us determine the position of the probe center using the bubble equation \eqref{eq:bubble-eqns} in the probe approximation (see Figure~\ref{fig:3ctr} for our convention for the distances $\hat{a},\hat{b},\hat{c}$ between centers).  As before, we write $\hat{a}=\hat{a}_0+\hat{a}'$ where $\hat{a}'$ is the small change in the distance by the addition of the probe from the unperturbed value $\hat{a}_0$. The bubble equations are
\begin{align}
    \frac{\Gamma^{12}_0}{\hat a}+\frac{\Gamma'^{12}}{\hat a}+\frac{\Gamma'^{13}}{\hat b}=\frac{\hat{\kappa}_3}{2}
    ,\qquad
    \frac{\Gamma'^{13}}{\hat b}+\frac{\Gamma'^{23}}{\hat c}=0 %\label{eq:second bubble GLMT}\,
    \label{eq:first bubble GLMT}
    ,
\end{align}
where we have split the inner product into the zeroth and first order terms in the probe charges, as $\Gamma^{12}=\ev{\Gamma^{(1)},\Gamma^{(2)}}=\Gamma^{12}_0+\Gamma'^{12}$.  
Note that $\Gamma'^{13}$ and $\Gamma'^{23}$ have no zeroth order term.
By comparing the zeroth and first order terms of the equations, we find the unperturbed distance
\begin{equation}
    \hat a_0
    =\frac{\hat{\kappa}_1\,\hat{\kappa}_2}{s^2(1+s)^2}~,
\end{equation}
and the relations
\begin{align}\begin{gathered}
    \frac{\hat{\kappa}_3}{2}\left(\frac{\hat a'}{\hat 
    a_0}\right)=\frac{\Gamma'^{13}}{\hat b}~,\\
    \frac{1}{\hat c}\left(\hat{\kappa}_1 \, \hat{Q}_1' + \hat{\kappa}_3 \, \hat{Q}_3' - \frac{\hat{\kappa}_{2}' \, \hat\sfq_2}{s} - 2 \, \hat{m}' \, s\right) + \frac{1}{\hat b}\left(-\hat{\kappa}_1 \, \hat{Q}_1' - \hat{\kappa}_3 \, \hat{Q}_3' + \frac{\hat{\kappa}_{2}' \, \hat\sfq_2}{1 + s} + 2 \, \hat{m}' \, (1 + s)\right) = 0~.
\end{gathered}\end{align}

The asymptotic charges are
\begin{align}\begin{split}
    &Q_1\,=\, 4\left(\frac{\hat{\sfq}_1}{s(s+1)}+\hat{Q}_1'\right)\,=\,\frac{g_s(\alpha')^3}{ V_4}\left(\frac{\kappa_2\kappa}{s(s+1)}+n_1'\right),\\
    &Q_5\,=\,4\left(\frac{\hat{\sfq}_2}{s(s+1)}\right)\,=\,g_s\alpha'\frac{\kappa_1\kappa}{s(s+1)}~,\\
    &Q_{\sst P}\,=\, 4\left(\frac{\hat{\sfq}_3}{s(s+1)}+\hat{Q}_3'\right)\,=\,\frac{g_s^2(\alpha')^4}{ R_y^2V_4}\left(\frac{\kappa_1\kappa_2}{s(s+1)}+n_p'\right).
\end{split}\end{align}
The angular momenta \eqref{eq:j-non-q} are given by
\begin{align}
    \hat{J}_L \,&=\, \tilde{m} \left(\frac{1}{2\,s^2}-\frac{1}{2\,(s+1)^2}\right)+\hat{m}'+\frac{\hat{\kappa}_1\,\hat{\kappa}_2'\,\hat{\kappa}_3}{s(s+1)},\qquad \\
    \vec{\hat{J}}_R\,&=\,\frac{\hat{\kappa}_3}{2} \hat{a} \, \vec{z} \quad \Rightarrow \quad     \hat{J}_R=\frac{\hat{\kappa}_3}{2}\hat a=\frac{Q_1 Q_5}{32 \hat{\kappa}_3}+\frac{\hat{\kappa}_3}{2} \hat{a}'-\frac{\hat{Q}_1' \hat{\kappa}_1 }{2 s (s+1)}\,,
\end{align}
where $\vec{z}=(0,0,1)$. 
The change in $\hat{J}_R$ caused to the addition of the probe is
\begin{align}
\hat{J}_R'=\frac{\hat{\kappa}_3}{2} \hat a'-\hat{Q}_1' \frac{\hat{\kappa}_1 }{2 s(s+1)}~.
\end{align}

Using the linearized bubble equations we obtain:
\begin{align}
\label{abc relation2}
    \hat{b}= \hat a_0\frac{ \Gamma_{13}'}{ \hat{J}_R'+\hat{Q}_1' \frac{Q_5}{ 8 \,\hat{\kappa}_3}},\qquad
    \hat{c}=\hat a_0\frac{ \Gamma_{32}'}{\hat{J}_R'+\hat{Q}_1' \frac{Q_5}{ {8}\,\hat{\kappa}_3}}\,.
\end{align}

%%%%%%%%%%%%%%%%%%%%%%%%%%%%%%%%%
\subsubsection{Position of the D1-P center}
%%%%%%%%%%%%%%%%%%%%%%%%%%%%%%%%%

As before, setting $\hat{r}\to \hat{b}$ in \eqref{thrhatunhat} and using the resulting relation \eqref{eq:cossq-abc}, we find the position of the D1-P center in terms of the quantized charges:
\begin{equation}
\label{eq:cossqtheta-st}
    \cos^2 {\theta}_* =\frac{1}{2}\left(1-\frac{\hat{m}'-\frac{\hat{\kappa}_2'Q_5}{8}}{ \hat{J}_R'+ \frac{Q_5}{8 \hat{\kappa}_3}\hat{Q}_1'}\right)=\frac{1}{2}\left(1-\frac{J_L'-\frac{\kappa_2'n_5}{2}}{ J_{R}'+ \frac{n_5n_1'}{2 \kappa}}\right),
\end{equation}
where the second term is given only in terms of (half) integer quantities. Again, the position is independent of spacetime moduli and depends solely on the quantized numbers of the probe.  Likewise, we find
\begin{align}
    {r}^2_*&=2 \hat{a}_0\left(-1+\frac{ \hat{m}' (2s+1)+\hat{\kappa}_2'(2 s+1)\frac{Q_5}{8}-(\hat{\kappa}_3 \hat{Q}_3'+\frac{Q_5 s(s+1)}{4\hat{\kappa}_3}\hat{Q}_1')}{ \hat{J}_R'+ \frac{Q_5}{8\hat{\kappa}_3} \hat{Q}_1'}\right)
\nn\\[.2cm]
    &=2 a_0\left(-1+\frac{  (2s+1)\left(J_L'+\kappa_2'\frac{n_5}{2}\right)-(\kappa n_p'+\frac{n_5 n_1' s(s+1)}{\kappa}}{ J_{R}'+ \frac{n_5 n_1'}{2\kappa}}\right)\,,
    \label{eq:rsq-st}
\end{align}
where we recall that the hatted quantities are the dimensionful charges, while the un-hatted quantities are the integer or half-integer quantum numbers.  Thus, again, the result is independent of spacetime moduli.

The result exactly matches that obtained from the worldsheet in the next section: The peak of the wavefunction of the momentum excitations is exactly at the same distance, and that was the main goal of this subsection.

Although we considered one probe D1-P center above, it can be straightforwardly generalized to solutions with an arbitrary number of probe centers.  We discuss such generalization in Appendix~\ref{sec:multiple_probes}.

%%%%%%%%%%%%%%%%%%%%%%%%%%%%%%%%%

\subsubsection{A D1-P excitation of the two-charge circular supertube background}
\label{ss:D1-P_in_LM}

%%%%%%%%%%%%%%%%%%%%%%%%%%%%%%%%%

As a byproduct of the above analysis, we can obtain a configuration involving a D1-P center added to a two-charge circular supertube by taking the appropriate vanishing-$s$ limit: $s\rightarrow 0$, $\kappa_1,\kappa_2 \rightarrow 0 $ holding fixed the finite ratios $\frac{\hat{\kappa}_1} {s},\,\frac{\hat{\kappa}_2}{s}$, as described below \eqref{k-quant-F1-P}. Denoting the finite limit of $\hat{\sfq}_1/s$ by $\hat{Q}_1$,
and the the finite limit of $\hat{\sfq}_2/s$ by $\hat{Q}_2$, we obtain: 
\begin{align}
    &\Gamma_{1}=(1,(0,-\hat{\kappa}_2',-\hat{\kappa}_3),(-\hat{\kappa}_2'\,\hat{\kappa}_3,0,0),0)~,
\nn\\
    & \Gamma_{2}=\left(0,(0,0,\hat{\kappa}_3),(\hat{Q}_1+\hat{\kappa}_2' \hat{\kappa}_3,\hat{Q}_2,0),\frac{\hat{Q}_1 \, \hat{Q}_2 \,}{2 \hat{\kappa}_3}+\frac{\hat{Q}_2 \hat{\kappa}_2'}{2}\right) ~,
\nn\\
    &\Gamma_{\text{D1-P}}=\left(0,\left(0,\hat{\kappa}_2',0\right),\left(\hat{Q}_1',0,\hat{Q}_3'\right),\hat{m}'\right)~,
\\
    &h=(0,(0,0,0),(0,0,1),0)~.
\nn
\end{align}
Let us summarize the resulting relevant quantities for the $s\rightarrow 0$ limit.
%with  $\alpha=-s$. 
The charges of the solution are
\begin{align}
		Q_1=4 \:\!(\hat{Q}_1+\hat{Q}_1')\,,\qquad
		Q_5=4 \;\! \hat{Q}_2\,~,\qquad
		Q_{\sst P}=4 \;\! \hat{Q}_3'\,,
\end{align}
and the position of the third center is given by: 
\begin{align}
    \cos^2 {\theta} =\frac{1}{2}\left(1-\frac{J_L'-\frac{\kappa_2'n_5}{2}}{ J_{R}'+ \frac{n_5n_1'}{2\kappa}}\right),\qquad
    {r}^2 =2 \hat{a}_0\left(-1+\frac{ \left( J_L'+\frac{\kappa_2' n_5 }{2}\right)-\kappa n_p'}{ J_R'+ \frac{n_5n_1'}{2 \kappa}}\right)\,.
    \label{eq:position in 5D with s=0}
\end{align}

%%%%%%%%%%%%%%%%%%%%%%%%%%%%%%%%%%%%%%%%%%
%%%%%%%%%%%%%%%%%%%%%%%%%%%%%%%%%%%%%%%%%%

\section{Worldsheet description of F1-P probes}
\label{sec:wound-strings}

The GLMT background of Section~\ref{sec:GLMT soln} in the NS-NS flux duality frame admits an exactly solvable worldsheet description in terms of gauged Wess-Zumino-Witten (WZW) models for the group coset~\rcite{Martinec:2017ztd}
\be
\label{cosets}
\frac\cG\cH \,=\, \frac{\sltwo\times\sutwo\times \bR_t\times\bS^1_y}{U(1)_L\times U(1)_R} 
\ee
times $\bT^4$ or $K3$, where $\cH$ gauges a pair of null isometries of $\cG$.
The null currents generating these isometries can be parametrized as
\begin{align}
\label{null-currents}
\cJ \,=\, J^3_\sl + l_2 J^3_\su + l_3 \,i\partial t + l_4\, i\partial y
~~&,~~~~
\bar\cJ = \bar J^3_\sl + r_2 \bar J^3_\su + r_3 \,i\bar\partial t + r_4\, i\bar\partial y ~.
\end{align}

The general three-charge spectral flowed circular supertube solutions of~\rcite{Giusto:2012yz} correspond to the coset models with parameters%
\footnote{In an NS5-brane decoupling limit, where the geometry is asymptotically that of the linear dilaton throat of NS5-branes.}
\begin{align}
\label{lrglmt}
\begin{split}
l_2 = 2s+1
~,~~~
r_2=1
~,~~~
l_3 &= r_3 =  -\bigg( \sfk R_y+\frac{n_5 s(s+1)}{\sfk R_y}\bigg) \;,
\\[.2cm]
l_4 = \sfk R_y - \frac{n_5 s(s+1)}{\sfk R_y}
~,~~~
r_4 &= -\sfk R_y - \frac{n_5 s(s+1)}{\sfk R_y} ~.
\end{split}
\end{align}
The circular NS5-F1 supertube background of~\rcite{Lunin:2001fv} corresponds to $s=0$.

%%%%%%%%%%%%%%%%%%%%%%%%%%%%%%%%%%%%%%%%%%
\subsection{Constraints on the string spectrum}
\label{sec:backgds}

Reparametrization invariance and local supersymmetry on the string worldsheet, together with the local gauge symmetry, lead to a set of physical state constraints on the string Hilbert space.  We adopt the notation of~\rcite{Martinec:2017ztd,Martinec:2018nco,Martinec:2019wzw,Martinec:2020gkv,Martinec:2022okx,Martinec:2025xoy}, which we will relate below to that of Section~\ref{sec:D1-P}.

We focus on the AdS$_3$ limit of these solutions, which sends the $y$-circle radius $R_y\to\infty$, holding fixed the rescaled energies $ER_y$ and $y$-momenta $P_yR_y$. One can take this limit of the background solutions by defining 
\begin{equation}
    \tilde{t} \,=\, \frac{t}{R_y} \,, \qquad
    \tilde{y} \,=\, \frac{y}{R_y} \,,
\end{equation}
and sending $R_y\to\infty $ at fixed $\tilde{t}$, $\tilde{y}$. The resulting solution is asymptotically AdS$_3 \times \mathbb{S}^3 \times \mathbb{T}^4$, and the six-dimensional part of the metric is given by~\eqref{adslim}.

We consider states with nonzero winding around the various circles~-- the azimuthal direction in $\sltwo$, the Euler angles in $\sutwo$, and the $y$-circle.%
\footnote{Since we work on the universal cover of $\sltwo$, there is no winding in its timelike coordinate, so $w=\bar w$.}
In this limit, we write the asymptotic energy $E$ and $y$-circle momenta $P_y,\bar P_y$ as
\be
\label{EPexpn}
E = w_y R_y + \frac{\vareps}{R_y}
~~,~~~~
P_y = w_y R_y + \frac{n_y}{R_y}
~~,~~~~
\bar P_y = - w_y R_y + \frac{n_y}{R_y} ~.
\ee
At leading order in large $R_y$, the zero-mode null gauging constraints on the left- and right-moving sector, $\cJ=\bar\cJ=0$, are respectively~\rcite{Martinec:2018nco,Martinec:2025xoy}
\begin{align}
\begin{split}
\label{genlargeRnull}
\sfk(\vareps-n_y) &= 2\Msl_{\sl,\tot} + (2s+1)2\Msu_{\su,\tot} -\frac{2s(s+1)}{\sfk}n_5w_y ~,
\\[.2cm]
\sfk(\vareps+n_y) &= 2\bMsl_{\sl,\tot} + 2\bMsu_{\su,\tot}  ~,
\end{split}
\end{align}
where we have defined
\be
\label{mtotdef}
\Msl_{\sl,\tot}=\Msl_\sl+\frac{n_5}2 w_\sl
~~,~~~~
\Msu_{\su,\tot}=\Msu_\su+\frac{n_5}2 w_\su
\ee
to be the eigenvalues of the zero mode of the total $J^3_\sl$ and $J^3_\su$ currents, including both bosonic and fermionic contributions, as well as worldsheet spectral flows parametrized by $w,w'$;
and similarly for $\bMsl_{\sl,\tot},\bMsu_{\su,\tot}$. 
The large-$R_y$ Virasoro constraints are 
\begin{align}
\begin{split}
\label{largeRvir}
0 &= \frac{- j_\sl ( j_\sl -1)+ j_\su( j_\su+1)}{n_5} - \Msl_\sl  w_\sl  + \Msu_\su  w_\su -\frac{n_5}4\Big( w_\sl ^2\tight-( w_\su )^2\Big) - \frac{w_y}{2}\big(\vareps \tight- n_y \big)
+ h^{~}_L ~,
\\[.3cm]
0 &= \frac{- j_\sl ( j_\sl -1)+ j_\su( j_\su+1)}{n_5} - \bMsl_\sl  w_\sl +\bMsu_\su  \bar w_\su -\frac{n_5}4\Big( w_\sl ^2\tight-( \bar w_\su )^2\Big) - \frac{w_y}{2}\big(\vareps \tight+ n_y \big) 
+ h^{~}_R ~,
\end{split}
\end{align}
where $ j_\sl , j_\su$ are the spins of the bosonic highest weight states for $\sltwo$ and $\sutwo$, and $h^{~}_L,h^{~}_R$ specify the non-zeromode excitation levels (for details, see~\rcite{Martinec:2018nco,Martinec:2020gkv,Martinec:2022okx,Martinec:2025xoy}).

%%%%%%%%%%%%%%%%%%%%%%%%%%%%%%%%%%%%
%%%%%%%%%%%%%%%%%%%%%%%%%%%%%%%%%%%%

\subsection{``Primitive" winding string states}
\label{sec:windup}

We now analyze the parts of the string spectrum that are of interest in the present work.
We look for a solution of the worldsheet constraints that is right-BPS, and has no excitations other than those required by the GSO projection: $h_L=h_R=0$.  The BPS condition on the right imposes
\be
\label{rightbps}
 j_\sl = j_\su+1, 
~~,~~~~
\bar  J_\sl  = \bar J_\su
~~,~~~~
\bMsl_\sl=J_\sl
~~,~~~~
\bMsu_\su=-\bar J_\su
~~,~~~~
 w_\sl  = - \bar w_\su   ~.
\ee
We are going to ignore various subtleties having to do with polarization states of the string, as subleading effects in the semi-classical limit of large $j$; thus for instance will ignore the distinction between the total spins $J_\sl,J_\su$ obtained from the tensor products of the center-of-mass bosonic spins $ j_\sl , j_\su$ and those of the polarization states.

The right null constraint~\eqref{genlargeRnull} then implies $\vareps=-n_y$, and thus that the right Virasoro constraint~\eqref{largeRvir} is also satisfied.  Substituting the left null constraint into the left-moving Virasoro constraint and regrouping terms, one finds
\be
\label{primitivity}
-\Big(n_y- \nfive\frac{s(s\tight+1)}\sfk  w_\sl \Big)\Big(w_y+\sfk  w_\sl \Big)
=
\Big((2s\tight+1) w_\sl + w_\su \Big)
\Big(\Msu_\su+\frac\nfive4
\big((2s\tight+1) w_\sl + w_\su \big)
\Big)
\ee
where we have written the result in terms of quantities that are invariant under large gauge  transformations~\rcite{Martinec:2018nco}: the $\cH$ spectral flow transformations, whose effect is to shift
\begin{align}
\label{gauge specflow}
\delta w_\sl &= q ~~, 
&\delta w_\su &= -(2s+1) q ~~,
&\delta \bar w_\su &= - q ~~,
\nn\\[.2cm]
\delta E &= -\bigg( \sfk R_y+\frac{n_5 s(s+1)}{\sfk R_y}\bigg) q ~~,
&\delta n_y \;&=\; -\nfive \frac{\sfm \sfn}{\sfk} q 
~~, 
&\delta w_y \;&=\; -\sfk q  ~~,
\end{align}
where $q\in\bZ$.  The oscillator modes, $\bT^4$ excitations, \etc, that we have set to zero in the Virasoro constraints constitute additional terms in~\eqref{primitivity} that when included result in a non-primitive probe solution.

Comparing to~\eqref{F1-P primitivity}, 
we see that the worldsheet constraints imply the primitivity condition, with the identifications 
\begin{align}
\label{quantnums}
n'_1 \,=\, w_y+\sfk w_\sl  
\quad~~~\,&,~~~~~
n'_P \,=\, -\Big( n_y - \nfive \frac{s(s\tight+1)}{\sfk}  w_\sl \Big) ~,
\nn\\[.2cm]
\kappa'_2 \,=\, \half\big((2s\tight+1) w_\sl + w_\su \big)
~~&,~~~~~
J_L' \,=\, \Msu_\su+\frac\nfive4\big((2s\tight+1) w_\sl + w_\su \big) ~,
\end{align}
where the LHS refers to the probe quantities defined in Section~\ref{sec:D1-P}, and the RHS are the factors in~\eqref{primitivity}.%
\footnote{We can use the large gauge transformation~\eqref{gauge specflow} to shift away $w_y$ in multiples of $\kappa$, and in particular to then restrict $w_y$ to the range $\{0,1,...,\kappa\tight-1$\}; $w_y$ then labels twisted sectors of the orbifold \eqref{eq:sug-orbact-GLMT}.}

The relations $\cos2\theta = -\Msu_\su/J_\su$ and $\cosh2\rho=\Msl_\sl/J_\sl$ (see Eqs.~\eqref{jmsu}, \eqref{thetamotion}, \eqref{jmsl}, and~\eqref{rhopm}) are properties of classical solutions of the $\sutwo$ and $\sltwo$ WZW models, as we saw from the analysis of Section~\ref{sec:wavefns}.  To use these relations in the gauged WZW model, however, we need to write gauge invariant expressions for the corresponding quantities.  We have  
\begin{align}
J'_L &\,=\, \Msu_\su + \frac\nfive2\Big(  w_\su  -(2s+1)\frac{w_y}{\kappa}\Big) ~,
\nn\\[.1cm]
\label{jlrp}
J'_R &\,=\, \bMsu_\su+\frac\nfive2\Big( \bar w_\su -\frac{w_y}{\kappa}\Big) ~,
\end{align}
where again the LHS refers to the probe quantities defined in Section~\ref{sec:D1-P}, and the RHS are the corresponding gauge-invariant worldsheet expressions.%
\footnote{While it might appear that this expression violates angular momentum quantization in twisted sectors $w_y\notin \kappa\bZ$, this is an artifact of the definition, in which the angular momentum of individual strings is compared to the angular momentum per winding of the background.  For further details, see~\rcite{Martinec:2025xoy}. }

The map of quantum numbers~\eqref{quantnums} shows that 
\be
\Msu_\su = J'_L-\frac\nfive2 \kappa'_2  ~;
\ee
using the BPS conditions~\eqref{rightbps} in the second line of~\eqref{jlrp},
we have
\be
\label{theta glmt}
-\cos 2\theta_* = -\frac{\Msu_\su}{J_\su} = \frac{ J'_L-\half \nfive \kappa'_2}{J'_R+\frac\nfive2 \frac{n'_1}\kappa} ~,
\ee
which matches~\eqref{eq:cossqtheta-st}.

On the other hand, the left $\sltwo$ spin, $\Msl_\sl$, is determined in terms of other data by the left null constraint,
\begin{align}
\label{Lnull}
-\Msl_\sl &= (2s\tight+1)\Msu_\su  +\frac\nfive2\big(  w_\sl  + (2s\tight+1) w_\su \big) -\frac\sfk2\big(\vareps-n_y\big) -\nfive
\frac{s(s\tight+1)}\kappa w_y ~,
\nn\\[.2cm]
&= (2s\tight+1)\Big( J'_L + \frac\nfive2 \kappa'_2\Big) - \sfk n'_P -\nfive\frac{s(s\tight+1)}\sfk n'_1  ~,
\end{align}
where we have again used the BPS condition.
We thus find the radial position of an F1-P probe in the GLMT background to be given by 
\begin{align}
\label{rho glmt}
\cosh2\rho_* = \frac{\Msl_\sl}{J_\sl} = \frac{ (2s\tight+1)\big( J'_L + \frac\nfive2 \kappa'_2\big) - \sfk n'_P -\nfive\frac{s(s+1)}\sfk n'_1  }{J'_R+\frac\nfive2 \frac{n'_1}\kappa}   ~.
\end{align}
Using the identification~\eqref{rhomap2}, we find a match with~\eqref{eq:rsq-st}.

Note that there are many possible wound string states located at the same point in the radial coordinate $\rho$ and polar angle $\theta$.  The positions in these coordinates are determined by $\Msl_\sl,\Msu_\su,J$ and is independent of the windings $w_y, w_\sl , w_\su , \bar w_\su $, subject only to the constraint~\eqref{primitivity}.  As we see from~\eqref{flowed gsu} (and its corresponding version for $\sltwo$), these strings wrap different cycles of the $y$-$\phi$-$\psi$ torus.

The relation between the bipolar coordinates associated to the underlying $\sltwo\times\sutwo$ geometry, and the Gibbons-Hawking coordinates used for the bubble equations of Section~\ref{sec:D1-P}, is obtained by combining \eqref{eq:coords1}, \eqref{eq:coords2} and the relation between $\rho$ and $r$ in~\eqref{eq:coords3},
\be
\frac{2\hat r}{\hat a}\cos\hat\theta = \cosh2\rho\,\cos2\theta-1
~~,~~~~
\frac{2\hat r}{\hat a}\sin\hat\theta = \sinh2\rho\,\sin2\theta ~.
\ee
These relations connect the quantum numbers of the unexcited wound string to the geometry of figure~\ref{fig:3ctr}, as follows.
Using \eqref{jmsu}, \eqref{thetapm}, \eqref{jmsl}, \eqref{rhopm} and \eqref{rightbps}, we identify the ratio between the separation $\hat b$ between centers 1 and 3 and the separation $\hat a$ between centers 1 and 2 to be
\be
\label{b over a}
\frac{\hat b}{\hat a} = \half\bigg(\frac{\Msl_\sl-\Msu_\su}{J}\bigg) ~.
\ee
Similarly, the angle $\hat\theta_0$ between the Taub-NUT/supertube axis and the Taub-NUT/probe axis is
\be
\cos\hat\theta_* = \frac{\Msl_\sl\Msu_\su-J^2}{J(\Msl_\sl - \Msu_\su)} ~.
\ee
We can also determine the separation $\hat c$ between the supertube and probe centers using the law of cosines,
\be
\label{c over a}
\frac{\hat c}{\hat a} = \bigg[1+ \bigg(\frac {\hat b}{\hat a}\bigg)^2 +2 \bigg(\frac {\hat b}{\hat a}\bigg)\cos\hat\theta_* \bigg]^{1/2}
=  \half \bigg(\frac{\Msl_\sl
+
\Msu_\su}{J}\bigg)~.
\ee
Using~\eqref{rho glmt}, \eqref{theta glmt}, and \eqref{rhomap}, we thus reproduce Eq.~\eqref{eq:cossq-abc}.%
\footnote{A simple example illustrates how the choice of quantum numbers specifies the location of the probe: Turning on $n=J-\sfM_\sl$ but setting $m=\sfM_\su+J=0$ in the single-mode superstratum yields $\hat\theta_*=\pi$; the centers are all collinear, with the probe at the supertube for $n=0$ and then moving away toward larger radius along the line passing through the center of space (the origin in the Gibbons-Hawking $\bR^3$ base), and the supertube center.  Turning off $n$ and dialing $0<m<2J$, the probe remains collinear with the other two centers but runs between them, until at $\sfM_\su=+J$ (corresponding to $m=2J$) and $n=0$, the probe is at the center of space.  If we then start dialing up $n$ with $\sfM_\su=+J$, the centers again remain collinear, and the probe now moves away from the origin in the direction opposite to the supertube center.}

It is quite remarkable that two quite disparate approaches lead to the same result.  On the one hand, the worldsheet calculation considers only the physical state constraints of a perturbative string, without regard for its back-reaction on the ambient spacetime.  On the other hand, the D1-P probe calculation considers only the geometry sourced by the probe, without regard to the effective theory on the brane.  Yet each consideration is sufficient to fix the probe location.

%%%%%%%%%%%%%%%%%%%%%%%%%%%%%%%%%%%%%%
\section{Superstrata, quantum effects and black holes}
\label{sec:BHregime}
%%%%%%%%%%%%%%%%%%%%%%%%%%%%%%%%%%%%%

Our technique to approximate the superstratum with a three-center solution allows us to see clearly some of the features of the superstratum geometry that are harder to see in the full smooth solution. For example, one can try to estimate the length of the AdS$_2$ throat of the superstratum, and to compare the result with the length of the AdS$_2$ throat of the corresponding extremal black-hole solutions.

In the classical extremal black-hole geometry this length is infinite, but it 
was shown that quantization of the moduli of microstate geometries \cite{Bena:2006kb, Bena:2007qc,deBoer:2008zn, deBoer:2009un}  limits the depth of throats. This leads to the correct mass-gap for the dual CFT but also  suggests that, despite  the macroscopic scales of the geometries, quantum effects are becoming dominant in such deep throats.

More recently, a similar conclusion was reached \rcite{Lin:2022rzw} using an Euclidean calculation in JT gravity, which also implies that below a certain maximal depth quantum effects become important and invalidate the extremal black-hole solution.

An alternative to  comparing throat lengths is to calculate the ratio of the redshifts between the top and the bottom of the AdS$_2$ throat, both in superstrata and in the ``corrected" black-hole solution \rcite{Lin:2022rzw}, and in the deepest multi-center solutions.

We focus on ``deep superstrata" which have a long AdS$_2$ throat and exist when  $J_R\ll N$.  We also consider large values of $k$ and $n$, so that the supergravity wave is highly localized and the effective description is accurate. 
The top of the AdS$_2$ throat of superstrata is where the superstratum radial coordinate is
\be
\label{AdS2 top}
r_{\rm top}^2 \;\approx\; Q_P \;\approx\; b^2\frac{n}{k} ~.
\ee
When $n\gg k$, the location of the momentum wave that supports the AdS$_2$ throat is far away from the other centers, at 
\be
\label{AdS2 bottom}
r_{\rm bottom}^2 \;\approx\; a^2 \frac{n}{k} ~.
\ee 
However, when $n$ is of the same order as $k$ or smaller, this equation needs to be changed. The distance from the momentum center to the other two centers of the effective superstratum solution becomes of order $a$ or smaller, so the AdS$_2$ region of the superstratum throat terminates when\footnote{The intuition behind this is very simple: the location in the $\IR^3$ base of the multi-center solution where the solution ceases being spherically-symmetric is where the AdS$_2$ region of the throat ends and the cap region begins.} :
 \be
 r_{\rm bottom}^2 \,\sim\, a^2 \,. 
 \ee
Hence, one can write concisely the expression for the location of the bottom of the superstratum AdS$_2$ throat as :
 \be
 \label{AdS2 bottom new}
 r_{\rm bottom}^2 \;\approx\; a^2 \max\left(\frac{n}{k} , 1 \right) \,.
 \ee

To compute the length of the AdS$_2$ portion of the throat we need to use the 
AdS$_2$ radial coordinate, ${\mathtt r} \propto r^2$, and we obtain 
\be
d_{\text{AdS$_2$}} \; \approx \; \int\limits_{~r_{\rm bottom}^2}^{~~r_{\rm top}^2} \!\frac{d{\mathtt r}}{{\mathtt r}} \;\approx \;
 \log 
{{r_{\rm top}^2} \over {r_{\rm bottom}^2}} \;.
\ee

Alternatively, we can compute the ratio of the redshifts between the top and the bottom of the AdS$_2$ throat 
\be
\Delta \;\equiv\;  \sqrt{g_{00}^{\rm top} \over g_{00}^{\rm bottom}} \;=\; {{\mathtt r}_{\rm AdS_2}^{\rm top} \over {\mathtt r}_{\rm AdS_2}^{\rm bottom}} \;=\; {r_{\rm top}^2 \over r_{\rm bottom}^2}\;.
\ee

When $b \gg a$, the ratio of $b$ and $a$ in the superstratum solution can be expressed in terms of the integer fivebrane and onebrane charges, $n_1$ and $n_5$, and  the half-quantized right-moving angular momentum, $J_R$. Using (\ref{ssReg}), (\ref{sschg1}), (\ref{eq:q-quant}), and (\ref{eq:j-quant}), one has:
\be
 \frac{b^2}{a^2} \;=\; 
{Q_1 Q_5 \over R_y \hat{J_R}}\;=\; 
 {n_1 n_5 \over 2 J_R} \,.
\ee

Note that for superstrata, the holographic dictionary can be extrapolated to small $J_R$, however when one approaches $J_R=1/2$, the extrapolation of the dual coherent states in the holographic CFT becomes a quantum superposition over a small number of states. Therefore, the holographic CFT indicates that $J_R=1/2$ sets the maximum length of the AdS$_2$ throat of superstrata. (See also the related discussion in~\cite{Bena:2018bbd}).

Since the quantized momentum charge of the superstratum is $n_p = n_1 n_5 \frac{n}{k}$, we can express the length of the throat as 
 \be
 \label{length-SS}
 d_{\rm AdS_2} \; \approx \; \log\left( {\sqrt{n_1 n_5 n_p\vphantom{d}}\over 2 J_R} \right)\,\min\left(\sqrt\frac{k}{n} ,~\sqrt\frac{n}{k}   \right)\,.
 \ee
 Alternatively, we can express the redshift difference as
 \be
\Delta \;\approx\;  {\sqrt{n_1 n_5 n_p\vphantom{d}}\over 2 J_R} \min\left(\sqrt\frac{k}{n}, \sqrt\frac{n}{k} \right).
\ee

We can now compare this AdS$_2$ length with the length of the AdS$_2$ throat of the supersymmetric black-hole solution where quantum effects  are supposed to invalidate the classical extremal black-hole geometry~\rcite{Lin:2022rzw}
 \be
 \label{length-BH}
 d^{\rm BH}_{\rm AdS_2} \; \approx \; \log  S \;
 \approx \; \log  \sqrt{n_1 n_5 n_p\vphantom{d}}\;  ,
 \ee
 where $S \sim \sqrt{n_1 n_5 n_p\vphantom{d}}$ is the entropy of the BPS black hole with these charges. The redshift difference corresponding to this throat length is simply
\be
\Delta \,\approx\, S \,.
\ee
 
Our result indicates that superstrata have AdS$_2$ throats that are always shorter than or equal to the AdS$_2$ throat length suggested by the calculations of~\rcite{Lin:2022rzw} using JT gravity. The equality happens only when  $n/k ={\cal O}(1)$ and $J_R={\cal O}(1)$.

It is not hard to see that the quantum effects that cut off the deepest scaling multi-center solutions also come in at a similar scale. This happens when $J_R \sim 1$ \cite{Bena:2006kb, Bena:2007qc,deBoer:2008zn, deBoer:2009un} and the distance between the GH centers of a scaling solutions, using the coordinates of the $\IR^3$ base of the solution is \rcite{Bena:2006kb}\footnote{We use the fact that the charges are proportional to the square of the dipole fluxes, $d_i$, and that in the D1-D5-P decoupling limit $Q_1, Q_5$ are larger than $Q_P$, so $d_3 > d_1, d_2$.}
\be
r^{\rm GH}_{\rm bottom} \;\approx\; {1 \over d_3} \;\approx\; {\sqrt{Q_p \over Q_1 Q_5}}\;.
\ee
Using the fact that $r^{\rm GH}_{\rm top} \approx Q_P$ and that the AdS$_2$ radial variable is the same as $r_{\rm GH}$, this gives 
\be
 \label{length-scaling}
 d^{\,\rm scaling}_{\rm AdS_2} \; \approx \; \log  \sqrt{n_1 n_5 n_p\vphantom{d}}\;  ,
 \ee

We thus have a remarkable convergence of the results of three calculations:
\begin{enumerate}[(i)]
\item \label{item:stratum}
The maximal AdS$_2$ throat length computed in a superstratum classical solution that is horizonless and smooth in ten-dimensions, \eqref{length-SS};
\item \label{item:JT}
The maximal AdS$_2$ throat length computed by estimating quantum effects in a classical Euclidean black-hole background in a two-dimensional effective theory;
\item \label{item:multictr}
The maximal AdS$_2$ throat length computed by quantizing multi-center solutions.
\end{enumerate}

These calculations are done in different theories:
(\ref{item:stratum}) superstrata, which keep all the bulk degrees of freedom of six-dimensional supergravity, but throw away all information about excitations that depend on the internal $T^4$ coordinates, as well as stringy modes (in particular those associated to the underlying fivebranes);
(\ref{item:JT}) JT gravity, which throws away almost all information about the theory in which the black hole is constructed and keeps only one light mode; (\ref{item:multictr}) quiver quantum mechanics, which throws away all information about higher dimensions and keeps only degrees of freedom corresponding to multi-center dynamics.
Hence, a priori, it was possible that these three different approximations could have given different estimates of the location where quantum effects become important, such that one cannot trust the classical solution. It is quite remarkable that they all indicate that classical solutions that have an AdS$_2$ throat longer than $\log S$ are problematic.

%%%%%%%%%%%%%%%%%%
%%%%%%%%%%%%%%%%%%%%
%%%%%%%%%%%%%%%%%%%%%%%%%%%%%%%%%%%%%%

%%%%%%%%%%%%%%%%%%%%%%%%%%%%%%%%%%%%%
\section{Discussion}
\label{sec:Conclusion}
%%%%%%%%%%%%%%%%%%%%%%%%%%%%%%%%%%%%%

The deep tension between General Relativity and Quantum Mechanics is strong evidence for the view that GR is an effective field theory, and the black-hole uniqueness theorems are a testament to the failure of GR to resolve black-hole microstructure.   The fact that GR has been stunningly successful in describing large-scale structure of both the universe and black-hole mergers is equally a testament to just how powerful an effective field theory can be when applied in its appropriate domain of validity. In this context, the microstate geometry program may be regarded as a milestone along the journey to finding much better effective field theories that can describe black-hole microstructure.

We take it as a given that one needs to use the full force of string theory to resolve the quantum properties, and structure, of a black hole.  This is also one of the starting points of the fuzzball paradigm, which posits that a complete description of the structure of a black hole must involve a complex, chaotic set of quantum string states, and that horizons and singularities are artifacts signaling the failure of effective field theory. Supergravity, as the low-energy limit of string theory, affords a much richer and more powerful effective field theory, as is evident from the huge range of microstate geometries, many of whose microscopic interpretations in the dual CFT have passed precision holographic tests.

Much of the research into microstate geometries has been driven by the desire to see, and account for, as much of the microstructure as possible. This has led to supergravity solutions with less and less symmetry and more and more intricate detail. This has given us a deeper understanding of the physical underpinnings and consequences of microstructure, ranging from energy gaps and brane fractionation, through tidal scrambling to our current discussion of momentum migration.   However, some of these solutions are so complex, and so lacking in symmetry, that it can be  very hard to probe them, or analyze their excitations.

In this paper, we have tried to address this challenge by developing a more systematic approach to effective supergravity descriptions of the intricate families of black-hole microstructure by simplifying, or averaging over, unnecessary detail while  retaining the physics of interest. This simplification and averaging will typically introduce singularities, or horizons, but one accepts them, \emph{knowing} that a full supergravity solution can resolve this behavior into microstructure.%  
\footnote{
In this respect, our philosophy resembles that in earlier work on two- and three-charge solutions~\cite{Lunin:2002fw,Lunin:2002bj,Balasubramanian:2005qu,Dabholkar:2006za,Chakrabarty:2021sff,Martinec:2022okx} in which certain high-frequency details of the solution were averaged over to simplify their description, at the expense of generating a singular profile or shockwave in the effective description.  In fact, although our discussion of a D1-P center in the GLMT background in section \ref{sec:D1-P_in_GLMT} did not involve such additional coarse-graining because we imposed the primitivity condition \eqref{F1-P primitivity}, it is a simple generalization to relax it so that $\hat{m}'$ is a free parameter satisfying the bound $\hat{m}'\le \hat{Q}_1'\hat{Q}_3'/(2\hat{\kappa}'_2)$. The non-primitivity  parameter $\delta=\hat{Q}_1'\hat{Q}_3'/(2\hat{\kappa}'_2)-\hat{m}'$ is related to the amount of high-frequency, small-amplitude fluctuations of the D1-P profile about the circular shape which have been coarse-grained.}

In particular, we have shown how several different detailed descriptions of black-hole microstructure can be reduced to a much simpler five-dimensional supergravity description. These five-dimensional geometries have some singular sources but we know how they can be resolved, in different limits, by the detailed descriptions of black-hole microstructure.  The five-dimensional description is much simpler but provides a  more  intuitive description of how the brane and momentum sources interact and how they are bound together  in the  gravitational back-reacted solution.  

Momentum migration provides a very good example of this. Many microstate geometries correspond to brane systems that carry momentum as transverse waves, or in fluctuating charge densities.  From the brane perspective, and in terms of the dual holographic CFT, these momentum waves lie on the underlying branes.  However, the back-reaction of the branes on a compact locus typically pinches-off that locus, creating a geometric transition to a geometry with a non-trivial cohomological cycle. As has been noted elsewhere, trying to keep the momentum localized on the original brane locus would create a singular geometry \cite{Niehoff:2013kia}. What happens instead is that the back-reacted geometry remains smooth, the momentum wave ``detaches'' from the brane locus, and its peak amplitude moves to a point some distance away from the original  brane locus.  We refer to this as ``momentum migration,''  and we have shown how the location of the peak amplitude can be  determined by the bubble equations of five-dimensional supergravity.  That is, the location of the peak of the momentum wave is typically bound to the other sources in the background, and can be determined by its charges and the charges and locations of the other sources.  Moreover, if a particular quantum number of the momentum wave source becomes large, the source itself becomes highly localized. 
 
A very natural question arises out of our work: what information is being retained and what is being lost in our effective descriptions? For superstrata, we have shown how the averaging of momentum modes washes out all the detailed oscillations along commuting isometry directions (the Cartan subalgebra directions), replacing the mode details by the mean-square ``bump-functions'' on the sphere and AdS directions.  These functions retain details of the mode numbers $(k,m,n)$ and the amplitude $b$, and the position of the peak on the sphere and AdS are determined by the ratios $m/k$ and $n/k$ respectively.  The value of $k$ then determines the width of the bump, or the extent to which the wave is localized. Our effective description thus seems to retain most of the information about the wave. 
However, in the large-mode-number limit, where the bump function becomes a delta function, the individual values of $(k,m,n)$ are lost and we only see their ratios (see for example, \eqref{sschg1}) in the effective multi-center solution.  Furthermore, one should remember that the basic superstratum has two independent classes of waves (the two holomorphic functions or three variables), corresponding to the $\ket{00}$ and $G^{+1}G^{+2}\ket{00}$ strands in CFT \cite{Ceplak:2018pws,Heidmann:2019zws}, as well as their generalizations built on $|(\dot A\dot B)\rangle$ strands~\rcite{Bakhshaei:2018vux}. Moreover, the newer vector superstrata \cite{Ceplak:2022pep,Martinec:2022okx,Ceplak:2022wri,Ceplak:2024dbj}, corresponding to excitations built on $G^{+ A}|\dot A\dot\alpha\rangle$ or $G^{+A}|\alpha \dot B\rangle$ strands in CFT, add yet more modes with similar bump functions.  
In principle, one could consider superstrata based on other strands \cite{Ceplak:2018pws,Shigemori:2020yuo}.  These details, and all the information about which particular fields are actually carrying the momentum, are lost. Moreover, it would prove rather challenging to de-convolve all the individual mode contributions in an effective geometry of a complex multi-mode superstratum.

Some of the major threads in our work here are the universal aspects of the localization of charge and momentum sources.  We have seen how it comes about in microstate geometries (Section \ref{sec:EffSS}), in wave-functions (Section \ref{sec:wavefns}), for string probes (Section \ref{sec:D1-P}), and in the exact description of string wavefunctions (Section \ref{sec:wound-strings}).  At a mundane mathematical level, this is because all these analyses devolve into some aspect of harmonic analysis on AdS$_3 \times S^3$, and this leads inexorably to the bump functions $\Delta_{k,m,n}$.  However, this observation misses the essential physical point that all of these approaches start from different approaches to black-hole microstructure, and the fact that they converge on the same results is remarkable.  For example, in microstate geometries, sources localize as a result of the bubble equations, which enforce the absence of CTCs; in the string worldsheet analysis, the same localization is the result of imposing the physical state conditions.

There is another remarkable aspect of this convergence of ideas which reinforces a very useful, and yet simple, physical picture of deep, scaling superstrata.

As the standard pictures of deep scaling superstrata (see Fig.~\ref{fig:supertstratum}) suggest, one can think of these geometries as if one had taken a smooth, global AdS$_3$ and cut a circular  disk  out of the bottom of the  AdS$_3$ bowl, and then glued a vertical, cylindrical pipe (the AdS$_2 \times S^1$ throat) to the hole, and then capped it off at the bottom by gluing that cut-out disk to the bottom of the pipe. The edge at which the cut is made is defined by the outermost of the momentum wave or the original supertube locus.  Indeed, this outermost feature   defines the edge of the bottom of the vertical pipe.

Our effective microstructure analysis can be applied to any depth of throat. The fact that the ``geographic''  features of all the charge sources are identical for shallow and deep geometries means that all this structure simply remains unmodified as it descends the throat, and that the outermost feature, whether it be the momentum wave or the singular momentum center of the three-center effective description, defines the edge of the cap at the bottom of the throat.

These considerations also lead to the results of Section \ref{sec:BHregime}, which confirm another universal aspect of the geometrization of black-hole microstructure:   AdS$_2$ throats longer than $\log S$ are problematic.

Another place in which effective superstrata have already been used implicitly is in studies of tidal disruption \rcite{Tyukov:2017uig,Bena:2020iyw, Bena:2018mpb, Martinec:2020cml,Ceplak:2021kgl} where the phenomenon depends on ultra-relativistic motion through non-trivial multipole moments.  The details of the microstructure were not needed to reveal this effect.

Apart from revealing some of the essential physics of microstructure, there is an obvious practical importance in finding simpler effective solutions: It makes analysis easier. However, in passing to effective microstructure we have may also appear to have ``let the genie out of the bottle'': we are once again allowing singular geometries, which begs the question, what singularities are now allowed, and what singularities must be still be forbidden?

While this is a very interesting general question, it goes far beyond the scope of the present work.  For the present, we can  offer a simple, practical prescription.  One will always get effective microstructure with allowable singularities if one performs an average of an existing microstate geometry. (As we did in Section \ref{sec:EffSS}.)  This will yield a simpler geometry but has the advantage that one knows how to resolve the singularities and this will reveal the limitations of the effective geometry. It would obviously be extremely interesting to find a much broader ``singularity repair kit'' that would allow a far broader definition of allowed effective geometries.

We hope, and expect, the ideas of effective microstructure to be important in the future.  
For example, in the classification of \rcite{Chang:2025rqy}, it has recently been suggested that many microstate geometries, like superstrata, are ``monotone'' BPS states and that stringy excitations around them could be ``fortuitous''. For a recent analysis of the role of boundary gravitons in the monotone/fortuitous classification, see \rcite{Hughes:2025car}. It would be interesting to explore these ideas using both superstrata and effective superstrata to find such fortuitous stringy excitations and determine the extent to which coarse graining into effective microstructure configurations, is compatible with fortuity.

More broadly, characterizing effective microstructure could be immensely important in the construction of templates, and the extraction of universal observable signals of microstructure.

%%%%%%%%%%%%%%%%%%%%%%%%%%%%%%%%%%%%%

\vspace{3em}
\noindent 
{\bf Acknowledgements:} \\
We thank Nejc {\v C}eplak, Roberto Emparan, Pierre Heidmann, Ang\`ele Lochet, Dimitrios Toulikas and the other participants in the {\it Black hole Microstructure  VI and~VII} conferences  for interesting and thought-provoking discussions. The work of IB and NPW was supported in part by the ERC Grant 787320 - QBH Structure. The work of IB was also supported in part by the ERC Grant 772408 - Stringlandscape. The work of EJM was supported in part by DOE grant DE-SC0009924. The work of MS was supported in part by MEXT
KAKENHI Grant Numbers 21H05184 and 24K00626.
The work of DT was supported by a Royal Society Tata University Research Fellowship.
The work of NPW was also supported in part by the DOE grant DE-SC0011687.

\newpage
\appendix

%%%%%%%%%%%%%%%%%%%%%%%%%%%%%%%%%
\section{Multiple D1-P light centers in the GLMT background}
%%%%%%%%%%%%%%%%%%%%%%%%%%%%%%%%%
\label{sec:multiple_probes}

In this Appendix, we generalize the discussion in Section \ref{sec:D1-P_in_GLMT} by considering multiple D1-P light centers in the GLMT background and show that they can be regarded as independent centers.

When two D1-P centers are present, the charge vectors and the asymptotic moduli vector are
\begin{align}
    \Gamma^{(1)}&=\biggl(s+1,\left(-\hat{\kappa}_1,-\hat{\kappa}_2-(s+1)(\hat{\kappa}_{2 (1)}'+\hat{\kappa}_{2 (2)}'),-\hat{\kappa}_3\right),
    \notag\\
    &
    \hspace{8mm}
    \Big(-\tfrac{\hat{\sfq}_1+(s+1)(\hat{\kappa}_{2 (1)}'+\hat{\kappa}_{2 (2)}')\hat{\kappa}_3}{s+1},-\tfrac{\hat{\sfq}_2}{s+1},
    -\tfrac{\hat{\sfq}_3+(s+1)(\hat{\kappa}_{2(1)}'+\hat{\kappa}_{2 (2)}')\hat{\kappa}_1 }{s+1}\Big),
    -\tfrac{\tilde{m}+(s+1)(\hat{\kappa}_{2 (1)}'+\hat{\kappa}_{2 (2)}')\hat{\kappa}_1 \hat{\kappa}_3}{2(s+1)^2}\biggr),
    \nn\\
    \Gamma^{(2)}
    &=\biggl(-s,
    \left(\hat{\kappa}_1,\hat{\kappa}_2+s(\hat{\kappa}_{2 (1)}'+\hat{\kappa}_{2 (2)}'),\hat{\kappa}_3\right), 
    \notag\\
    &\hspace{8mm}\left(\tfrac{\hat{\sfq}_1+s (\hat{\kappa}_{2 (1)}'+\hat{\kappa}_{2 (2)}')\hat{\kappa}_3}{s},
    \tfrac{\hat{\sfq}_2}{s},\tfrac{\hat{\sfq}_3+s \hat{\kappa}_1\,(\hat{\kappa}_{2 (1)}'+\hat{\kappa}_{2 (2)}')}{s}\right),
    \tfrac{\tilde{m}+s \hat{\kappa}_1\,(\hat{\kappa}_{2(1)}'+\hat{\kappa}_{2 (2)}')\,\hat{\kappa}_3}{2\,s^2}\biggr),\nn\\
    \Gamma_{\text{D1-P}}^{(3)}&=\left(0,\left(0,\hat{\kappa}_{2 (1)}',0\right),\left(\hat{Q}_{1 (1)}',0,\hat{Q}_{3 (1)}'\right),\hat{m}_{(1)}'\right),\nn\\
    \Gamma_{\text{D1-P}}^{(4)}&=\left(0,\left(0,\hat{\kappa}_{2 (2)}',0\right),\left(\hat{Q}_{1 (2)}',0,\hat{Q}_{3 (2)}'\right),\hat{m}_{(2)}'\right),\nn\\
    h&=(0,(0,0,0),(0,0,1),0)\,.
\end{align}
The conventions for the distances between the centers are given as below:
\begin{center}
\begin{tikzpicture}[scale=0.6]

\coordinate (A) at (0, 6);
\coordinate (B) at (2,4);
\coordinate (C) at (0, 1);
\coordinate (D) at (-1,4.5);

\draw[thick] (A) -- (B) -- (C) -- cycle;
\draw[thick] (A) -- (D) -- (C) -- cycle;
\draw[thick] (A) -- (D) -- (B) -- cycle;

\fill[black] (A) circle (3pt);
\fill[black] (B) circle (3pt);
\fill[black] (C) circle (3pt);
\fill[black] (D) circle (3pt);

\node[anchor=south] at (A) {$s+1$, (1)};
\node[anchor=west] at (C) {$-s$, (2)};
\node[anchor=west] at (B) {D1-P, (3)};
\node[anchor=east] at (D) {D1-P, (4)};

\node[anchor=south] at ($ (A)!0.5!(B) $) {$\hat{b}$};
\node[anchor=west] at ($ (B)!0.5!(C) $) {$\hat{c}$};
\node[anchor=west,yshift=-12] at ($ (C)!0.5!(A) $) {$\hat{a}$};
\node[anchor=south,xshift=-5,yshift=-3] at ($ (D)!0.5!(A) $) {$\hat{d}$};
\node[anchor=east] at ($ (D)!0.5!(C) $) {$\hat{e}$};
\node[anchor=north,yshift=1,xshift=3] at ($ (D)!0.5!(B) $) {$\hat{f}$};

\end{tikzpicture}
\end{center}

The relevant bubble equations are
\begin{align}
\begin{gathered}
    \frac{\Gamma^{12}_0}{\hat{a}}+\frac{\Gamma'^{12}}{\hat{a}}+\frac{\Gamma'^{13}}{\hat{b}}+\frac{\Gamma'^{14}}{\hat{d}}=\frac{\hat{\kappa}_3}{2}\;,\\
    \frac{\Gamma'^{31}}{\hat{b}}+\frac{\Gamma'^{32}}{\hat{c}}+\frac{\Gamma'^{34}}{\hat{f}}\,=\,0\;,\qquad
    \frac{\Gamma'^{41}}{\hat{d}}+\frac{\Gamma'^{42}}{\hat{e}}+\frac{\Gamma'^{43}}{\hat{f}}\,=\,0\,,
\end{gathered}
\end{align}
where the notation follows that in \eqref{eq:first bubble GLMT}. We assume the charges of the D1-P centers to be much smaller than those of the background. Then the zeroth-order terms in the first equation give
\begin{align}
    \frac{\Gamma'^{12}_0}{\hat a'}=\frac{\hat{\kappa}_3}{2}
\end{align}
while the first-order terms give
\begin{equation}
\label{eq:Bubble equation 1 two probes}
    -\frac{\hat{\kappa}_3 }{2}\frac{\hat{a}'}{\hat{a}_0}+\frac{\Gamma'^{12}}{\hat{a}_0}+\frac{\Gamma'^{13}}{\hat{b}}+\frac{\Gamma'^{14}}{\hat{d}}=0\,.
\end{equation}

From the above equations, we find the angular momentum added by presence of the extra D1-P centers:
\begin{equation}
    \hat J_R'=\frac{\hat{\kappa}_3}{2} \hat{a}'-(\hat{Q}_{1(1)}'+\hat{Q}_{1 (2)}')\frac{\hat{\kappa}_1 \hat{\kappa}_3}{2 s(s+1)}\,.
\end{equation}
We can naturally split the change in $\hat{a}$ as $\hat{a}'=\hat{a}'_{(1)}+\hat{a}'_{(2)} $. Then the bubble equation \eqref{eq:Bubble equation 1 two probes} can be nicely separated into independent equations:
\begin{align}
    -\frac{\hat{\kappa}_3}{2} \frac{\hat{a}'_{(1)}}{\hat{a}_0}+\frac{\Gamma'^{13}}{\hat{b}}=0\,,\qquad
     - \frac{\hat{\kappa}_3}{2} \frac{\hat{a}_{(2)}}{\hat{a}_0}+\frac{\Gamma'^{14}}{\hat{d}}=0\,.
\end{align}
It is important to note that, at the order we are working in, $\Gamma'^{14}$ does not receive contributions from the first D1-P center and $\Gamma'^{13}$ does not get contribution from the second D1-P center.  This means that the equations for the two centers completely decouple:
\begin{align}
    \hat{J}_{R(1)}'=
    \frac{\hat{\kappa}_3}{2} \hat{a}'_{(1)}-\hat{Q}_{1(1)}'\frac{\hat{\kappa}_1 \hat{\kappa}_3}{2 s(s+1)},
        \qquad
    \hat{J}_{R(2)}'=
    \frac{\hat{\kappa}_3}{2} \hat{a}'_{(2)}-\hat{Q}_{1(2)}'\frac{\hat{\kappa}_1 \hat{\kappa}_3}{2 s(s+1)}\,.
\end{align}
Since all the equations can be split as if we had two independent centers, we obtain exactly the same equations for the positions as \eqref{eq:cossqtheta-st} and \eqref{eq:rsq-st}:
\begin{equation}
    \cos^2 \tilde{\theta}_{*(i)} =\frac{1}{2}\left(1-\frac{\hat{m}_{(i)}'-\frac{\hat{\kappa}_{2(i)}'\hat{Q}_5}{8}}{ \hat{J}_{R(i)}'+ \frac{{Q}_5}{8 \hat{\kappa}_3}\hat{Q}_{1(i)}'}\right),
\end{equation}
\begin{equation}
    \tilde{r}_{(i)}^2=2 \hat{a}_0\left(-1+\frac{ \hat{m}'_{(i)} (2s+1)+\hat{\kappa}_{2(i)}'(2 s+1)\frac{{Q}_5}{8}-(\hat{\kappa}_3 \hat{Q}_{3(i)}'+\frac{{Q}_5 s(s+1)}{4\hat{\kappa}_3}\hat{Q}_{1(i)}')}{\hat{J}_{R(i)}'+\frac{{Q}_5}{8 \hat{\kappa}_3}\hat{Q}_{1(i)}' }\right)\,,
\end{equation}
where $i=1,2$. The same analysis would work for an arbitrary number of light D1-P centers.

%%%%%%%%%%%%%%%%%%%%%%%%%%%%%%%%%%%%%%%%%%%%%%%%%%
%%%%%%%%%%%%%%%%%%%%%%%%%%%%%%%%%%%%%%%%%%%%%%%%%%

%%%%%%%%%%%%%%%%%%%%%%%%%%%%%%%%%%%%%%%%%%%%%%%%%%%%%
\newpage

\begin{adjustwidth}{-1mm}{-1mm} % to adjust the L and R margins 

\bibliographystyle{utphys}      

\bibliography{references}       % calls file "microstates.bib"

\providecommand{\href}[2]{#2}\begingroup\raggedright\begin{thebibliography}{100}

\bibitem{Bena:2022rna}
I.~Bena, E.~J. Martinec, S.~D. Mathur, and N.~P. Warner, ``{Fuzzballs and
  Microstate Geometries: Black-Hole Structure in String Theory},''
  \href{http://arxiv.org/abs/2204.13113}{{\ttfamily arXiv:2204.13113
  [hep-th]}}.

\bibitem{Mathur:2009hf}
S.~D. Mathur, ``{The information paradox: A pedagogical introduction},''
  \href{http://dx.doi.org/10.1088/0264-9381/26/22/224001}{{\em Class. Quant.
  Grav.} {\bfseries 26} (2009) 224001},
\href{http://arxiv.org/abs/0909.1038}{{\ttfamily arXiv:0909.1038 [hep-th]}}.
%%CITATION = 0909.1038;%%.

\bibitem{Bena:2007kg}
I.~Bena and N.~P. Warner, ``{Black holes, black rings and their microstates},''
  \href{http://dx.doi.org/10.1007/978-3-540-79523-0}{{\em Lect. Notes Phys.}
  {\bfseries 755} (2008) 1--92},
\href{http://arxiv.org/abs/hep-th/0701216}{{\ttfamily arXiv:hep-th/0701216}}.
%%CITATION = HEP-TH/0701216;%%.

\bibitem{Bena:2025pcy}
I.~Bena and N.~P. Warner, ``{Microstate Geometries},''
  \href{http://arxiv.org/abs/2503.17310}{{\ttfamily arXiv:2503.17310
  [hep-th]}}.

\bibitem{Bena:2015bea}
I.~Bena, S.~Giusto, R.~Russo, M.~Shigemori, and N.~P. Warner, ``{Habemus
  Superstratum! A constructive proof of the existence of superstrata},''
  \href{http://dx.doi.org/10.1007/JHEP05(2015)110}{{\em JHEP} {\bfseries 05}
  (2015) 110},
\href{http://arxiv.org/abs/1503.01463}{{\ttfamily arXiv:1503.01463 [hep-th]}}.
%%CITATION = ARXIV:1503.01463;%%.

\bibitem{Bena:2016agb}
I.~Bena, E.~Martinec, D.~Turton, and N.~P. Warner, ``{Momentum Fractionation on
  Superstrata},'' \href{http://dx.doi.org/10.1007/JHEP05(2016)064}{{\em JHEP}
  {\bfseries 05} (2016) 064},
\href{http://arxiv.org/abs/1601.05805}{{\ttfamily arXiv:1601.05805 [hep-th]}}.
%%CITATION = ARXIV:1601.05805;%%.

\bibitem{Bena:2016ypk}
I.~Bena, S.~Giusto, E.~J. Martinec, R.~Russo, M.~Shigemori, D.~Turton, and
  N.~P. Warner, ``{Smooth horizonless geometries deep inside the black-hole
  regime},'' \href{http://dx.doi.org/10.1103/PhysRevLett.117.201601}{{\em Phys.
  Rev. Lett.} {\bfseries 117} no.~20, (2016) 201601},
\href{http://arxiv.org/abs/1607.03908}{{\ttfamily arXiv:1607.03908 [hep-th]}}.
%%CITATION = ARXIV:1607.03908;%%.

\bibitem{Bena:2017xbt}
I.~Bena, S.~Giusto, E.~J. Martinec, R.~Russo, M.~Shigemori, D.~Turton, and
  N.~P. Warner, ``{Asymptotically-flat supergravity solutions deep inside the
  black-hole regime},'' \href{http://dx.doi.org/10.1007/JHEP02(2018)014}{{\em
  JHEP} {\bfseries 02} (2018) 014},
\href{http://arxiv.org/abs/1711.10474}{{\ttfamily arXiv:1711.10474 [hep-th]}}.
%%CITATION = ARXIV:1711.10474;%%.

\bibitem{Bakhshaei:2018vux}
E.~Bakhshaei and A.~Bombini, ``{Three-charge superstrata with internal
  excitations},'' \href{http://dx.doi.org/10.1088/1361-6382/ab01bc}{{\em Class.
  Quant. Grav.} {\bfseries 36} no.~5, (2019) 055001},
  \href{http://arxiv.org/abs/1811.00067}{{\ttfamily arXiv:1811.00067
  [hep-th]}}.

\bibitem{Ceplak:2018pws}
N.~{\v C}eplak, R.~Russo, and M.~Shigemori, ``{Supercharging Superstrata},''
  \href{http://dx.doi.org/10.1007/JHEP03(2019)095}{{\em JHEP} {\bfseries 03}
  (2019) 095}, \href{http://arxiv.org/abs/1812.08761}{{\ttfamily
  arXiv:1812.08761 [hep-th]}}.

\bibitem{Heidmann:2019zws}
P.~Heidmann and N.~P. Warner, ``{Superstratum Symbiosis},''
  \href{http://dx.doi.org/10.1007/JHEP09(2019)059}{{\em JHEP} {\bfseries 09}
  (2019) 059}, \href{http://arxiv.org/abs/1903.07631}{{\ttfamily
  arXiv:1903.07631 [hep-th]}}.

\bibitem{Ganchev:2022exf}
B.~Ganchev, A.~Houppe, and N.~P. Warner, ``{Elliptical and Purely NS
  Superstrata},'' \href{http://arxiv.org/abs/2207.04060}{{\ttfamily
  arXiv:2207.04060 [hep-th]}}.

\bibitem{Ceplak:2022pep}
N.~\v{C}eplak, ``{Vector Superstrata},''
  \href{http://dx.doi.org/10.1007/JHEP08(2023)047}{{\em JHEP} {\bfseries 08}
  (2023) 047}, \href{http://arxiv.org/abs/2212.06947}{{\ttfamily
  arXiv:2212.06947 [hep-th]}}.

\bibitem{Ganchev:2023sth}
B.~Ganchev, S.~Giusto, A.~Houppe, R.~Russo, and N.~P. Warner,
  ``{Microstrata},'' \href{http://dx.doi.org/10.1007/JHEP10(2023)163}{{\em
  JHEP} {\bfseries 10} (2023) 163},
  \href{http://arxiv.org/abs/2307.13021}{{\ttfamily arXiv:2307.13021
  [hep-th]}}.

\bibitem{Ceplak:2024dbj}
N.~{\v{C}}eplak and S.~D. Hampton, ``{Vector superstrata. Part II},''
  \href{http://dx.doi.org/10.1007/JHEP10(2024)011}{{\em JHEP} {\bfseries 10}
  (2024) 011}, \href{http://arxiv.org/abs/2405.05341}{{\ttfamily
  arXiv:2405.05341 [hep-th]}}.

\bibitem{Kanitscheider:2006zf}
I.~Kanitscheider, K.~Skenderis, and M.~Taylor, ``{Holographic anatomy of
  fuzzballs},'' \href{http://dx.doi.org/10.1088/1126-6708/2007/04/023}{{\em
  JHEP} {\bfseries 04} (2007) 023},
\href{http://arxiv.org/abs/hep-th/0611171}{{\ttfamily arXiv:hep-th/0611171}}.
%%CITATION = HEP-TH/0611171;%%.

\bibitem{Giusto:2015dfa}
S.~Giusto, E.~Moscato, and R.~Russo, ``{AdS$_{3}$ holography for 1/4 and 1/8
  BPS geometries},'' \href{http://dx.doi.org/10.1007/JHEP11(2015)004}{{\em
  JHEP} {\bfseries 11} (2015) 004},
\href{http://arxiv.org/abs/1507.00945}{{\ttfamily arXiv:1507.00945 [hep-th]}}.
%%CITATION = ARXIV:1507.00945;%%.

\bibitem{Giusto:2019qig}
S.~Giusto, S.~Rawash, and D.~Turton, ``{AdS$_{3}$ holography at dimension
  two},'' \href{http://dx.doi.org/10.1007/JHEP07(2019)171}{{\em JHEP}
  {\bfseries 07} (2019) 171}, \href{http://arxiv.org/abs/1904.12880}{{\ttfamily
  arXiv:1904.12880 [hep-th]}}.

\bibitem{Rawash:2021pik}
S.~Rawash and D.~Turton, ``{Supercharged AdS$_{3}$ Holography},''
  \href{http://dx.doi.org/10.1007/JHEP07(2021)178}{{\em JHEP} {\bfseries 07}
  (2021) 178}, \href{http://arxiv.org/abs/2105.13046}{{\ttfamily
  arXiv:2105.13046 [hep-th]}}.

\bibitem{Ganchev:2021ewa}
B.~Ganchev, S.~Giusto, A.~Houppe, and R.~Russo, ``{$\hbox {AdS}_3$ holography
  for non-BPS geometries},''
  \href{http://dx.doi.org/10.1140/epjc/s10052-022-10133-2}{{\em Eur. Phys. J.
  C} {\bfseries 82} no.~3, (2022) 217},
  \href{http://arxiv.org/abs/2112.03287}{{\ttfamily arXiv:2112.03287
  [hep-th]}}.

\bibitem{Martinec:2017ztd}
E.~J. Martinec and S.~Massai, ``{String Theory of Supertubes},''
  \href{http://dx.doi.org/10.1007/JHEP07(2018)163}{{\em JHEP} {\bfseries 07}
  (2018) 163}, \href{http://arxiv.org/abs/1705.10844}{{\ttfamily
  arXiv:1705.10844 [hep-th]}}.

\bibitem{Martinec:2018nco}
E.~J. Martinec, S.~Massai, and D.~Turton, ``{String dynamics in NS5-F1-P
  geometries},'' \href{http://dx.doi.org/10.1007/JHEP09(2018)031}{{\em JHEP}
  {\bfseries 09} (2018) 031}, \href{http://arxiv.org/abs/1803.08505}{{\ttfamily
  arXiv:1803.08505 [hep-th]}}.

\bibitem{Martinec:2019wzw}
E.~J. Martinec, S.~Massai, and D.~Turton, ``{Little Strings, Long Strings, and
  Fuzzballs},'' \href{http://dx.doi.org/10.1007/JHEP11(2019)019}{{\em JHEP}
  {\bfseries 11} (2019) 019}, \href{http://arxiv.org/abs/1906.11473}{{\ttfamily
  arXiv:1906.11473 [hep-th]}}.

\bibitem{Martinec:2020gkv}
E.~J. Martinec, S.~Massai, and D.~Turton, ``{Stringy Structure at the BPS
  Bound},'' \href{http://dx.doi.org/10.1007/JHEP12(2020)135}{{\em JHEP}
  {\bfseries 12} (2020) 135}, \href{http://arxiv.org/abs/2005.12344}{{\ttfamily
  arXiv:2005.12344 [hep-th]}}.

\bibitem{Bufalini:2021ndn}
D.~Bufalini, S.~Iguri, N.~Kovensky, and D.~Turton, ``{Black hole microstates
  from the worldsheet},'' \href{http://dx.doi.org/10.1007/JHEP08(2021)011}{{\em
  JHEP} {\bfseries 08} (2021) 011},
  \href{http://arxiv.org/abs/2105.02255}{{\ttfamily arXiv:2105.02255
  [hep-th]}}.

\bibitem{Bufalini:2022wyp}
D.~Bufalini, S.~Iguri, N.~Kovensky, and D.~Turton, ``{Worldsheet Correlators in
  Black Hole Microstates},''
  \href{http://dx.doi.org/10.1103/PhysRevLett.129.121603}{{\em Phys. Rev.
  Lett.} {\bfseries 129} no.~12, (2022) 121603},
  \href{http://arxiv.org/abs/2203.13828}{{\ttfamily arXiv:2203.13828
  [hep-th]}}.

\bibitem{Bufalini:2022wzu}
D.~Bufalini, S.~Iguri, N.~Kovensky, and D.~Turton, ``{Worldsheet computation of
  heavy-light correlators},''
  \href{http://dx.doi.org/10.1007/JHEP03(2023)066}{{\em JHEP} {\bfseries 03}
  (2023) 066}, \href{http://arxiv.org/abs/2210.15313}{{\ttfamily
  arXiv:2210.15313 [hep-th]}}.

\bibitem{Martinec:2022okx}
E.~J. Martinec, S.~Massai, and D.~Turton, ``{On the BPS Sector in AdS3/CFT2
  Holography},'' \href{http://dx.doi.org/10.1002/prop.202300015}{{\em Fortsch.
  Phys.} {\bfseries 71} no.~4-5, (2023) 2300015},
  \href{http://arxiv.org/abs/2211.12476}{{\ttfamily arXiv:2211.12476
  [hep-th]}}.

\bibitem{Martinec:2025xoy}
E.~J. Martinec, S.~Massai, and D.~Turton, ``{Cogito, ergo - strings:
  Supersymmetric ergoregions and their stringy excitations},''
  \href{http://arxiv.org/abs/2508.05420}{{\ttfamily arXiv:2508.05420
  [hep-th]}}.

\bibitem{Martinec:2024emf}
E.~J. Martinec and Y.~Zigdon, ``{BPS fivebrane stars. Part III. Effective
  actions},'' \href{http://dx.doi.org/10.1007/JHEP03(2025)074}{{\em JHEP}
  {\bfseries 03} (2025) 074}, \href{http://arxiv.org/abs/2411.16630}{{\ttfamily
  arXiv:2411.16630 [hep-th]}}.

\bibitem{Martinec:2023xvf}
E.~J. Martinec and Y.~Zigdon, ``{BPS fivebrane stars. Part I. Expectation
  values of observables},''
  \href{http://dx.doi.org/10.1007/JHEP02(2024)033}{{\em JHEP} {\bfseries 02}
  (2024) 033}, \href{http://arxiv.org/abs/2311.09155}{{\ttfamily
  arXiv:2311.09155 [hep-th]}}.

\bibitem{Martinec:2023gte}
E.~J. Martinec and Y.~Zigdon, ``{BPS fivebrane stars. Part II. Fluctuations},''
  \href{http://dx.doi.org/10.1007/JHEP02(2024)034}{{\em JHEP} {\bfseries 02}
  (2024) 034}, \href{http://arxiv.org/abs/2311.09157}{{\ttfamily
  arXiv:2311.09157 [hep-th]}}.

\bibitem{MZ4}
E.~J. Martinec and Y.~Zigdon, ``To appear,''.

\bibitem{Martinec:2015pfa}
E.~J. Martinec and B.~E. Niehoff, ``{Hair-brane Ideas on the Horizon},''
  \href{http://dx.doi.org/10.1007/JHEP11(2015)195}{{\em JHEP} {\bfseries 11}
  (2015) 195},
\href{http://arxiv.org/abs/1509.00044}{{\ttfamily arXiv:1509.00044 [hep-th]}}.
%%CITATION = ARXIV:1509.00044;%%.

\bibitem{Bena:2006kb}
I.~Bena, C.-W. Wang, and N.~P. Warner, ``{Mergers and Typical Black Hole
  Microstates},'' \href{http://dx.doi.org/10.1088/1126-6708/2006/11/042}{{\em
  JHEP} {\bfseries 11} (2006) 042},
\href{http://arxiv.org/abs/hep-th/0608217}{{\ttfamily arXiv:hep-th/0608217}}.
%%CITATION = HEP-TH/0608217;%%.

\bibitem{Bena:2007qc}
I.~Bena, C.-W. Wang, and N.~P. Warner, ``{Plumbing the Abyss: Black Ring
  Microstates},'' \href{http://dx.doi.org/10.1088/1126-6708/2008/07/019}{{\em
  JHEP} {\bfseries 07} (2008) 019},
\href{http://arxiv.org/abs/0706.3786}{{\ttfamily arXiv:0706.3786 [hep-th]}}.
%%CITATION = 0706.3786;%%.

\bibitem{Bena:2018bbd}
I.~Bena, P.~Heidmann, and D.~Turton, ``{AdS$_{2}$ holography: mind the cap},''
  \href{http://dx.doi.org/10.1007/JHEP12(2018)028}{{\em JHEP} {\bfseries 12}
  (2018) 028},
\href{http://arxiv.org/abs/1806.02834}{{\ttfamily arXiv:1806.02834 [hep-th]}}.
%%CITATION = ARXIV:1806.02834;%%.

\bibitem{Sen:2009bm}
A.~Sen, ``{Two Charge System Revisited: Small Black Holes or Horizonless
  Solutions?},'' \href{http://dx.doi.org/10.1007/JHEP05(2010)097}{{\em JHEP}
  {\bfseries 05} (2010) 097},
\href{http://arxiv.org/abs/0908.3402}{{\ttfamily arXiv:0908.3402 [hep-th]}}.
%%CITATION = ARXIV:0908.3402;%%.

\bibitem{Mathur:2018tib}
S.~D. Mathur and D.~Turton, ``{The fuzzball nature of two-charge black hole
  microstates},'' \href{http://dx.doi.org/10.1016/j.nuclphysb.2019.114684}{{\em
  Nucl. Phys. B} {\bfseries 945} (2019) 114684},
  \href{http://arxiv.org/abs/1811.09647}{{\ttfamily arXiv:1811.09647
  [hep-th]}}.

\bibitem{Bena:2006is}
I.~Bena, C.-W. Wang, and N.~P. Warner, ``{The foaming three-charge black
  hole},'' \href{http://dx.doi.org/10.1103/PhysRevD.75.124026}{{\em Phys. Rev.}
  {\bfseries D75} (2007) 124026},
\href{http://arxiv.org/abs/hep-th/0604110}{{\ttfamily arXiv:hep-th/0604110}}.
%%CITATION = HEP-TH/0604110;%%.

\bibitem{Bena:2005va}
I.~Bena and N.~P. Warner, ``{Bubbling supertubes and foaming black holes},''
  \href{http://dx.doi.org/10.1103/PhysRevD.74.066001}{{\em Phys. Rev.}
  {\bfseries D74} (2006) 066001},
\href{http://arxiv.org/abs/hep-th/0505166}{{\ttfamily arXiv:hep-th/0505166}}.
%%CITATION = HEP-TH/0505166;%%.

\bibitem{Berglund:2005vb}
P.~Berglund, E.~G. Gimon, and T.~S. Levi, ``{Supergravity microstates for BPS
  black holes and black rings},''
  \href{http://dx.doi.org/10.1088/1126-6708/2006/06/007}{{\em JHEP} {\bfseries
  0606} (2006) 007},
\href{http://arxiv.org/abs/hep-th/0505167}{{\ttfamily arXiv:hep-th/0505167
  [hep-th]}}.
%%CITATION = HEP-TH/0505167;%%.

\bibitem{Denef:2000nb}
F.~Denef, ``{Supergravity flows and D-brane stability},''
  \href{http://dx.doi.org/10.1088/1126-6708/2000/08/050}{{\em JHEP} {\bfseries
  0008} (2000) 050},
\href{http://arxiv.org/abs/hep-th/0005049}{{\ttfamily arXiv:hep-th/0005049
  [hep-th]}}.
%%CITATION = HEP-TH/0005049;%%.

\bibitem{Denef:2002ru}
F.~Denef, ``{Quantum quivers and Hall / hole halos},''
  \href{http://dx.doi.org/10.1088/1126-6708/2002/10/023}{{\em JHEP} {\bfseries
  0210} (2002) 023},
\href{http://arxiv.org/abs/hep-th/0206072}{{\ttfamily arXiv:hep-th/0206072
  [hep-th]}}.
%%CITATION = HEP-TH/0206072;%%.

\bibitem{Bates:2003vx}
B.~Bates and F.~Denef, ``{Exact solutions for supersymmetric stationary black
  hole composites},'' \href{http://dx.doi.org/10.1007/JHEP11(2011)127}{{\em
  JHEP} {\bfseries 1111} (2011) 127},
\href{http://arxiv.org/abs/hep-th/0304094}{{\ttfamily arXiv:hep-th/0304094
  [hep-th]}}.
%%CITATION = HEP-TH/0304094;%%.

\bibitem{Denef:2007vg}
F.~Denef and G.~W. Moore, ``{Split states, entropy enigmas, holes and halos},''
  \href{http://dx.doi.org/10.1007/JHEP11(2011)129}{{\em JHEP} {\bfseries 11}
  (2011) 129},
\href{http://arxiv.org/abs/hep-th/0702146}{{\ttfamily arXiv:hep-th/0702146
  [hep-th]}}.
%%CITATION = HEP-TH/0702146;%%.

\bibitem{Bena:2017fvm}
I.~Bena, P.~Heidmann, and P.~F. Ramirez, ``{A systematic construction of
  microstate geometries with low angular momentum},''
  \href{http://dx.doi.org/10.1007/JHEP10(2017)217}{{\em JHEP} {\bfseries 10}
  (2017) 217},
\href{http://arxiv.org/abs/1709.02812}{{\ttfamily arXiv:1709.02812 [hep-th]}}.
%%CITATION = ARXIV:1709.02812;%%.

\bibitem{Avila:2017pwi}
J.~Avila, P.~F. Ramirez, and A.~Ruiperez, ``{One Thousand and One Bubbles},''
  \href{http://dx.doi.org/10.1007/JHEP01(2018)041}{{\em JHEP} {\bfseries 01}
  (2018) 041},
\href{http://arxiv.org/abs/1709.03985}{{\ttfamily arXiv:1709.03985 [hep-th]}}.
%%CITATION = ARXIV:1709.03985;%%.

\bibitem{Rawash:2022sum}
S.~Rawash and D.~Turton, ``{Evolutionary Algorithms for Multi-Center
  Solutions},'' \href{http://dx.doi.org/10.1002/prop.202300255}{{\em Fortsch.
  Phys.} {\bfseries 72} no.~2, (2024) 2300255},
  \href{http://arxiv.org/abs/2212.08585}{{\ttfamily arXiv:2212.08585
  [hep-th]}}.

\bibitem{Balasubramanian:2006gi}
V.~Balasubramanian, E.~G. Gimon, and T.~S. Levi, ``{Four Dimensional Black Hole
  Microstates: From D-branes to Spacetime Foam},''
  \href{http://dx.doi.org/10.1088/1126-6708/2008/01/056}{{\em JHEP} {\bfseries
  0801} (2008) 056},
\href{http://arxiv.org/abs/hep-th/0606118}{{\ttfamily arXiv:hep-th/0606118
  [hep-th]}}.
%%CITATION = HEP-TH/0606118;%%.

\bibitem{deBoer:2009un}
J.~de~Boer, S.~El-Showk, I.~Messamah, and D.~Van~den Bleeken, ``{A bound on the
  entropy of supergravity?},''
  \href{http://dx.doi.org/10.1007/JHEP02(2010)062}{{\em JHEP} {\bfseries 02}
  (2010) 062},
\href{http://arxiv.org/abs/0906.0011}{{\ttfamily arXiv:0906.0011 [hep-th]}}.
%%CITATION = 0906.0011;%%.

\bibitem{Bena:2010gg}
I.~Bena, N.~Bobev, S.~Giusto, C.~Ruef, and N.~P. Warner, ``{An
  Infinite-Dimensional Family of Black-Hole Microstate Geometries},''
  \href{http://dx.doi.org/10.1007/JHEP03(2011)022,
  10.1007/JHEP04(2011)059}{{\em JHEP} {\bfseries 1103} (2011) 022},
  \href{http://arxiv.org/abs/1006.3497}{{\ttfamily arXiv:1006.3497 [hep-th]}}.

\bibitem{Shigemori:2019orj}
M.~Shigemori, ``{Counting Superstrata},''
  \href{http://dx.doi.org/10.1007/JHEP10(2019)017}{{\em JHEP} {\bfseries 10}
  (2019) 017}, \href{http://arxiv.org/abs/1907.03878}{{\ttfamily
  arXiv:1907.03878 [hep-th]}}.

\bibitem{Mayerson:2020acj}
D.~R. Mayerson and M.~Shigemori, ``{Counting D1-D5-P microstates in
  supergravity},'' \href{http://dx.doi.org/10.21468/SciPostPhys.10.1.018}{{\em
  SciPost Phys.} {\bfseries 10} no.~1, (2021) 018},
  \href{http://arxiv.org/abs/2010.04172}{{\ttfamily arXiv:2010.04172
  [hep-th]}}.

\bibitem{deLange:2015gca}
P.~de~Lange, D.~R. Mayerson, and B.~Vercnocke, ``{Structure of Six-Dimensional
  Microstate Geometries},''
  \href{http://dx.doi.org/10.1007/JHEP09(2015)075}{{\em JHEP} {\bfseries 09}
  (2015) 075},
\href{http://arxiv.org/abs/1504.07987}{{\ttfamily arXiv:1504.07987 [hep-th]}}.
%%CITATION = ARXIV:1504.07987;%%.

\bibitem{Mayerson:2020tcl}
D.~R. Mayerson, R.~A. Walker, and N.~P. Warner, ``{Microstate Geometries from
  Gauged Supergravity in Three Dimensions},''
  \href{http://dx.doi.org/10.1007/JHEP10(2020)030}{{\em JHEP} {\bfseries 10}
  (2020) 030}, \href{http://arxiv.org/abs/2004.13031}{{\ttfamily
  arXiv:2004.13031 [hep-th]}}.

\bibitem{Ganchev:2021iwy}
B.~Ganchev, A.~Houppe, and N.~P. Warner, ``{New superstrata from
  three-dimensional supergravity},''
  \href{http://dx.doi.org/10.1007/JHEP04(2022)065}{{\em JHEP} {\bfseries 04}
  (2022) 065}, \href{http://arxiv.org/abs/2110.02961}{{\ttfamily
  arXiv:2110.02961 [hep-th]}}.

\bibitem{Balasubramanian:2000rt}
V.~Balasubramanian, J.~de~Boer, E.~Keski-Vakkuri, and S.~F. Ross,
  ``{Supersymmetric conical defects: Towards a string theoretic description of
  black hole formation},''
  \href{http://dx.doi.org/10.1103/PhysRevD.64.064011}{{\em Phys. Rev.}
  {\bfseries D64} (2001) 064011},
\href{http://arxiv.org/abs/hep-th/0011217}{{\ttfamily arXiv:hep-th/0011217}}.
%%CITATION = HEP-TH/0011217;%%.

\bibitem{Maldacena:2000dr}
J.~M. Maldacena and L.~Maoz, ``{De-singularization by rotation},'' {\em JHEP}
  {\bfseries 12} (2002) 055,
\href{http://arxiv.org/abs/hep-th/0012025}{{\ttfamily arXiv:hep-th/0012025}}.
%%CITATION = HEP-TH/0012025;%%.

\bibitem{Mateos:2001qs}
D.~Mateos and P.~K. Townsend, ``{Supertubes},''
  \href{http://dx.doi.org/10.1103/PhysRevLett.87.011602}{{\em Phys. Rev. Lett.}
  {\bfseries 87} (2001) 011602},
\href{http://arxiv.org/abs/hep-th/0103030}{{\ttfamily arXiv:hep-th/0103030}}.
%%CITATION = HEP-TH/0103030;%%.

\bibitem{Emparan:2001ux}
R.~Emparan, D.~Mateos, and P.~K. Townsend, ``{Supergravity supertubes},'' {\em
  JHEP} {\bfseries 07} (2001) 011,
\href{http://arxiv.org/abs/hep-th/0106012}{{\ttfamily arXiv:hep-th/0106012}}.
%%CITATION = HEP-TH/0106012;%%.

\bibitem{Lunin:2001jy}
O.~Lunin and S.~D. Mathur, ``{AdS/CFT duality and the black hole information
  paradox},'' \href{http://dx.doi.org/10.1016/S0550-3213(01)00620-4}{{\em Nucl.
  Phys.} {\bfseries B623} (2002) 342--394},
\href{http://arxiv.org/abs/hep-th/0109154}{{\ttfamily arXiv:hep-th/0109154}}.
%%CITATION = HEP-TH/0109154;%%.

\bibitem{Lunin:2002iz}
O.~Lunin, J.~M. Maldacena, and L.~Maoz, ``{Gravity solutions for the D1-D5
  system with angular momentum},''
\href{http://arxiv.org/abs/hep-th/0212210}{{\ttfamily arXiv:hep-th/0212210}}.
%%CITATION = HEP-TH/0212210;%%.

\bibitem{Niehoff:2013kia}
B.~E. Niehoff and N.~P. Warner, ``{Doubly-Fluctuating BPS Solutions in Six
  Dimensions},'' \href{http://dx.doi.org/10.1007/JHEP10(2013)137}{{\em JHEP}
  {\bfseries 1310} (2013) 137},
\href{http://arxiv.org/abs/1303.5449}{{\ttfamily arXiv:1303.5449 [hep-th]}}.
%%CITATION = ARXIV:1303.5449;%%.

\bibitem{Guo:2024pvv}
B.~Guo, S.~D. Hampton, and N.~P. Warner, ``{Inscribing geodesic circles on the
  face of the superstratum},''
  \href{http://dx.doi.org/10.1007/JHEP05(2024)224}{{\em JHEP} {\bfseries 05}
  (2024) 224}, \href{http://arxiv.org/abs/2401.17366}{{\ttfamily
  arXiv:2401.17366 [hep-th]}}.

\bibitem{Lunin:2004uu}
O.~Lunin, ``{Adding momentum to D1-D5 system},''
  \href{http://dx.doi.org/10.1088/1126-6708/2004/04/054}{{\em JHEP} {\bfseries
  04} (2004) 054},
\href{http://arxiv.org/abs/hep-th/0404006}{{\ttfamily arXiv:hep-th/0404006}}.
%%CITATION = HEP-TH/0404006;%%.

\bibitem{Giusto:2004id}
S.~Giusto, S.~D. Mathur, and A.~Saxena, ``{Dual geometries for a set of
  3-charge microstates},''
  \href{http://dx.doi.org/10.1016/j.nuclphysb.2004.09.001}{{\em Nucl. Phys.}
  {\bfseries B701} (2004) 357--379},
\href{http://arxiv.org/abs/hep-th/0405017}{{\ttfamily arXiv:hep-th/0405017}}.
%%CITATION = HEP-TH/0405017;%%.

\bibitem{Giusto:2004ip}
S.~Giusto, S.~D. Mathur, and A.~Saxena, ``{3-charge geometries and their CFT
  duals},'' \href{http://dx.doi.org/10.1016/j.nuclphysb.2005.01.009}{{\em Nucl.
  Phys.} {\bfseries B710} (2005) 425--463},
\href{http://arxiv.org/abs/hep-th/0406103}{{\ttfamily arXiv:hep-th/0406103}}.
%%CITATION = HEP-TH/0406103;%%.

\bibitem{Giusto:2012yz}
S.~Giusto, O.~Lunin, S.~D. Mathur, and D.~Turton, ``{D1-D5-P microstates at the
  cap},'' \href{http://dx.doi.org/10.1007/JHEP02(2013)050}{{\em JHEP}
  {\bfseries 1302} (2013) 050},
\href{http://arxiv.org/abs/1211.0306}{{\ttfamily arXiv:1211.0306 [hep-th]}}.
%%CITATION = ARXIV:1211.0306;%%.

\bibitem{Lin:2022rzw}
H.~W. Lin, J.~Maldacena, L.~Rozenberg, and J.~Shan, ``{Holography for people
  with no time},'' \href{http://dx.doi.org/10.21468/SciPostPhys.14.6.150}{{\em
  SciPost Phys.} {\bfseries 14} no.~6, (2023) 150},
  \href{http://arxiv.org/abs/2207.00407}{{\ttfamily arXiv:2207.00407
  [hep-th]}}.

\bibitem{Gauntlett:2002nw}
J.~P. Gauntlett, J.~B. Gutowski, C.~M. Hull, S.~Pakis, and H.~S. Reall, ``{All
  supersymmetric solutions of minimal supergravity in five- dimensions},''
  \href{http://dx.doi.org/10.1088/0264-9381/20/21/005}{{\em Class. Quant.
  Grav.} {\bfseries 20} (2003) 4587--4634},
  \href{http://arxiv.org/abs/hep-th/0209114}{{\ttfamily arXiv:hep-th/0209114}}.

\bibitem{Gutowski:2004yv}
J.~B. Gutowski and H.~S. Reall, ``{General supersymmetric AdS(5) black
  holes},'' \href{http://dx.doi.org/10.1088/1126-6708/2004/04/048}{{\em JHEP}
  {\bfseries 04} (2004) 048},
  \href{http://arxiv.org/abs/hep-th/0401129}{{\ttfamily arXiv:hep-th/0401129}}.

\bibitem{Bena:2004de}
I.~Bena and N.~P. Warner, ``{One ring to rule them all ... and in the darkness
  bind them?},'' {\em Adv. Theor. Math. Phys.} {\bfseries 9} (2005) 667--701,
\href{http://arxiv.org/abs/hep-th/0408106}{{\ttfamily arXiv:hep-th/0408106}}.
%%CITATION = HEP-TH/0408106;%%.

\bibitem{Giusto:2012gt}
S.~Giusto and R.~Russo, ``{Adding new hair to the 3-charge black ring},''
  \href{http://dx.doi.org/10.1088/0264-9381/29/8/085006}{{\em
  Class.Quant.Grav.} {\bfseries 29} (2012) 085006},
\href{http://arxiv.org/abs/1201.2585}{{\ttfamily arXiv:1201.2585 [hep-th]}}.
%%CITATION = ARXIV:1201.2585;%%.

\bibitem{Gauntlett:2004wh}
J.~P. Gauntlett and J.~B. Gutowski, ``{Concentric black rings},''
  \href{http://dx.doi.org/10.1103/PhysRevD.71.025013}{{\em Phys. Rev. D}
  {\bfseries 71} (2005) 025013},
  \href{http://arxiv.org/abs/hep-th/0408010}{{\ttfamily arXiv:hep-th/0408010}}.

\bibitem{Bena:2005ni}
I.~Bena, P.~Kraus, and N.~P. Warner, ``{Black rings in Taub-NUT},''
  \href{http://dx.doi.org/10.1103/PhysRevD.72.084019}{{\em Phys. Rev.}
  {\bfseries D72} (2005) 084019},
\href{http://arxiv.org/abs/hep-th/0504142}{{\ttfamily arXiv:hep-th/0504142}}.
%%CITATION = HEP-TH/0504142;%%.

\bibitem{Bena:2008wt}
I.~Bena, N.~Bobev, and N.~P. Warner, ``{Spectral Flow, and the Spectrum of
  Multi-Center Solutions},''
  \href{http://dx.doi.org/10.1103/PhysRevD.77.125025}{{\em Phys. Rev.}
  {\bfseries D77} (2008) 125025},
\href{http://arxiv.org/abs/0803.1203}{{\ttfamily arXiv:0803.1203 [hep-th]}}.
%%CITATION = 0803.1203;%%.

\bibitem{Bena:2022fzf}
I.~Bena, N.~\v{C}eplak, S.~D. Hampton, A.~Houppe, D.~Toulikas, and N.~P.
  Warner, ``{Themelia: the irreducible microstructure of black holes},''
  \href{http://arxiv.org/abs/2212.06158}{{\ttfamily arXiv:2212.06158
  [hep-th]}}.

\bibitem{Bena:2008dw}
I.~Bena, N.~Bobev, C.~Ruef, and N.~P. Warner, ``{Supertubes in Bubbling
  Backgrounds: Born-Infeld Meets Supergravity},''
  \href{http://dx.doi.org/10.1088/1126-6708/2009/07/106}{{\em JHEP} {\bfseries
  07} (2009) 106},
\href{http://arxiv.org/abs/0812.2942}{{\ttfamily arXiv:0812.2942 [hep-th]}}.
%%CITATION = 0812.2942;%%.

\bibitem{Giusto:2013rxa}
S.~Giusto, L.~Martucci, M.~Petrini, and R.~Russo, ``{6D microstate geometries
  from 10D structures},''
  \href{http://dx.doi.org/10.1016/j.nuclphysb.2013.08.018}{{\em Nucl.Phys.}
  {\bfseries B876} (2013) 509--555},
\href{http://arxiv.org/abs/1306.1745}{{\ttfamily arXiv:1306.1745 [hep-th]}}.
%%CITATION = ARXIV:1306.1745;%%.

\bibitem{Gutowski:2003rg}
J.~B. Gutowski, D.~Martelli, and H.~S. Reall, ``{All supersymmetric solutions
  of minimal supergravity in six dimensions},''
  \href{http://dx.doi.org/10.1088/0264-9381/20/23/008}{{\em Class. Quant.
  Grav.} {\bfseries 20} (2003) 5049--5078},
\href{http://arxiv.org/abs/hep-th/0306235}{{\ttfamily arXiv:hep-th/0306235}}.
%%CITATION = HEP-TH/0306235;%%.

\bibitem{Cariglia:2004kk}
M.~Cariglia and O.~A.~P. Mac~Conamhna, ``{The General form of supersymmetric
  solutions of N=(1,0) U(1) and SU(2) gauged supergravities in
  six-dimensions},'' \href{http://dx.doi.org/10.1088/0264-9381/21/13/006}{{\em
  Class. Quant. Grav.} {\bfseries 21} (2004) 3171--3196},
\href{http://arxiv.org/abs/hep-th/0402055}{{\ttfamily arXiv:hep-th/0402055
  [hep-th]}}.
%%CITATION = HEP-TH/0402055;%%.

\bibitem{Bena:2011dd}
I.~Bena, S.~Giusto, M.~Shigemori, and N.~P. Warner, ``{Supersymmetric Solutions
  in Six Dimensions: A Linear Structure},''
  \href{http://dx.doi.org/10.1007/JHEP03(2012)084}{{\em JHEP} {\bfseries 1203}
  (2012) 084},
\href{http://arxiv.org/abs/1110.2781}{{\ttfamily arXiv:1110.2781 [hep-th]}}.
%%CITATION = ARXIV:1110.2781;%%.

\bibitem{Bena:2017geu}
I.~Bena, E.~Martinec, D.~Turton, and N.~P. Warner, ``{M-theory Superstrata and
  the MSW String},'' \href{http://dx.doi.org/10.1007/JHEP06(2017)137}{{\em
  JHEP} {\bfseries 06} (2017) 137},
\href{http://arxiv.org/abs/1703.10171}{{\ttfamily arXiv:1703.10171 [hep-th]}}.
%%CITATION = ARXIV:1703.10171;%%.

\bibitem{Giusto:2004kj}
S.~Giusto and S.~D. Mathur, ``{Geometry of D1-D5-P bound states},''
  \href{http://dx.doi.org/10.1016/j.nuclphysb.2005.09.037}{{\em Nucl. Phys.}
  {\bfseries B729} (2005) 203--220},
\href{http://arxiv.org/abs/hep-th/0409067}{{\ttfamily arXiv:hep-th/0409067}}.
%%CITATION = HEP-TH/0409067;%%.

\bibitem{Jejjala:2005yu}
V.~Jejjala, O.~Madden, S.~F. Ross, and G.~Titchener, ``{Non-supersymmetric
  smooth geometries and D1-D5-P bound states},''
  \href{http://dx.doi.org/10.1103/PhysRevD.71.124030}{{\em Phys. Rev.}
  {\bfseries D71} (2005) 124030},
\href{http://arxiv.org/abs/hep-th/0504181}{{\ttfamily arXiv:hep-th/0504181}}.
%%CITATION = HEP-TH/0504181;%%.

\bibitem{Heidmann:2019xrd}
P.~Heidmann, D.~R. Mayerson, R.~Walker, and N.~P. Warner, ``{Holomorphic Waves
  of Black Hole Microstructure},''
  \href{http://dx.doi.org/10.1007/JHEP02(2020)192}{{\em JHEP} {\bfseries 02}
  (2020) 192}, \href{http://arxiv.org/abs/1910.10714}{{\ttfamily
  arXiv:1910.10714 [hep-th]}}.

\bibitem{Shigemori:2020yuo}
M.~Shigemori, ``{Superstrata},''
  \href{http://dx.doi.org/10.1007/s10714-020-02698-8}{{\em Gen. Rel. Grav.}
  {\bfseries 52} no.~5, (2020) 51},
  \href{http://arxiv.org/abs/2002.01592}{{\ttfamily arXiv:2002.01592
  [hep-th]}}.

\bibitem{Martinec:2023zha}
E.~J. Martinec, ``{AdS$_{3}$ orbifolds, BTZ black holes, and holography},''
  \href{http://dx.doi.org/10.1007/JHEP10(2023)016}{{\em JHEP} {\bfseries 10}
  (2023) 016}, \href{http://arxiv.org/abs/2307.02559}{{\ttfamily
  arXiv:2307.02559 [hep-th]}}.

\bibitem{Park:2015gka}
M.~Park and M.~Shigemori, ``{Codimension-2 solutions in five-dimensional
  supergravity},'' \href{http://dx.doi.org/10.1007/JHEP10(2015)011}{{\em JHEP}
  {\bfseries 10} (2015) 011},
\href{http://arxiv.org/abs/1505.05169}{{\ttfamily arXiv:1505.05169 [hep-th]}}.
%%CITATION = ARXIV:1505.05169;%%.

\bibitem{Lunin:2001fv}
O.~Lunin and S.~D. Mathur, ``{Metric of the multiply wound rotating string},''
  \href{http://dx.doi.org/10.1016/S0550-3213(01)00321-2}{{\em Nucl. Phys.}
  {\bfseries B610} (2001) 49--76},
\href{http://arxiv.org/abs/hep-th/0105136}{{\ttfamily arXiv:hep-th/0105136}}.
%%CITATION = HEP-TH/0105136;%%.

\bibitem{deBoer:2008zn}
J.~de~Boer, S.~El-Showk, I.~Messamah, and D.~Van~den Bleeken, ``{Quantizing N=2
  Multicenter Solutions},''
  \href{http://dx.doi.org/10.1088/1126-6708/2009/05/002}{{\em JHEP} {\bfseries
  05} (2009) 002},
\href{http://arxiv.org/abs/0807.4556}{{\ttfamily arXiv:0807.4556 [hep-th]}}.
%%CITATION = 0807.4556;%%.

\bibitem{Lunin:2002fw}
O.~Lunin and S.~D. Mathur, ``{Rotating deformations of AdS(3) x S(3), the
  orbifold CFT and strings in the pp-wave limit},''
  \href{http://dx.doi.org/10.1016/S0550-3213(02)00677-6}{{\em Nucl. Phys.}
  {\bfseries B642} (2002) 91--113},
\href{http://arxiv.org/abs/hep-th/0206107}{{\ttfamily arXiv:hep-th/0206107}}.
%%CITATION = HEP-TH/0206107;%%.

\bibitem{Lunin:2002bj}
O.~Lunin, S.~D. Mathur, and A.~Saxena, ``{What is the gravity dual of a chiral
  primary?},'' \href{http://dx.doi.org/10.1016/S0550-3213(03)00081-6}{{\em
  Nucl. Phys.} {\bfseries B655} (2003) 185--217},
\href{http://arxiv.org/abs/hep-th/0211292}{{\ttfamily arXiv:hep-th/0211292}}.
%%CITATION = HEP-TH/0211292;%%.

\bibitem{Balasubramanian:2005qu}
V.~Balasubramanian, P.~Kraus, and M.~Shigemori, ``{Massless black holes and
  black rings as effective geometries of the D1-D5 system},''
  \href{http://dx.doi.org/10.1088/0264-9381/22/22/010}{{\em Class. Quant.
  Grav.} {\bfseries 22} (2005) 4803--4838},
  \href{http://arxiv.org/abs/hep-th/0508110}{{\ttfamily arXiv:hep-th/0508110}}.

\bibitem{Dabholkar:2006za}
A.~Dabholkar, N.~Iizuka, A.~Iqubal, A.~Sen, and M.~Shigemori, ``{Spinning
  strings as small black rings},''
  \href{http://dx.doi.org/10.1088/1126-6708/2007/04/017}{{\em JHEP} {\bfseries
  04} (2007) 017}, \href{http://arxiv.org/abs/hep-th/0611166}{{\ttfamily
  arXiv:hep-th/0611166}}.

\bibitem{Chakrabarty:2021sff}
B.~Chakrabarty, S.~Rawash, and D.~Turton, ``{Shockwaves in black hole
  microstate geometries},''
  \href{http://dx.doi.org/10.1007/JHEP02(2022)202}{{\em JHEP} {\bfseries 02}
  (2022) 202}, \href{http://arxiv.org/abs/2112.08378}{{\ttfamily
  arXiv:2112.08378 [hep-th]}}.

\bibitem{Ceplak:2022wri}
N.~\v{C}eplak, S.~Hampton, and N.~P. Warner, ``{Linearizing the BPS equations
  with vector and tensor multiplets},''
  \href{http://dx.doi.org/10.1007/JHEP03(2023)145}{{\em JHEP} {\bfseries 03}
  (2023) 145}, \href{http://arxiv.org/abs/2204.07170}{{\ttfamily
  arXiv:2204.07170 [hep-th]}}.

\bibitem{Tyukov:2017uig}
A.~Tyukov, R.~Walker, and N.~P. Warner, ``{Tidal Stresses and Energy Gaps in
  Microstate Geometries},''
  \href{http://dx.doi.org/10.1007/JHEP02(2018)122}{{\em JHEP} {\bfseries 02}
  (2018) 122},
\href{http://arxiv.org/abs/1710.09006}{{\ttfamily arXiv:1710.09006 [hep-th]}}.
%%CITATION = ARXIV:1710.09006;%%.

\bibitem{Bena:2020iyw}
I.~Bena, A.~Houppe, and N.~P. Warner, ``{Delaying the Inevitable: Tidal
  Disruption in Microstate Geometries},''
  \href{http://dx.doi.org/10.1007/JHEP02(2021)103}{{\em JHEP} {\bfseries 02}
  (2021) 103}, \href{http://arxiv.org/abs/2006.13939}{{\ttfamily
  arXiv:2006.13939 [hep-th]}}.

\bibitem{Bena:2018mpb}
I.~Bena, E.~J. Martinec, R.~Walker, and N.~P. Warner, ``{Early Scrambling and
  Capped BTZ Geometries},''
  \href{http://dx.doi.org/10.1007/JHEP04(2019)126}{{\em JHEP} {\bfseries 04}
  (2019) 126},
\href{http://arxiv.org/abs/1812.05110}{{\ttfamily arXiv:1812.05110 [hep-th]}}.
%%CITATION = ARXIV:1812.05110;%%.

\bibitem{Martinec:2020cml}
E.~J. Martinec and N.~P. Warner, ``{The Harder They Fall, the Bigger They
  Become: Tidal Trapping of Strings by Microstate Geometries},''
  \href{http://dx.doi.org/10.1007/JHEP04(2021)259}{{\em JHEP} {\bfseries 04}
  (2021) 259}, \href{http://arxiv.org/abs/2009.07847}{{\ttfamily
  arXiv:2009.07847 [hep-th]}}.

\bibitem{Ceplak:2021kgl}
N.~Ceplak, S.~Hampton, and Y.~Li, ``{Toroidal tidal effects in microstate
  geometries},'' \href{http://dx.doi.org/10.1007/JHEP03(2022)021}{{\em JHEP}
  {\bfseries 03} (2022) 021}, \href{http://arxiv.org/abs/2106.03841}{{\ttfamily
  arXiv:2106.03841 [hep-th]}}.

\bibitem{Chang:2025rqy}
C.-M. Chang, Y.-H. Lin, and H.~Zhang, ``{Fortuity in the D1-D5 system},''
  \href{http://arxiv.org/abs/2501.05448}{{\ttfamily arXiv:2501.05448
  [hep-th]}}.

\bibitem{Hughes:2025car}
M.~R.~R. Hughes and M.~Shigemori, ``{Fortuity within Supergravity},''
  \href{http://arxiv.org/abs/2505.14888}{{\ttfamily arXiv:2505.14888
  [hep-th]}}.

\end{thebibliography}\endgroup

\end{adjustwidth}
%%%%%%%%%%%%%%%%%%%%%%%%%%%%%%%%%%%%%%%%%%%%%%%%%%%%%

%%%%%%%%%%%%%%%%%%%%%%%%%%%%%%%%%%%%%

\end{document}